\documentclass[10pt]{article}
\usepackage[left=1.5cm,right=1.5cm]{geometry}
\usepackage[numbers,square]{natbib}
\usepackage[T1]{fontenc}
\usepackage{tikz-cd}
\usepackage[english]{babel}
\usepackage{layout}
\usepackage{amsmath}
\usepackage{amssymb,amstext, amsthm, simplewick, amsfonts}
\usepackage{framed}
\usepackage{bm}
\usepackage{amsfonts}
\usepackage{color} 
\usepackage{braket}
\usepackage{multicol} 
\usepackage{epsfig}
\usepackage{cancel}
\usepackage{caption}
\usepackage{epsf}
\usepackage{feynmf} 
\usepackage{frontespizio}
\usepackage{rotating}
\usepackage{subfigure}
\usepackage{pstricks}
\usepackage{type1ec}
\usepackage{lettrine}
\usepackage{bbold}
\usepackage{calligra}
\usepackage{tikz}
\usepackage{float}
\usepackage{subfigure}
\usepackage{slashed}
\usepackage{mathrsfs}
\usepackage{curve2e}
\usepackage{setspace}
\usepackage{indentfirst}
\usepackage{emptypage}
\usepackage{relsize}
\usepackage{mathrsfs}
\usepackage{stackengine}
\usepackage{calc}
\usepackage{hyperref}
\usepackage{comment}
\hypersetup{
	colorlinks,
	citecolor=green,
	filecolor=black,
	linkcolor=blue,
	urlcolor=black
}

\newcommand{\3}[1]{C_{
		\ifthenelse{\equal{\ThreePt}{\empty}}{#1}{
			\ifthenelse{\equal{#1}{\empty}}{\ThreePt}{\ThreePt,#1}}}}
\newcommand{\redef}[1]{{C'}_{
		\ifthenelse{\equal{\ThreePt}{\empty}}{#1}{
			\ifthenelse{\equal{#1}{\empty}}{\ThreePt}{\ThreePt,#1}}}}
\newcommand{\ren}[1]{C_{
		\ifthenelse{\equal{\ThreePt}{\empty}}{#1}{
			\ifthenelse{\equal{#1}{\empty}}{\ThreePt}{\ThreePt,#1}}}}
\newcommand{\sd}[1]{D_{
		\ifthenelse{\equal{\ThreePt}{\empty}}{#1}{
			\ifthenelse{\equal{#1}{\empty}}{\ThreePt}{\ThreePt,#1}}}}

\numberwithin{equation}{section} 
\makeatletter
\@addtoreset{equation}{section}
\newcommand{\bea}{\begin{eqnarray}}
	\newcommand{\eea}{\end{eqnarray}}
\makeatother
\newcommand{\beqa}{\begin{eqnarray}}
	\newcommand{\eeqa}{\end{eqnarray}}

\newcommand{\nn}{\nonumber}
\let\a=\alpha   \let\b=\beta      
         
        \let\m=\mu
\let\n=\nu

\def\si{\sigma}
\newcommand{\bann}{\begin{eqnarray*}}
	\newcommand{\eann}{\end{eqnarray*}}

\newcommand{\bmi}{\begin{minipage}}
	\newcommand{\emi}{\end{minipage}}

\newcommand{\la}{\langle}
\newcommand{\ra}{\rangle}

\newcommand{\be}{\begin{equation}}
	\newcommand{\ee}{\end{equation}}

\newcommand{\beq}{\begin{equation}}
	\newcommand{\eeq}{\end{equation}}
\newcommand{\secref}[1]{Section~\ref{#1}}		

\newcommand{\ThreePt}{\empty}
\usepackage{accents}
\makeatletter
\newcommand{\xLine}[2][]{\ext@arrow 0359\Rightarrowfill@{#1}{#2}}
\makeatother
\xdefinecolor{darkgreen}{RGB}{102, 204, 70}
\xdefinecolor{darkblue}{RGB}{0, 0, 153}
\usepackage{amssymb}
\usepackage{pifont}
\newcommand{\bes}{\begin{subequations}}
	\newcommand{\ees}{\end{subequations}}

\usepackage{tikz-feynman}
\usetikzlibrary{arrows}
\usetikzlibrary{snakes}
\usetikzlibrary{decorations,decorations.pathmorphing,decorations.markings}

\tikzset{graviton/.style={decorate, decoration={snake}, double}}
\tikzset{gluon/.style={decorate, decoration={coil, segment length=8, aspect=1.2, amplitude=3 }}}
\newcommand{\mel}[3]{\bra{#1}{#2}\ket{#3}}

\newcommand{\KBessel}{%
	K_0\!\left(bQ\sqrt{(\alpha_1-1)(\alpha_2-1)}\right)
	-
	K_0\!\left(bQ\sqrt{\alpha_1\alpha_2 }\right)
}

\newcommand{\Den}{\alpha_1+\alpha_2-1}

\newcommand{\Cz}{%
	C_0\!\left(
	( 1-\alpha_1-\alpha_2+\alpha_1\alpha_2)Q^2,\,
	\alpha_1\alpha_2 Q^2,\,
	Q^2,\,
	0,0,0
	\right)
}

\newcommand{\Lq}{\log\!\left(\frac{\mu^2}{Q^2}\right)}
\newcommand{\La}{\log\!\left(\frac{\mu^2}{\alpha_1\alpha_2 Q^2}\right)}
\newcommand{\Lb}{%
	\log\!\left(
	\frac{\mu^2}{
		(1-\alpha_1-\alpha_2+\alpha_1\alpha_2)Q^2
	}
	\right)
}

\begin{document}

	\begin{center}
	\vspace{1.5cm}
	\begin{center}
		\vspace{1.5cm}
		{\Large \bf The Gravitational Form Factor of the Pion in Perturbative QCD\\}
		\vspace{0.3cm} 
		{\Large \bf with a Dilaton Interaction}
		
		\vspace{0.3cm}
		
		\vspace{1cm}
		{\bf $^{1,2}$Claudio Corian\`o, $^3$Hsiang-Nan Li,  $^1$Dario Melle and $^1$Leonardo Torcellini\\}
		\vspace{1cm}
		{\it  $^1$Dipartimento di Matematica e Fisica, Universit\`{a} del Salento \\
			and INFN Sezione di Lecce,Via Arnesano 73100 Lecce, Italy\\
			National Center for HPC, Big Data and Quantum Computing\\
			$^2$CNR nanotec, Lecce\\
			$^3$Institute of Physics, Academia Sinica, Taipei, Taiwan 115, Republic of China
			}
		
		\vspace{0.5cm}

\begin{abstract}
	We investigate the pion gravitational form factors (GFFs) at intermediate and large momentum transfer within the framework of QCD factorization. Our analysis centers on the non-Abelian $TJJ$ correlator, which couples the local QCD energy-momentum tensor to two external gluon fields and explicitly encodes the perturbative effects of the trace anomaly. We demonstrate how this quantum anomaly induces a scalar, dilaton-like contribution to the hard-scattering kernel. To ensure field-theoretic consistency, a careful separation of the quark and gluon sectors is performed, accounting for the modifications introduced by gauge-fixing terms and Slavnov-Taylor identities on the off-shell gluonic structure. To obtain realistic phenomenological predictions and regulate soft-gluon endpoint divergences in the hard kernel, we implement the Sudakov resummation framework coupled with a Gaussian model for the pion's transverse-momentum-dependent wave function. We show that the resulting anomaly-induced corrections significantly modify the behavior of the pion GFFs at large momentum transfer, leaving a unique imprint on the trace sector and the $D$-term.
\end{abstract}

		\end{center}

\end{center}
\newpage

\section{Introduction}

Gravitational form factors (GFFs) provide a fundamental characterization of hadron structure, encoding the distributions of energy, momentum, pressure, shear forces, and angular momentum carried by quarks and gluons inside hadrons. Although they are defined through matrix elements of the QCD energy--momentum tensor (EMT) and therefore do not correspond to direct gravitational probes, they furnish one of the most comprehensive field-theoretic descriptions of the internal dynamics of hadrons. In particular, they offer access to the mechanical structure of strongly interacting bound states, complementing the information obtained from electromagnetic and axial form factors and extending the partonic description of hadrons beyond charge and spin observables.

Experimentally, GFFs can be accessed through hard exclusive scattering processes. A primary channel in this context is deeply virtual Compton scattering (DVCS), in which a high--energy electron scatters off a hadron by exchanging a highly virtual photon that subsequently converts into a real photon,
\begin{equation}
eN \to e^\prime N^\prime \gamma .
\label{chza:eq:DVCSprocess}
\end{equation}
Over the past decade, the central role of GFFs in elucidating hadron structure has motivated an extensive experimental program. First extractions of the proton quark~\cite{Burkert:2018bqq} and gluon~\cite{Duran:2022xag} GFFs have been achieved through DVCS measurements and exclusive \(J/\psi\) photoproduction, respectively. In comparison, the determination of pion GFFs remains more challenging, although initial phenomenological constraints on the pion quark GFFs have been obtained from high--precision data collected at the Belle experiment at KEKB~\cite{Belle:2015oin,Savinov:2013hda,Kumano:2017lhr}. Significant progress is expected in the near future from ongoing and planned experimental programs, including the upgraded \(12\;\mathrm{GeV}\) Jefferson Lab~\cite{JeffersonLabHallA:2022pnx,CLAS:2022syx} and the forthcoming Electron--Ion Collider~\cite{AbdulKhalek:2021gbh}, which will substantially extend the kinematic reach and precision of DVCS and related exclusive measurements. These developments provide strong motivation for refined theoretical analyses of GFFs, especially in the hard-scattering regime where QCD factorization and symmetry-based methods can be applied systematically. In this context, related neutral-current processes can also be investigated and further enlarge the phenomenological reach of the framework~\cite{Amore:2004ng}.

Perturbative studies of hadronic GFFs at large momentum transfer have also been developed directly within QCD factorization.  In particular, the gluon GFFs of the pion and the nucleon were analyzed in~\cite{Tong:2021ctu}, and a broader perturbative calculation of GFFs at large momentum transfer was presented in~\cite{Tong:2022zax}.  These works provide useful benchmarks for the large-\(Q^2\) behavior of quark and gluon contributions and form part of the theoretical background for the factorized treatment considered below.

DVCS occupies a special position in the kinematic landscape of strong-interaction processes. At moderate momentum transfer, it interpolates between a soft regime, where descriptions based on QCD sum rules are applicable and the reaction is largely governed by the Feynman mechanism, namely by the overlap of initial- and final-state hadronic wave functions~\cite{Coriano:1993mr}, and a genuinely hard regime at larger virtualities, where factorization theorems isolate short-distance kernels from long-distance hadronic matrix elements. This dual character makes DVCS an especially powerful probe of hadron structure across different distance scales, allowing one to connect nonperturbative physics with perturbatively calculable amplitudes in a controlled manner.

A systematic and model-independent connection between DVCS observables and EMT form factors is provided by generalized parton distributions (GPDs)~\cite{Ji:1996ek,Radyushkin:1996nd,Radyushkin:1996ru,Ji:1996nm,Collins:1996fb,Radyushkin:1997ki,Vanderhaeghen:1998uc}. GPDs parameterize nonforward light-cone matrix elements of quark and gluon operators and thus interpolate between ordinary parton distributions and elastic form factors. Their Mellin moments encode local operator matrix elements, among which the EMT plays a central role. In the nucleon case, the second Mellin moments of the unpolarized GPDs \(H^a(x,\xi,t)\) and \(E^a(x,\xi,t)\) are directly related to the EMT form factors \(A^a(t)\), \(B^a(t)\), and \(D^a(t)\),
\begin{align}
\int_{-1}^{1} dx \, x\, H^a(x,\xi,t)
&=
A^a(t)+\xi^2 D^a(t),
\\
\int_{-1}^{1} dx \, x\, E^a(x,\xi,t)
&=
B^a(t)-\xi^2 D^a(t).
\label{chza:Eq:GPD-Mellin}
\end{align}
Here \(t=q^2\) denotes the invariant momentum transfer and \(\xi\) is the longitudinal momentum asymmetry, or skewness, parameter. Physically, the GPDs \(H\) and \(E\) describe the amplitude for removing from the nucleon a parton carrying a light-cone momentum fraction \(x-\xi\) of the average nucleon momentum \(P\) and reinserting it with fraction \(x+\xi\), while the hadron absorbs the momentum transfer \(q\). Through their Mellin moments, GPDs therefore provide direct access to the GFFs and to the internal mechanical and angular-momentum structure of hadrons.

While much of the phenomenological discussion of GFFs has been developed in terms of GPDs and related exclusive observables, the perturbative analysis of EMT matrix elements at large momentum transfer raises a more specific field-theoretic question: how the EMT couples to colored constituents in QCD in the hard regime. This problem naturally leads to the study of the non-Abelian \(TJJ\) correlator, which governs the coupling of the EMT to two gauge currents, or equivalently to two external gluon fields in the background-field formulation. The importance of this correlator is twofold. First, it controls the short-distance structure entering hard exclusive amplitudes in which the EMT couples to partonic constituents. Second, it encodes the quantum breaking of scale invariance through the trace anomaly and thus provides the perturbative origin of anomaly-mediated contributions.

In the conformal limit, the trace of the EMT vanishes at the classical level in massless QCD, but acquires a nonvanishing expectation value after renormalization, proportional to the QCD beta function. In momentum space this effect is captured by a specific form factor whose spectral density satisfies a nontrivial sum rule. In suitable kinematic limits, and in particular for massless quarks, this structure reduces to an effective pole contribution that can be interpreted as the exchange of a scalar state induced by the anomaly. This effective scalar exchange, often described as dilaton-like, provides a natural bridge between the perturbative description of the trace anomaly and the factorized treatment of hadronic GFFs.

A systematic analysis of this mechanism requires a tensor decomposition of the non-Abelian \(TJJ\) correlator into transverse--traceless, longitudinal, and trace sectors. While analogous decompositions are standard in conformal field theory, their implementation in gauge theories is more subtle because gauge fixing and BRST symmetry modify the off-shell structure of the correlator. As a consequence, the quark and gluon sectors must be treated separately at one loop. The quark contribution is structurally close to the Abelian case and obeys ordinary vector Ward identities on the external gluon lines. The gluon sector, by contrast, is constrained by Slavnov--Taylor identities and contains additional local terms that are absent in a purely conformal treatment. The anomaly emerges only after these distinct sectors are combined consistently.


In this work, we investigate the role of these anomaly-induced contributions in the perturbative description of the pion GFFs at large momentum transfer. Specifically, we embed the non-Abelian $TJJ$ correlator, which corrects, at one-loop level, the vertex coupling of the local energy-momentum tensor to two external gluons, into the standard hard-scattering framework for exclusive hadronic processes. By factorizing the pion EMT matrix elements into a convolution of nonperturbative pion distribution amplitudes (DAs) and perturbatively calculable hard-scattering kernels, we isolate the dynamical effects of the trace anomaly. This approach allows us to explicitly track how the anomaly-induced, dilaton-like effective scalar exchange propagates into the physical form factors in the asymptotic, large-$Q^2$ regime. In this sense, the present analysis provides a concrete realization, in the pion case, of the general field-theoretic framework developed in~\cite{Coriano:2024qbr}, establishing a direct link between a fundamental field-theoretic anomaly and observable mechanical properties of a pseudo-Goldstone boson. The remainder of this paper is structured as follows. 

In Section~\ref{chza:aaaasec2}, we define the kinematics of the hard exclusive process and review the definition of the pion GFFs $A_\pi(t)$ and $D_\pi(t)$. After a brief review in Section~\ref{chza:sec:emt_conformal} of the QCD energy-momentum tensor,  Section~\ref{chza:seven} is dedicated to the field-theoretic structure of the off-shell non-Abelian $TJJ$ correlator at one-loop order; here, we detail the separation between the quark and gluon loop contributions and highlight how gauge-fixing terms and Slavnov-Taylor identities affect the longitudinal and trace sectors. In Section~\ref{chza:contjjtra}, we compute the hard-scattering kernels, explicitly incorporating the $TJJ$ anomaly vertex into the pion's partonic subprocesses. 
In Section~\ref{chza:phenomesec1}, we analyze the Sudakov resummation framework for the pion form factor at large momentum transfer. This resummation is essential to handle the soft-gluon divergences that arise in the endpoint regions of the pion DAs, providing a reliable and stable perturbative expansion by incorporating transverse-momentum effects. In Section~\ref{chza:phenomesec2}, we introduce a concrete phenomenological model for the pion wave function based on a Gaussian distribution in transverse momentum. This specific ansatz allows us to construct a realistic nonperturbative input and establish a complete phenomenological framework to quantify our anomaly-induced corrections and evaluate their numerical impact on the GFFs.
Finally, in Section~\ref{chza:numericalsec}, we present the analysis of our results, discussing the numerical impact of the trace anomaly on the $Q^2$ evolution of the pion form factors. The paper also includes three appendices. Appendix~\ref{chza:rules} collects 
the Feynman rules and vertex expressions employed in the diagrammatic 
evaluation of the amplitudes. Appendix~\ref{chza:ff} provides the explicit 
form factors of the off-shell non-Abelian $TJJ$ correlator 
in the transverse-traceless sector, including the full dependence on the 
loop integral coefficients for both the quark and gluon contributions. 
Appendix~\ref{chza:kernelexpb} contains the complete hard-scattering kernel 
expressions in impact-parameter space, covering both the tree-level 
contributions and the corrections induced by the $TJJ$ insertion for 
each GFF.

We find that the trace anomaly introduces a distinct behavior in the trace sector of the form factors, which significantly alters the behavior of the pion $D$-term at large momentum transfer compared to purely traditional, anomaly-free hard-scattering predictions. In particular, the effective scalar exchange induced by the anomaly leaves a unique imprint on the internal pressure and shear force distributions within the pion, demonstrating that quantum scale violation plays an indispensable structural role even in the deeply perturbative, high-energy domain.

\section{The pion form factor in the limit of large momentum transfer}\label{chza:aaaasec2}

A systematic and controlled description of hadronic form factors within QCD becomes possible in the asymptotic regime of large spacelike momentum transfer,
\begin{equation}
Q^2 = -q^2 \to \infty, \qquad q^\mu = P_2^\mu - P_1^\mu,
\end{equation}
where short-distance dynamics dominates and long-distance nonperturbative effects can be factorized into universal hadronic quantities. In this regime, the interaction resolves the hadron at partonic distances of order $1/Q$, allowing for a perturbative treatment of the hard scattering subprocess.

As a prototypical example, we consider the GFFs of the pion, defined through the matrix element of the QCD energy-momentum tensor $ T^{\mu\nu}$,
\begin{equation}\label{chza:eq:GFF_definition_improved}
\langle \pi(P_2)| {T}^{\mu\nu} | \pi(P_1) \rangle
= 2 P^\mu P^\nu A^\pi(q^2)
+ \frac{1}{2} \left( q^\mu q^\nu - g^{\mu\nu}q^2 \right) D^\pi(q^2),
\end{equation}
where
\begin{equation}
P = \frac{P_1 + P_2}{2},
\end{equation}
and $A^\pi(q^2)$ and $D^\pi(q^2)$ encode, respectively, the momentum and mechanical (pressure and shear) distributions inside the pion.

To make the scaling properties manifest, it is convenient to work in the Breit frame, where the momentum transfer is purely spatial. Neglecting the pion mass in the large-$Q^2$ limit, one may choose the kinematics
\begin{equation}
P_1^\mu = \left( \frac{Q}{2}, 0, 0, -\frac{Q}{2} \right), \qquad 
P_2^\mu = \left( \frac{Q}{2}, 0, 0, \frac{Q}{2} \right), \qquad 
q^\mu = (0, 0, 0, Q),
\end{equation}
so that $P_1^2 = P_2^2 \simeq 0$. Introducing light-cone vectors,
\begin{equation}
n^\mu = \frac{1}{\sqrt{2}}(1,0,0,-1), \qquad 
\bar n^\mu = \frac{1}{\sqrt{2}}(1,0,0,1), \qquad 
n^2 = \bar n^2 = 0, \qquad n \cdot \bar n = 1,
\end{equation}
the external momenta can be written as
\begin{equation}
P_1^\mu = \frac{Q}{2} n^\mu, \qquad 
P_2^\mu = \frac{Q}{2} \bar n^\mu,
\label{chza:eq:lightcone_p_improved}
\end{equation}
making explicit the collinear nature of the incoming and outgoing states.

In this frame, the pion is probed in a highly boosted configuration, and its dominant Fock component consists of a collinear quark--antiquark pair carrying longitudinal momentum fractions of order unity. Denoting by $k_1, k_2$ ($k_1', k_2'$) the parton momenta in the initial (final) state, one has the scaling behavior
\begin{equation}
k_i^\mu \sim \alpha_i P_1^\mu + k_{i\perp}^\mu, \qquad 
k_i^{\prime \mu} \sim \bar\alpha_i P_2^\mu + k_{i\perp}^\mu,
\end{equation}
with
\begin{equation}
\alpha_i \sim \mathcal{O}(1), \qquad 
|k_\perp| \ll Q.
\end{equation}
The transverse momenta remain of order $\Lambda_{\mathrm{QCD}}$ and are therefore power suppressed in the hard scattering subprocess.

This hierarchy of scales underlies the factorization of the matrix element into a short-distance kernel and universal nonperturbative DAs. At asymptotically large $Q^2$, higher Fock components involving additional soft gluons or sea quarks are suppressed by powers of $\Lambda_{\mathrm{QCD}}/Q$, and the dominant contribution arises from the valence quark--antiquark configuration.

Physically, the insertion of the energy-momentum tensor corresponds to a localized interaction occurring over a time scale $\Delta t \sim 1/Q$. During this short interval, the partons behave as quasi-free and collinear, and the graviton probes a frozen configuration of the pion. This justifies the use of perturbative QCD for the hard subprocess, while the long-distance structure is encoded in the pion DA.

Within this framework, the GFF can be computed using the hard-scattering mechanism. The momentum transfer injected by the graviton must be redistributed among the constituents through the exchange of a hard gluon carrying virtuality of order $Q^2$, ensuring the recombination of the partons into the final pion state. This mechanism is directly analogous to the perturbative treatment of the electromagnetic form factor, with the crucial difference that the energy-momentum tensor probes the full energy flow inside the hadron.

Neglecting transverse momenta at leading power, one adopts the collinear approximation
\begin{equation}
k_1 = \alpha_1 P_1, \qquad k'_1 = \bar{\alpha}_1 P_2, \qquad
k_2 = \alpha_2 P_1, \qquad k'_2 = \bar{\alpha}_2 P_2,
\label{chza:eq:collinear_approx_improved}
\end{equation}
with $\bar{\alpha}_i = 1 - \alpha_i$. The pion states can then be expanded in terms of valence Fock components as
\begin{align}
\label{chza:wf}
|\pi(P_1)\rangle &= \int_0^1 d\alpha_1\, \tilde{\phi}_\pi(\alpha_1)\,
| u(\alpha_1 P_1)\, \bar{d}(\bar{\alpha}_1 P_1) \rangle, \\
\langle \pi(P_2)| &= \int_0^1 d\alpha_2\, \tilde{\phi}_\pi(\alpha_2)\,
\langle u(\alpha_2 P_2)\, \bar{d}(\bar{\alpha}_2 P_2) |,
\end{align}
The light-cone momentum fractions \(\alpha_{1,2}\in[0,1]\) label the momentum carried by the quark in each pion DA.
where $\tilde{\phi}_\pi(\alpha)$ is the leading-twist pion DA, normalized as
\begin{equation}
\int_0^1 d\alpha\, \tilde{\phi}_\pi(\alpha) = 1.
\end{equation}

The matrix element can then be expressed, within collinear factorization, as a convolution of the DAs with a perturbatively calculable hard kernel,
\begin{equation}
\langle \pi(P_2)| {T}^{\mu\nu} | \pi(P_1) \rangle
=  \frac{i\,f_\pi^2 C_F g_s^2}{N_C}\int_0^1 d\alpha_1 \int_0^1 d\alpha_2 \,
\tilde{\phi}_\pi(\alpha_1)\,
T_H^{\mu\nu}(\alpha_1,\alpha_2,Q)\,
\tilde{\phi}_\pi(\alpha_2).
\label{chza:eq:factorization_master}
\end{equation}

At leading order in $\alpha_s(Q)$, the hard kernel $T_H^{\mu\nu}$ is determined by the tree-level diagrams involving a single hard gluon exchange. The perturbative expansion is controlled by the running coupling evaluated at the scale $Q$, which is small in the asymptotic regime.

The explicit computation proceeds by inserting the QCD interaction vertices and performing Wick contractions of quark and gluon fields. The resulting amplitude can be reduced to a set of tensor structures consistent with Lorentz symmetry,
\begin{equation}
T_H^{\mu\nu} =
F_1 P_1^\mu P_2^\nu
+ F_2 P_2^\mu P_1^\nu
+ F_3 P_1^\mu P_1^\nu
+ F_4 P_2^\mu P_2^\nu
+ F_5 g^{\mu\nu},
\end{equation}
where the scalar functions $F_i(\alpha_1,\alpha_2,Q^2)$ are obtained from the evaluation of the diagrams.

Matching this decomposition onto Eq.~\eqref{chza:eq:GFF_definition_improved} allows one to extract the GFFs as convolution integrals over the pion DA,
\begin{align}
A_\pi(q^2) &= \int_0^1 d\alpha_1 \int_0^1 d\alpha_2\,
\tilde{\phi}_\pi(\alpha_1)\,
\mathcal{A}(\alpha_1,\alpha_2,Q^2)\,
\tilde{\phi}_\pi(\alpha_2), \\
D_\pi(q^2) &= \int_0^1 d\alpha_1 \int_0^1 d\alpha_2\,
\tilde{\phi}_\pi(\alpha_1)\,
\mathcal{D}(\alpha_1,\alpha_2,Q^2)\,
\tilde{\phi}_\pi(\alpha_2),
\end{align}
where $\mathcal{A}$ and $\mathcal{D}$ are determined by appropriate linear combinations of the $F_i$.

This result establishes the leading-power factorization of the pion GFFs at large momentum transfer, showing explicitly that their asymptotic behavior is governed by short-distance QCD dynamics convoluted with universal nonperturbative DAs.

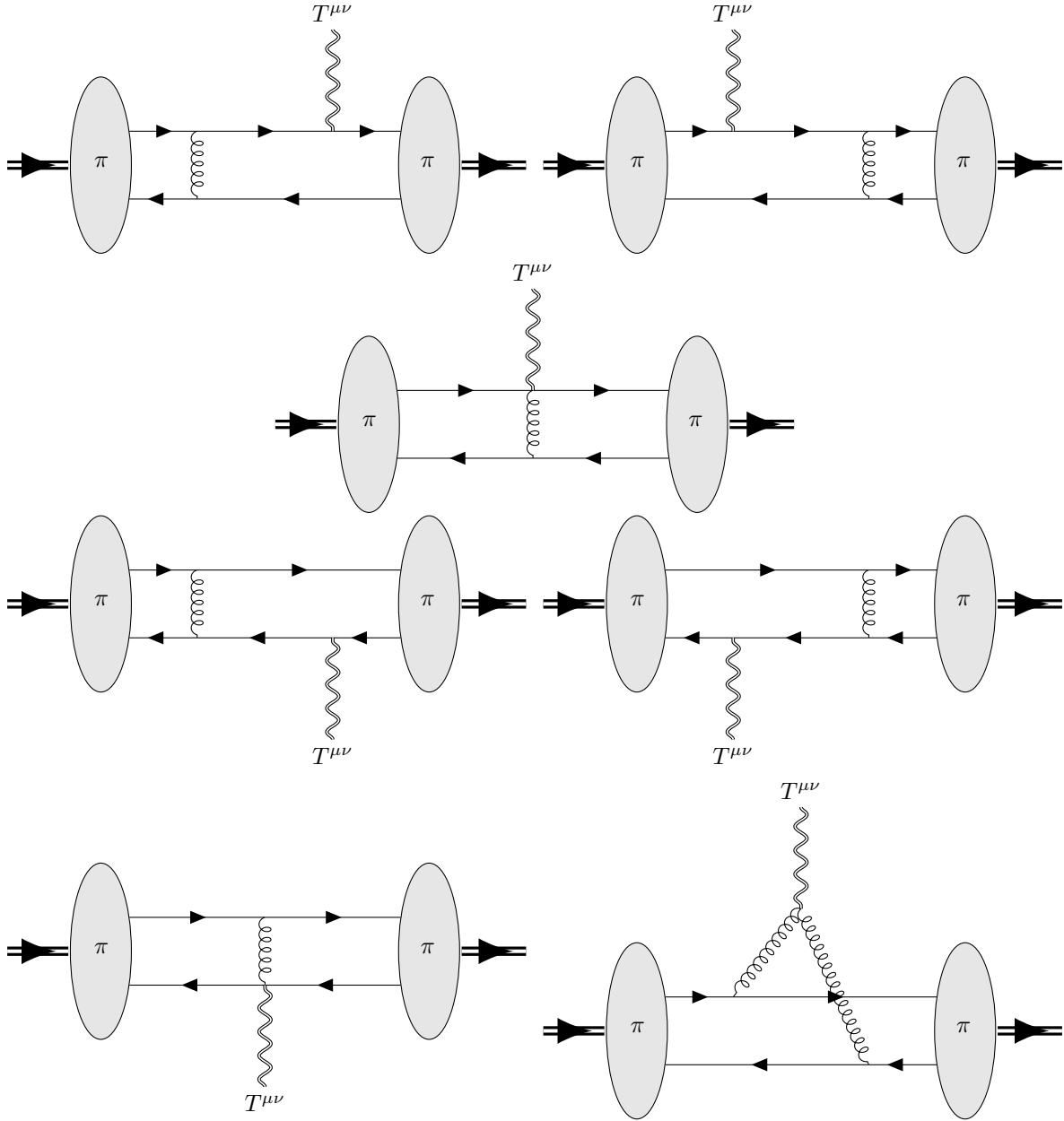
\begin{figure}
	\centering
	\begin{tikzpicture}
		\begin{feynman}
			\vertex (i1);
			\vertex[right=1cm of i1] (a1);
			\vertex[right=2cm of a1] (b1);
			\vertex[right=3cm of a1] (o1) ;
			\vertex[below=1cm of i1] (i2) ;
			\vertex[right=1cm of i2] (a2);
			\vertex[right=3cm of a2] (o2) ;
			
			\vertex[above=1cm of a1] (t2);
			\vertex[above=2cm of a2] (t3);
			
			\vertex[above=1.5cm of b1] (i3) {$T^{\mu\nu}$};
			
			\diagram* { 
				(i1)  -- [fermion] (a1) -- [fermion] (b1)   --[fermion] (o1),
				
				(i2)  -- [anti fermion] (a2)  --[anti fermion] (o2),

				(i3)  -- [graviton] (b1),
				(a1)-- [gluon] (a2)
			};
			
			\node[] at (-0.35,-0.45) {$\pi\ $};
			\node[] at (4.46,-0.45) {$\pi\ $};

			\filldraw[fill=black, fill opacity=0.1] (-0.42,-0.5) ellipse (.45cm and 1.3cm);
			\filldraw[fill=black, fill opacity=0.1] (4.42 ,-0.5) ellipse (.45cm and 1.3cm);
			
			\draw[double, double distance=1.5,line width=1.2, postaction={decorate, decoration={
					markings,
					mark=at position 0.8 with  {\arrow{ latex} } } } ] (-1.8,-0.5) -- node [above] {} (-0.9,-0.5);
			
			\draw[double, double distance=1.5,line width=1.2, postaction={decorate, decoration={
					markings,
					mark=at position 0.8 with  {\arrow{ latex} } } } ] (4.9,-0.5) -- node [above] {} (5.85,-0.5);
		\end{feynman}
	\end{tikzpicture}
	\begin{tikzpicture}
		\begin{feynman}
			\vertex (i1);
			\vertex[right=1cm of i1] (a1);
			\vertex[right=2cm of a1] (b1);
			\vertex[right=3cm of a1] (o1) ;
			\vertex[below=1cm of i1] (i2) ;
			\vertex[right=3cm of i2] (a2);
			\vertex[right=1cm of a2] (o2) ;
			
			\vertex[above=1cm of a1] (t2);
			\vertex[above=2cm of a2] (t3);
			
			\vertex[above=1.5cm of a1] (i3) {$T^{\mu\nu}$};
			
			\diagram* { 
				(i1)  -- [fermion] (a1) -- [fermion] (b1)   --[fermion] (o1),
				
				(i2)  -- [anti fermion] (a2)  --[anti fermion] (o2),

				(i3)  -- [graviton] (a1),
				(b1)-- [gluon] (a2)
			};
			
			\node[] at (-0.35,-0.45) {$\pi\ $};
			\node[] at (4.46,-0.45) {$\pi\ $};

			\filldraw[fill=black, fill opacity=0.1] (-0.42,-0.5) ellipse (.45cm and 1.3cm);
			\filldraw[fill=black, fill opacity=0.1] (4.42 ,-0.5) ellipse (.45cm and 1.3cm);
			
			\draw[double, double distance=1.5,line width=1.2, postaction={decorate, decoration={
					markings,
					mark=at position 0.8 with  {\arrow{ latex} } } } ] (-1.8,-0.5) -- node [above] {} (-0.9,-0.5);
			
			\draw[double, double distance=1.5,line width=1.2, postaction={decorate, decoration={
					markings,
					mark=at position 0.8 with  {\arrow{ latex} } } } ] (4.9,-0.5) -- node [above] {} (5.85,-0.5);
		\end{feynman}
	\end{tikzpicture}
	\begin{tikzpicture}
		\begin{feynman}
			\vertex (i1);
			\vertex[right=2cm of i1] (a1);
			\vertex[right=2cm of a1] (b1);
			\vertex[right=3cm of a1] (o1) ;
			\vertex[below=1cm of i1] (i2) ;
			\vertex[right=2cm of i2] (a2);
			\vertex[right=2cm of a2] (o2) ;
			
			\vertex[above=1cm of a1] (t2);
			\vertex[above=2cm of a2] (t3);
			
			\vertex[above=1.5cm of a1] (i3) {$T^{\mu\nu}$};
			
			\diagram* { 
				(i1)  -- [fermion] (a1) -- [fermion] (b1)  ,
				
				(i2)  -- [anti fermion] (a2)  --[anti fermion] (o2),

				(i3)  -- [graviton] (a1),
				(a1)-- [gluon] (a2)
			};
			
			\node[] at (-0.35,-0.45) {$\pi\ $};
			\node[] at (4.46,-0.45) {$\pi\ $};

			\filldraw[fill=black, fill opacity=0.1] (-0.42,-0.5) ellipse (.45cm and 1.3cm);
			\filldraw[fill=black, fill opacity=0.1] (4.42 ,-0.5) ellipse (.45cm and 1.3cm);
			
			\draw[double, double distance=1.5,line width=1.2, postaction={decorate, decoration={
					markings,
					mark=at position 0.8 with  {\arrow{ latex} } } } ] (-1.8,-0.5) -- node [above] {} (-0.9,-0.5);
			
			\draw[double, double distance=1.5,line width=1.2, postaction={decorate, decoration={
					markings,
					mark=at position 0.8 with  {\arrow{ latex} } } } ] (4.9,-0.5) -- node [above] {} (5.85,-0.5);
		\end{feynman}
	\end{tikzpicture}

	\begin{tikzpicture}
		\begin{feynman}
			\vertex (i1);
			\vertex[right=1cm of i1] (a1);
			\vertex[right=1cm of a2] (b2);
			\vertex[right=3cm of a1] (o1) ;
			\vertex[below=1cm of i1] (i2) ;
			\vertex[right=1cm of i2] (a2);
			\vertex[right=3cm of a2] (o2) ;
			
			\vertex[above=1cm of a1] (t2);
			\vertex[above=2cm of a2] (t3);
			
			\vertex[below=1.5cm of b2] (i3) {$T^{\mu\nu}$};
			
			\diagram* { 
				(i1)  -- [fermion] (a1)    --[fermion] (o1),
				
				(i2)  -- [anti fermion] (a2) -- [anti fermion] (b2) --[anti fermion] (o2),

				(i3)  -- [graviton] (b2),
				(a1)-- [gluon] (a2)
			};
			
			\node[] at (-0.35,-0.45) {$\pi\ $};
			\node[] at (4.46,-0.45) {$\pi\ $};

			\filldraw[fill=black, fill opacity=0.1] (-0.42,-0.5) ellipse (.45cm and 1.3cm);
			\filldraw[fill=black, fill opacity=0.1] (4.42 ,-0.5) ellipse (.45cm and 1.3cm);
			
			\draw[double, double distance=1.5,line width=1.2, postaction={decorate, decoration={
					markings,
					mark=at position 0.8 with  {\arrow{ latex} } } } ] (-1.8,-0.5) -- node [above] {} (-0.9,-0.5);
			
			\draw[double, double distance=1.5,line width=1.2, postaction={decorate, decoration={
					markings,
					mark=at position 0.8 with  {\arrow{ latex} } } } ] (4.9,-0.5) -- node [above] {} (5.85,-0.5);
		\end{feynman}
	\end{tikzpicture}
	\begin{tikzpicture}
		\begin{feynman}
			\vertex (i1);
			\vertex[right=3cm of i1] (a1);
			\vertex[right=2cm of a2] (b2);
			\vertex[right=1cm of a1] (o1) ;
			\vertex[below=1cm of i1] (i2) ;
			\vertex[right=1cm of i2] (a2);
			\vertex[right=3cm of a2] (o2) ;
			
			\vertex[above=1cm of a1] (t2);
			\vertex[above=2cm of a2] (t3);
			
			\vertex[below=1.5cm of a2] (i3) {$T^{\mu\nu}$};
			
			\diagram* { 
				(i1)  -- [fermion] (a1)    --[fermion] (o1),
				
				(i2)  -- [anti fermion] (a2) -- [anti fermion] (b2) --[anti fermion] (o2),

				(i3)  -- [graviton] (a2),
				(a1)-- [gluon] (b2)
			};
			
			\node[] at (-0.35,-0.45) {$\pi\ $};
			\node[] at (4.46,-0.45) {$\pi\ $};

			\filldraw[fill=black, fill opacity=0.1] (-0.42,-0.5) ellipse (.45cm and 1.3cm);
			\filldraw[fill=black, fill opacity=0.1] (4.42 ,-0.5) ellipse (.45cm and 1.3cm);
			
			\draw[double, double distance=1.5,line width=1.2, postaction={decorate, decoration={
					markings,
					mark=at position 0.8 with  {\arrow{ latex} } } } ] (-1.8,-0.5) -- node [above] {} (-0.9,-0.5);
			
			\draw[double, double distance=1.5,line width=1.2, postaction={decorate, decoration={
					markings,
					mark=at position 0.8 with  {\arrow{ latex} } } } ] (4.9,-0.5) -- node [above] {} (5.85,-0.5);
		\end{feynman}
	\end{tikzpicture}
	\begin{tikzpicture}
		\begin{feynman}
			\vertex (i1);
			\vertex[right=2cm of i1] (a1);
			\vertex[right=2cm of a1] (b1);
			\vertex[right=3cm of a1] (o1) ;
			\vertex[below=1cm of i1] (i2) ;
			\vertex[right=2cm of i2] (a2);
			\vertex[right=2cm of a2] (o2) ;
			
			\vertex[above=1cm of a1] (t2);
			\vertex[above=2cm of a2] (t3);
			
			\vertex[below=1.5cm of a2] (i3) {$T^{\mu\nu}$};
			
			\diagram* { 
				(i1)  -- [fermion] (a1) -- [fermion] (b1)  ,
				
				(i2)  -- [anti fermion] (a2)  --[anti fermion] (o2),

				(i3)  -- [graviton] (a2),
				(a1)-- [gluon] (a2)
			};
			
			\node[] at (-0.35,-0.45) {$\pi\ $};
			\node[] at (4.46,-0.45) {$\pi\ $};

			\filldraw[fill=black, fill opacity=0.1] (-0.42,-0.5) ellipse (.45cm and 1.3cm);
			\filldraw[fill=black, fill opacity=0.1] (4.42 ,-0.5) ellipse (.45cm and 1.3cm);
			
			\draw[double, double distance=1.5,line width=1.2, postaction={decorate, decoration={
					markings,
					mark=at position 0.8 with  {\arrow{ latex} } } } ] (-1.8,-0.5) -- node [above] {} (-0.9,-0.5);
			
			\draw[double, double distance=1.5,line width=1.2, postaction={decorate, decoration={
					markings,
					mark=at position 0.8 with  {\arrow{ latex} } } } ] (4.9,-0.5) -- node [above] {} (5.85,-0.5);
		\end{feynman}
	\end{tikzpicture}
	\begin{tikzpicture}
		\begin{feynman}
			\vertex (i1);
			\vertex[right=1cm of i1] (a1);
			\vertex[right=3cm of a1] (o1) ;
			\vertex[below=1cm of i1] (i2) ;
			\vertex[right=3cm of i2] (a2);
			\vertex[right=1cm of a2] (o2) ;
			
			\vertex[above=1cm of a1] (t2);
			\vertex[above=2cm of a2] (t3);
			\vertex[] (t1) at (2,1.3) ;
			\vertex[above=1.5cm of t1] (i3) {$T^{\mu\nu}$};
			
			\diagram* { 
				(i1)  -- [fermion] (a1)   --[fermion] (o1),
				
				(i2)  -- [anti fermion] (a2)  --[anti fermion] (o2),

				(i3)  -- [graviton] (t1),
				(t1) -- [gluon] (a1) ,
				(t1) -- [gluon] (a2),
			};
			
			\node[] at (-0.35,-0.45) {$\pi\ $};
			\node[] at (4.46,-0.45) {$\pi\ $};

			\filldraw[fill=black, fill opacity=0.1] (-0.42,-0.5) ellipse (.45cm and 1.3cm);
			\filldraw[fill=black, fill opacity=0.1] (4.42 ,-0.5) ellipse (.45cm and 1.3cm);
			
			\draw[double, double distance=1.5,line width=1.2, postaction={decorate, decoration={
					markings,
					mark=at position 0.8 with  {\arrow{ latex} } } } ] (-1.8,-0.5) -- node [above] {} (-0.9,-0.5);
			
			\draw[double, double distance=1.5,line width=1.2, postaction={decorate, decoration={
					markings,
					mark=at position 0.8 with  {\arrow{ latex} } } } ] (4.9,-0.5) -- node [above] {} (5.85,-0.5);
		\end{feynman}
	\end{tikzpicture}
	\caption{Diagrams of the hard scattering contribution to the pion GFF corresponding to the order $\mathcal{O}(\alpha_s)$.}
 \label{chza:GFFdiagramsTree}
\end{figure}
To compute the hadronic matrix element, we insert the energy-momentum tensor and the gluon exchanges at second order in $g_s$, therefore
\begin{align}
	T^{\mu\nu} &= \int d^4x\, d^4y\, \langle \pi(P_2) |
	T \left\{
	ig_s\, \bar{u}(x) \slashed{A}^a(x) \frac{\lambda^a}{2} u(x)\, 
	\hat{T}^{\mu\nu}(0)\,
	ig_s\, \bar{d}(y) \slashed{A}^b(y) \frac{\lambda^b}{2} d(y)
	\right\}
	| \pi(P_1) \rangle,
\end{align}
where $\hat{T}^{\mu\nu}(0)$ is the local energy-momentum tensor insertion point, and the gluon fields mediate the hard scattering between quarks.

To proceed with the evaluation of the hadronic matrix element, we perform Wick contractions of the gluon and quark fields. 
As for the $u$-quark fields appearing in the matrix element involving the energy-momentum tensor $T^{\mu\nu}$, there are two types of contractions corresponding to the two hard-scattering diagrams involving
\begin{align}
	\langle 0 | T \{ u^j_\alpha(x) \bar{u}^i_\beta(0) \} | 0 \rangle &\\
	\langle 0 | T \{ u^j_\alpha(0) \bar{u}^i_\beta(x) \} | 0 \rangle &
	\label{chza:eq:quark_props}
\end{align}
corresponding respectively to diagrams in Fig. \ref{chza:GFFdiagramsTree}, and similarly for the $d$ quark. We can insert the valence Fock states of the pion in the collinear approximation, using \eqref{chza:wf},
and evaluating matrix elements of field operators on the pion state, we obtain
\begin{align}
	\langle 0 | \bar{d}_n(y)\, u_l(0) | \pi(P_1) \rangle 
	&= \frac{f_\pi}{4} \int_0^1 d\alpha_1\, e^{-i\bar{\alpha}_1 P_1 \cdot y}\, \tilde{\phi}_\pi(\alpha_1)\, (\slashed{P}_1 \gamma_5)_{l n} , \\
	\langle \pi(P_2)| \bar{u}_i(x)\, d_m(y) | 0 \rangle 
	&= \frac{f_\pi}{4} \int_0^1 d\alpha_2\, e^{i\alpha_2 P_2 \cdot x + i\bar{\alpha}_2 P_2 \cdot y}\, \tilde{\phi}_\pi(\alpha_2)\, (\slashed{P}_2 \gamma_5)_{i m},
\end{align}
here \(f_\pi\) is the pion decay constant.

These expressions originate from the Fourier transform of the light-cone expansion of quark fields, with momenta \( \alpha_i P_i \) and \( \bar{\alpha}_i P_i \), and reproduce the correct Dirac structure for a pseudoscalar meson. Higher-order terms in $y^\mu$ are neglected because the typical distance between quarks is $\mathcal{O}(1/Q)$ at large $Q$.

These results allow us to carry out the full diagrammatic calculation of the GFFs $A^\pi(q^2)$  and  $ D^\pi(q^2)$ within the collinear factorization framework at the tree level. 
The evaluation of the diagrams in Fig. \ref{chza:GFFdiagramsTree} is straightforward once the previously introduced identities are employed. In momentum space, the result can be expressed as
	\begin{equation}\label{kerfact}
	\mel{\pi(P_2)}{T^{\mu\nu}}{\pi(P_1)} = \frac{i\,f_\pi^2 C_F g_s^2}{N_C}\int_0^1 d\alpha_1 \int_0^1 d\alpha_2 \, 
	\tilde{\phi}_\pi(\alpha_1)\, K_\pi^{\mu\nu}(\alpha_1,\alpha_2)\, \tilde{\phi}_\pi(\alpha_2),
\end{equation}
with \(C_F=(N_C^2-1)/(2N_C)\).
The kernel $K_\pi^{\mu\nu}(\alpha_1,\alpha_2)$ can be decomposed in terms of form factors as
\begin{equation}\label{decompkpi}
	K_\pi^{\mu\nu}(\alpha_1,\alpha_2) = F_1\, P_1^{\mu } P_2^{\nu } 
	+ F_2\, P_2^{\mu } P_1^{\nu } 
	+ F_3\, P_1^{\mu } P_1^{\nu } 
	+ F_4\, P_2^{\mu } P_2^{\nu } 
	+ F_5\, g^{\mu\nu},
\end{equation}
with
\begin{equation}
	F_1=F_2=-4\frac{{\alpha_1} (2 \alpha_2-1)-\alpha_2+1}{(\alpha_1-1) \alpha_1 (\alpha_2-1) \alpha_2 \,q^2},
\end{equation}

\begin{equation}
	F_3=F_4=-2\frac{{\alpha_1} (2 \alpha_2-1)-\alpha_2+1}{(\alpha_1-1) \alpha_1 (\alpha_2-1) \alpha_2 \,q^2},
\end{equation}
\begin{equation}
	F_5=\frac{{\alpha_1} (2 \alpha_2-1)-\alpha_2+1}{(\alpha_1-1) \alpha_1 (\alpha_2-1) \alpha_2}.
\end{equation}

The direct computation of the form factors $F_i$ immediately reveals the symmetry under the exchange $\mu \leftrightarrow \nu$ as well as $P_1 \leftrightarrow P_2$, which is associated with the conditions
\begin{equation}
	F_1 = F_2 \ \ \ \ \ \ \ \ F_3 = F_4(\alpha_1\leftrightarrow\alpha_2).
\end{equation}
It also reveals that in the limit of large momentum transfer ($Q^2 \gg m_\pi^2$), which corresponds to the chiral (massless pion) limit, the calculation exhibits explicit gauge invariance at the amplitude level. In this regime, the unphysical gauge-dependent terms proportional to the longitudinal components of the gauge propagators are naturally canceled out. This tree-level gauge independence ensures a robust baseline for the factorization framework before radiative corrections are taken into account.

Finally, the relations connecting the form factors $F_i$ with the GFFs $A_\pi$ and $D_\pi$ are
\begin{equation}\label{chza:atree}
	A_\pi= \frac{f_\pi^2 C_F g_s^2}{N_C}\int_0^1 d\alpha_1 \int_0^1 d\alpha_2\, \tilde{\phi}_\pi(\alpha_1)\,
	(F_1+F_3)\,
	\tilde{\phi}_\pi(\alpha_2),
\end{equation}
\begin{equation}\label{chza:dtree}
	D_\pi=-\frac{f_\pi^2 C_F g_s^2}{N_C}\int_0^1 d\alpha_1 \int_0^1 d\alpha_2\, \tilde{\phi}_\pi(\alpha_1)\,
	\frac{2 F_5}{q^2}\,
	\tilde{\phi}_\pi(\alpha_2).
\end{equation}

One important question concerns the convergence of the factorization formula already at leading order. In performing the integrals over the DAs in the full interval $0 < \alpha_{1,2} < 1$, we tacitly assume that, at the endpoint $\alpha_i \to 1$, the function $\tilde{\phi}_\pi(\alpha)$ vanishes at least linearly in $\bar{\alpha} = 1 - \alpha$, ensuring the finiteness of the integrals. In fact, the DA at finite factorization scale can be written as a Gegenbauer expansion,
\begin{equation}
	\tilde\phi_{\pi}(\a,\mu) = 6\a(1-\a) \left[
	1 + \sum_{n=2,4,\dots} a_n(\mu)\, C^{3/2}_n(2\a-1)
	\right],
\end{equation}
with the Gegenbauer moments
\begin{equation}
	a_n(\mu)
	=
	a_n(\mu_0)
	\left(
	\frac{\alpha_s(\mu)}{\alpha_s(\mu_0)}
	\right)^{\gamma_n},
\end{equation}
where the anomalous dimensions $\gamma_n$ are
\begin{equation}
	\gamma_n
	=C_F	\left[	-3+ 4 \sum_{j=1}^{n+1} \frac{1}{j}-\frac{2}{(n+1)(n+2)}\right].
\end{equation}
The running coupling 
\begin{equation}\label{chza:evolalph}
	\alpha_s(\mu) =\frac{4\pi}{\beta_0\log\frac{\mu^2}{\Lambda_{\rm QCD}^2}},
\end{equation}  
with the QCD scale $\Lambda_{\rm QCD}=0.25$ GeV and the QCD beta function at one-loop level
\begin{equation}\label{chza:eq:beta0def}
	\beta_0=\frac{11C_A-2n_f}{3},
\end{equation}
where \(C_A=N_C=3\) and \(n_f\) is the number of active quark flavors.
Eq. \eqref{chza:evolalph} is evaluated at the largest hard scale in the convolution, and reduces to
\begin{equation}
	\tilde\phi^{\text{as}}_{\pi}(\a) = 6\a(1-\a),
\end{equation}
at asymptotically large scales. Therefore, the convergence of the integral in Eqs \eqref{chza:atree} and \eqref{chza:dtree} is  straightforward.

However, if we include power-suppressed corrections, for example effects of order $\mathcal{O}(k_\perp/Q^2)$ that were neglected in the collinear approximation, we encounter divergent integrals in the factorization approach. To regularize these divergences, various extensions to the factorization framework have been proposed, such as incorporating the transverse-momentum dependence of the pion DAs
\begin{equation}
	\tilde{\phi}_\pi(\alpha) \rightarrow \tilde{\phi}_\pi(\alpha, {k}_\perp).
\end{equation}

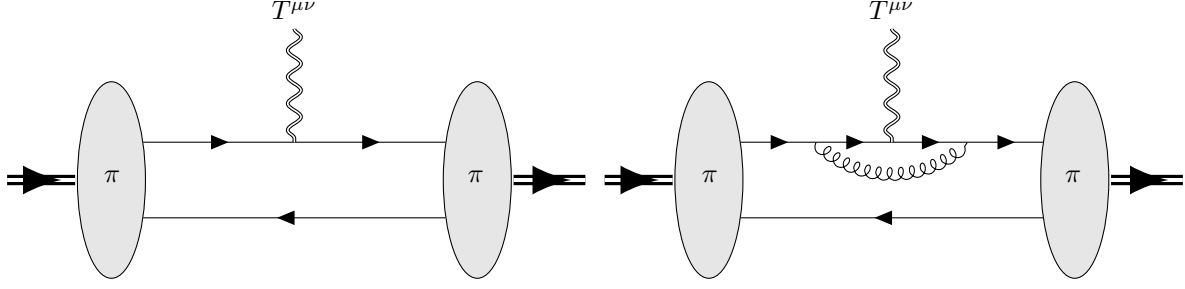
\begin{figure}
	\centering
	\begin{tikzpicture}
		\begin{feynman}
			\vertex (i1);
			\vertex[right=1cm of i1] (ax2);
			\vertex[right=3cm of i1] (ax3);
			\vertex[right=2cm of i1] (a1);
			\vertex[right=2cm of a1] (b1);
			\vertex[right=3cm of a1] (o1) ;
			\vertex[below=1cm of i1] (i2) ;
			
			\vertex[below=1cm of b1] (o2) ;
			
			\vertex[above=1cm of a1] (t2);
			\vertex[above=2cm of a2] (t3);
			
			\vertex[above=1.5cm of a1] (i3) {$T^{\mu\nu}$};
			
			\diagram* { 
				(i1)  -- [fermion] (a1) -- [fermion]  (b1)  ,
				
				(i2)    --[anti fermion] (o2),

				(i3)  -- [graviton] (a1),
				
			};
			
			\node[] at (-0.35,-0.45) {$\pi\ $};
			\node[] at (4.46,-0.45) {$\pi\ $};

			\filldraw[fill=black, fill opacity=0.1] (-0.42,-0.5) ellipse (.45cm and 1.3cm);
			\filldraw[fill=black, fill opacity=0.1] (4.42 ,-0.5) ellipse (.45cm and 1.3cm);
			
			\draw[double, double distance=1.5,line width=1.2, postaction={decorate, decoration={
					markings,
					mark=at position 0.8 with  {\arrow{ latex} } } } ] (-1.8,-0.5) -- node [above] {} (-0.9,-0.5);
			
			\draw[double, double distance=1.5,line width=1.2, postaction={decorate, decoration={
					markings,
					mark=at position 0.8 with  {\arrow{ latex} } } } ] (4.9,-0.5) -- node [above] {} (5.85,-0.5);
		\end{feynman}
	\end{tikzpicture}
	\begin{tikzpicture}
		\begin{feynman}
			\vertex (i1);
			\vertex[right=1cm of i1] (ax2);
			\vertex[right=3cm of i1] (ax3);
			\vertex[right=2cm of i1] (a1);
			\vertex[right=2cm of a1] (b1);
			\vertex[right=3cm of a1] (o1) ;
			\vertex[below=1cm of i1] (i2) ;
			
			\vertex[below=1cm of b1] (o2) ;
			
			\vertex[above=1cm of a1] (t2);
			\vertex[above=2cm of a2] (t3);
			
			\vertex[above=1.5cm of a1] (i3) {$T^{\mu\nu}$};
			
			\diagram* { 
				(i1)   -- [fermion] (ax2) -- [fermion] (a1) -- [fermion] (ax3) -- [fermion] (b1)  ,
				
				(i2)    --[anti fermion] (o2),

				(i3)  -- [graviton] (a1),
				(ax2)-- [gluon, out=-45, in=-135] (ax3)
			};
			
			\node[] at (-0.35,-0.45) {$\pi\ $};
			\node[] at (4.46,-0.45) {$\pi\ $};

			\filldraw[fill=black, fill opacity=0.1] (-0.42,-0.5) ellipse (.45cm and 1.3cm);
			\filldraw[fill=black, fill opacity=0.1] (4.42 ,-0.5) ellipse (.45cm and 1.3cm);
			
			\draw[double, double distance=1.5,line width=1.2, postaction={decorate, decoration={
					markings,
					mark=at position 0.8 with  {\arrow{ latex} } } } ] (-1.8,-0.5) -- node [above] {} (-0.9,-0.5);
			
			\draw[double, double distance=1.5,line width=1.2, postaction={decorate, decoration={
					markings,
					mark=at position 0.8 with  {\arrow{ latex} } } } ] (4.9,-0.5) -- node [above] {} (5.85,-0.5);
		\end{feynman}
	\end{tikzpicture}
	\caption{Soft-overlap contribution diagrams to the pion form factor. The left panel shows the lowest-order contribution, while the right panel shows one of the diagrams contributing to the $\mathcal{O}(\a_s)$ correction.}
	\label{chza:softover}
\end{figure}

\section{Energy--momentum tensor and conformal properties}
\label{chza:sec:emt_conformal}
The stress-tensor operator is defined by varying the gauge-fixed QCD action with respect to the metric,
\begin{equation}
T_{\mu\nu}(x)
=
-\left.
\frac{2}{\sqrt{-g}}
\frac{\delta S_{\mathrm{QCD}}}{\delta g^{\mu\nu}(x)}
\right|_{g=\eta}.
\label{chza:eq:emt_def_short}
\end{equation}
Equivalently, in the weak-field expansion \(g_{\mu\nu}=\eta_{\mu\nu}+\kappa h_{\mu\nu}\),
\begin{equation}
S_{\mathrm{QCD}}[g]
=
S_{\mathrm{QCD}}[\eta]
-\frac{\kappa}{2}\int d^4x\,h_{\mu\nu}(x)T^{\mu\nu}(x)
 +\mathcal O(h^2).
\label{chza:eq:linear_emt_short}
\end{equation}
In flat space we use
\begin{equation}
\mathcal L_{\mathrm{QCD}}
=
-\frac{1}{4}F_{\rho\sigma}^aF^{a\,\rho\sigma}
+
\frac{i}{2}\bar\psi\gamma^\rho\overleftrightarrow{D}_\rho\psi
-
m\bar\psi\psi
-
\frac{1}{2\xi}(\partial^\rho A_\rho^a)^2
+
\partial^\rho\bar c^{\,a}D_\rho^{ab}c^b,
\label{chza:eq:qcd_lagrangian_short}
\end{equation}
with
\begin{equation}
F_{\mu\nu}^a
=
\partial_\mu A_\nu^a-\partial_\nu A_\mu^a+g_s f^{abc}A_\mu^bA_\nu^c,
\qquad
D_\mu^{ab}c^b=\partial_\mu c^a+g_s f^{acb}A_\mu^c c^b .
\label{chza:eq:field_strength_short}
\end{equation}
The explicit Hilbert tensor entering the perturbative \(TJJ\) vertex is then
\begin{align}
T_{\mu\nu}
&=
-\eta_{\mu\nu}\mathcal L_{\mathrm{QCD}}
-
F_{\mu\rho}^aF_{\nu}^{a\,\rho}
-
\frac{1}{\xi}\eta_{\mu\nu}\,\partial^\rho\!\left(A_\rho^a\,\partial^\sigma A_\sigma^a\right)
\nonumber\\
&\quad
+
\frac{1}{\xi}
\left[
A_\nu^a\,\partial_\mu(\partial^\sigma A_\sigma^a)
+
A_\mu^a\,\partial_\nu(\partial^\sigma A_\sigma^a)
\right]
\nonumber\\
&\quad
+
\frac{i}{4}
\Big[
\bar\psi\gamma_\mu(\overrightarrow{\partial}_\nu-i g_s A_\nu^a t^a)\psi
-
\bar\psi(\overleftarrow{\partial}_\nu+i g_s A_\nu^a t^a)\gamma_\mu\psi
\nonumber\\
&\hspace{2.4cm}
+
\bar\psi\gamma_\nu(\overrightarrow{\partial}_\mu-i g_s A_\mu^a t^a)\psi
-
\bar\psi(\overleftarrow{\partial}_\mu+i g_s A_\mu^a t^a)\gamma_\nu\psi
\Big]
\nonumber\\
&\quad
+
\partial_\mu \bar c^{\,a}\,D_\nu^{ab}c^b
+
\partial_\nu \bar c^{\,a}\,D_\mu^{ab}c^b .
\label{chza:eq:emt_explicit_short}
\end{align}

In the classical massless limit the gauge-invariant part of the tensor is conserved and traceless, up to equations of motion and gauge-fixing terms.  Renormalization breaks this Weyl invariance.  In physical matrix elements the massless trace is governed by the gauge-invariant anomaly term
\begin{equation}
T^\mu_{\ \mu}
=
\frac{\beta(g_s)}{2g_s}\,
F_{\rho\sigma}^aF^{a\,\rho\sigma}
\,+\,
\sum_f (1+\gamma_m)m_f\bar\psi_f\psi_f
\,+\,
\hbox{BRST-exact and equation-of-motion terms}.
\label{chza:eq:qcd_trace_anomaly_short}
\end{equation}
Thus, in the chiral limit, the conformal breaking relevant for the pion hard kernel is controlled by the QCD beta function.  In the form-factor calculation below the non-Abelian \(TJJ\) correlator is used as a one-loop stress-tensor insertion in the two-gluon hard subgraph.  The quark, gluon and ghost loop sectors combine into a renormalized off-shell \(TJJ\) vertex; its trace part contains precisely the beta-function coefficient that is later projected onto the pion GFFs.

 \section{Tensor-sector decomposition of the non-Abelian $TJJ$ correlator}
\label{chza:seven}
The second ingredient is the sector decomposition of the off-shell \(TJJ\) vertex.  We separate the correlator into transverse--traceless, longitudinal or semi-local, and trace sectors, treating the quark and gluon loop contributions separately before combining them in the renormalized result.  This is the minimal organization needed for the subsequent matching onto the pion form factors.

The general form of the non-Abelian $\langle TJJ \rangle$ correlator in $d$ dimensions can be constructed through a decomposition into transverse, longitudinal and trace terms \cite{Bzowski:2013sza}, exploiting its symmetries. The perturbative QCD correlator contains additional longitudinal terms with respect to the abstract conformal field theory (CFT) solution of \cite{Bzowski:2018fql}. These terms are tied to gauge fixing, BRST invariance and the Slavnov--Taylor identities, and they vanish or reduce in the appropriate on-shell limits.

We consider the decomposition of the operators $T$ and $J$ in terms of their transverse-traceless part and longitudinal (local) one. Following \cite{Bzowski:2013sza} we define
\begin{align}
	T^{\mu_i\nu_i}(p_i)&\equiv t^{\mu_i\nu_i}(p_i)+t_{loc}^{\mu_i\nu_i}(p_i),\label{chza:decT}\\
	J^{a_i \, \mu_i}(p_i)&\equiv j^{a_i \, \mu_i}(p_i)+j_{loc}^{a_i \, \mu_i}(p_i),\label{chza:decJ}
\end{align}
where
\begin{align}
	\label{chza:loct}
	t^{\mu_i\nu_i}(p_i)&=\Pi^{\mu_i\nu_i}_{\alpha_i\beta_i}(p_i)\,T^{\alpha_i \beta_i}(p_i), &&t_{loc}^{\mu_i\nu_i}(p_i)=\Sigma^{\mu_i\nu_i}_{\alpha_i\beta_i}(p_i)\,T^{\alpha_i \beta_i}(p_i),\\
	j^{a_i \, \mu_i}(p_i)&=\pi^{\mu_i}_{\alpha_i}(p_i)\,J^{a_i \, \alpha_i }(p_i), &&\hspace{1ex}j_{loc}^{a_i \, \mu_i}(p_i)=\frac{p_i^{\mu_i}\,p_{i\,\alpha_i}}{p_i^2}\,J^{a_i \, \alpha_i}(p_i).
\end{align}
Having introduced the transverse-traceless ($\Pi$), transverse $(\pi)$, longitudinal ($\Sigma$) projectors, given respectively by 
\begin{align}
	\label{chza:prozero}
	&\pi^{\mu}_{\alpha}  = \delta^{\mu}_{\alpha} - \frac{p^{\mu} p_{\alpha}}{p^2}, \\
	&\Pi^{\mu \nu}_{\alpha \beta}  = \frac{1}{2} \left( \pi^{\mu}_{\alpha} \pi^{\nu}_{\beta} + \pi^{\mu}_{\beta} \pi^{\nu}_{\alpha} \right) - \frac{1}{d - 1} \pi^{\mu \nu}\pi_{\alpha \beta}\label{chza:TTproj}, \\
	&\Sigma^{\mu_i\nu_i}_{\alpha_i\beta_i}=\frac{p_{i\,\beta_i}}{p_i^2}\Big[2\delta^{(\nu_i}_{\alpha_i}p_i^{\mu_i)}-\frac{p_{i\alpha_i}}{(d-1)}\left(\delta^{\mu_i\nu_i}+(d-2)\frac{p_i^{\mu_i}p_i^{\nu_i}}{p_i^2}\right)\Big]+\frac{\pi^{\mu_i\nu_i}(p_i)}{(d-1)}\delta_{\alpha_i\beta_i}\equiv\mathscr{T}^{\mu_i\nu_i}_{\alpha_i}p_{i\,\beta_i} +\frac{\pi^{\mu_i\nu_i}(p_i)}{(d-1)}\delta_{\alpha_i\beta_i}\label{chza:Lproj}.
\end{align}
with
\begin{equation} \label{chza:a:T}
	\mathscr{T}_{\mu\nu \alpha} ({p}) = \frac{1}{p^2} \left[ 2 p_{(\mu} \delta_{\nu)\alpha} - \frac{p_\alpha}{d-1} \left( \delta_{\mu\nu} + (d-2) \frac{p_\mu p_\nu}{p^2} \right) \right]
\end{equation}	
and
\begin{equation}\label{chza:Idecomp}
\delta_{\mu(\alpha}\delta_{\beta)\nu}=\Pi_{\mu\nu\alpha\beta}({p})+\mathscr{T}_{\mu\nu (\alpha}({p})\,p_{\beta)}+\frac{1}{d-1}\pi_{\mu\nu}({p})\delta_{\alpha\beta}.
\end{equation}
Turning to the $TJJ$ case, we can divide the three-point function into two parts: the \emph{transverse-traceless} part and the \emph{semi-local} part (indicated by subscript $loc$) expressible through the transverse and trace Ward identities. These parts are obtained by using the projectors $\Pi$ and $\Sigma$, previously defined. 
We can then decompose the full three-point function as 

\begin{align}
	\label{chza:dec1}
	\braket{T^{\m \n}\,J^{a \, \a}\,J^{b \, \b}}&=
	\braket{t^{\m \n }\,j^{a \, \a}\,j^{b \, \b}}+\braket{T^{\m \n}\,J^{a \, \a}\,j_{loc}^{b \, \b}}+\braket{T^{\m \n}\,j_{loc}^{a \, \a}\,J^{b \, \b}}+\braket{t_{loc}^{\m \n}\,J^{a \, \a}\,J^{b \, \b}}\notag\\
	&\quad-\braket{T^{\m \n }\,j_{loc}^{a \, \a}\,j_{loc}^{b \, \b}}-\braket{t_{loc}^{\m \n}\,j_{loc}^{a \, \a}\,J^{b \, \b}}-\braket{t_{loc}^{\m \n}\,J^{a \, \a}\,j_{loc}^{b \, \b}}+\braket{t_{loc}^{\m \n }\,j_{loc}^{a \, \a}\,j_{loc}^{b \, \b}}.
\end{align}
In a CFT approach, all the terms on the right-hand side of the decomposition, apart from the first one, may be computed by means of transverse and trace Ward identities, with the anomaly induced by the renormalization of the theory by a single counterterm, proportional to the square of the field strength $(F^2)$. 

\subsection{The decomposition of the correlator in the quark sector}
\label{chza:eight}
Using the projectors $\Pi$ and $\pi$ one can write the most general form of the transverse-traceless part as
\begin{equation}
	{\braket{t^{\m \n}(q)\,j^{a \, \a}(p_1)\,j^{b \, b }(p_2)}}_q =\Pi^{\m \n}_{\m_1 \n_1}(q)\pi^{\a}_{\a_1 }(p_1)\pi^{\b}_{\b_1}(p_2)\,\,X^{a b \, \m_1 \n_1 \a_1 \b_1}_q,
\end{equation}
where $X^{a b \, \m_1 \n_1 \a_1 \b_1}_q$ is a general tensor of rank four built out of the metric and momenta. We can enumerate all possible tensors that can appear in $X^{a b \, \m_1 \n_1 \a_1 \b_1}$ while preserving the symmetry of the 
correlator 
\begin{align}
	\langle t^{\mu \nu }(q)j^{a \, \a}(p_1)j^{b \, \b}(p_2)\rangle_q & =
	{\Pi_1}^{\mu \nu}_{\m_1 \n_1}{\pi_2}^{\a}_{\a_1}{\pi_3}^{\b}_{\b_1}
	\left( A_{1}^{(q) a b} \ p_1^{\m_1 }p_1^{\n_1}p_2^{\a_1}q^{\b_1} + 
	A_{2}^{(q) a b}\ \delta^{\a_1 \b_1} p_1^{\m_1}p_1^{\n_1} + 
	A_{3}^{(q) a b}\ \delta^{\m_1\a_1}p_1^{\n_1}q^{\b_1}\right. \notag\\
	& \left. + 
	A_{3}^{(q) a b}(p_1\leftrightarrow p_2)\delta^{\m_1\b_1}p_1^{\n_1}p_2^{\a_1}
	+ A_{4}^{(q) a b}\  \delta^{\m_1\b_1}\delta^{\a_1\n_1}\right)\label{chza:DecompTJJ}
\end{align}
with the reconstruction taking the form \cite{Bzowski:2013sza}
	\begin{align}
		& \la T_{\mu_1 \nu_1}(q) J^{\mu_2 a_2}({p}_1) J^{\mu_3 a_3}({p}_2) \ra_q = \la t_{\mu_1 \nu_1}(q) j^{\mu_2 a_2}({p}_1) j^{\mu_3 a_3}({p}_2) \ra_q \nn\\[0.5ex]
		& \qquad +  2 \mathscr{T}_{\mu_1 \nu_1}^{\quad\,\,\alpha}(q)\Big[\delta_{[\alpha}^{\mu_3}p_{2\beta]}  \la J^{\mu_2 a_2}({p}_1) J^{\beta a_3}(-{p}_1) \ra_q
		+\delta_{[\alpha}^{\mu_2}p_{1\beta]}  \la J^{\mu_3 a_3}({p}_2) J^{\beta a_2}(-{p}_2) \ra_q\Big]
		\nn\\
		& \qquad + \frac{1}{d - 1}\,\pi_{\mu_1 \nu_1}(q) \mathcal{A}^{\mu_2 \mu_3 a_2 a_3}_q, \label{chza:tjjdec}
	\end{align}
where $\mathscr{T}_{\mu_1 \nu_1\alpha}$ is defined in \eqref{chza:a:T} and $\langle JJ\rangle_q$ is the 2-point function of the gluon with a virtual quark, namely
\beq
\la J^{\alpha a}({p}_1) J^{\beta b}(-{p}_1) \ra_q=
\frac{2}{3}n_f  \,  \frac{g_s^2}{16\pi ^2} {B}_0\left({{p_1}}^2\right) {p_1}^2   \delta ^{a b}\pi^{\alpha\beta}(p_1),
\eeq 
where \(B_0(p^2)\) is the standard scalar two-point Passarino--Veltman function in dimensional regularization.
The transverse Ward identities are 
\begin{align}
	& q^{\nu_1} \la T_{\mu_1 \nu_1}(q) J^{\mu_2 a_2}({p}_1) J^{\mu_3 a_3}({p}_2) \ra_q  \nn\\[0.5ex]
	& \qquad = \: 2 \delta^{\mu_3}_{[\mu_1} p_{2\alpha]} \la J^{\mu_2 a_2}({p}_1) J^{\alpha a_3}(-{p}_1) \ra_q + 2 \delta^{\mu_2}_{[\mu_1} p_{1\alpha]} \la J^{\alpha a_2}({p}_2) J^{\mu_3 a_3}(-{p}_2) \ra_q, \\[1ex]\label{chza:TWI_TJJ_2}
	& p_{1 \mu_2} \la T_{\mu_1 \nu_1}(q) J^{\mu_2 a_2}({p}_1) J^{\mu_3 a_3}({p}_2) \ra_q = 0, \\[1ex]
\end{align}
while the anomalous trace Ward identity is given by
\begin{align}
	& \la T(q) J^{\mu_2 a_2}({p}_1) J^{\mu_3 a_3}({p}_2) \ra_q = \mathcal{A}^{\mu_2 \mu_3 a_2 a_3}_q=-{2 \over 3} \, \,  \, \, \frac{g_s^2}{16\pi^2} \,   n_f \delta^{a_2a_3}u^{\mu_2 \mu_3}(p_1,p_2),
\end{align}
where $u$ stands for the second derivative of the field strength squared $F^2$ which in momentum space is
\begin{equation}\label{chza:defu}
	u^{\a \b}(p_1,p_2)= (p_1\cdot p_2) g^{\a\b}-p_2^\a p_1^\b
\end{equation}
The form factors of the transverse part of the correlator can be explicitly defined as
\begin{equation}\label{chza:aiq}
	A_{j }^{(q) a b}= -\,   \frac{g_s^2}{16\pi^2}\, \bar{A}_j^{{(q)}a b}, \qquad j=1,2,3, 4
\end{equation}
where the label $(q)$ refers to the contribution from a single quark loop in the perturbative expansion; a direct computation gives
\begin{eqnarray}
	\label{chza:Abar}
	\bar{A}_1^{(q) ab}  &=& \frac{n_f\delta^{ab}}{48 \left(p_1^2 p_2^2-(p_1 \cdot p_2)^2 \right)^4} \Big[ A_{10} + A_{11}  B_0 (p_1^2) + A_{12} B_0(p_2^2) + A_{13} B_0(q^2) + A_{14} C_0 (p_1^2 , p_2^2 , q^2 ) \Big] \notag \\
	\bar{A}_2^{(q) ab}  &=& -\frac{\delta ^{ab}}{144 \left(p_1^2 p_2^2-(p_1 \cdot p_2)^2 \right)^3} \Big[ A^{(q)}_{20} + A^{(q)}_{21}  B_0 (p_1^2) + A^{(q)}_{22} B_0(p_2^2)  + A^{(q)}_{23} B_0(q^2) + A^{(q)}_{24} C_0 (p_1^2 , p_2^2 , q^2 ) \Big] \notag \\
	\bar{A}_3^{(q) ab}  &=& \frac{\delta ^{ab}}{72 \left(p_1^2 p_2^2-(p_1 \cdot p_2)^2 \right)^3} \Big[A^{(q)}_{31}  B_0 (p_1^2) + A^{(q)}_{32} B_0(p_2^2)  + A^{(q)}_{33} B_0(q^2) + A^{(q)}_{34} C_0 (p_1^2 , p_2^2 , q^2 ) \Big] \notag \\
	\bar{A}_4^{(q) ab}  &=& -\frac{ \delta^{ab}}{72 \left(p_1^2 p_2^2-(p_1 \cdot p_2)^2 \right)^2} \Big[A^{(q)}_{40} + A^{(q)}_{41}  B_0 (p_1^2) + A^{(q)}_{42} B_0(p_2^2) + A^{(q)}_{43} B_0(q^2) + A^{(q)}_{44} C_0 (p_1^2 , p_2^2 , q^2 ) \Big]. \notag \\
\end{eqnarray}
The expressions of the $A_{ij}$ are given in Appendix \ref{chza:ff}, while $C_0(p_1^2,p_2^2,q^2)$ is the massless three-point scalar integral whose explicit expression is given in Eq.~\eqref{chza:c0formula}.  The explicit formulas of $\bar A_i^{(q)ab}$ can be isolated from the general 
expressions for $A_{ij}$ in Appendix \ref{chza:ff}  by selecting the $n_f$ dependent terms in $A_{ij}$.

\subsection{The decomposition of the correlator in the gluon sector}
\label{chza:nine}
The one-loop decomposition of the gluon sector follows \eqref{chza:dec1}, 
but its final expression is modified compared to 
\eqref{chza:tjjdec}, which is affected by new trace contributions not present in the quark sector. This sector provides a contribution of the form
	\begin{equation}
		\begin{aligned}
			& \langle T^{\mu \nu}(q)  J^{ a\alpha}(p_1) J^{ b\beta}(p_2)\rangle_g=\langle t^{\mu \nu}(q)  j^{ a\alpha}(p_1) j^{ b\beta}(p_2)\rangle_g +\langle t^{\mu \nu }(q)j_{loc}^{a  \a}(p_1)j^{b  \b}(p_2)\rangle_g +\langle t^{\mu \nu }(q)j^{a  \a}(p_1)j_{loc}^{b  \b}(p_2)\rangle_g\\
			& \qquad+2 \mathscr{T}^{\mu \nu \rho }(q)\left[\delta_{[\rho}^{\beta} p_{2 \sigma ]}\langle J^{ a\alpha}({p}_1) J^{ b\sigma}(-{p}_1)\rangle_g +\delta_{[\rho}^{\alpha} p_{1 \sigma]}\langle J^{ b\beta}({p}_2) J^{ a\sigma}(-p_2)\rangle\right]_g  
			+\frac{1}{d-1} \pi^{\mu \nu}(q) \left[\mathcal{A}^{\alpha \beta a b}_g+\mathcal{B}^{\alpha \beta a b}_g\right]
		\end{aligned}
		\label{chza:res}
	\end{equation}
There are additional local terms of the form 
\beq
\langle t^{\mu \nu }(q)j^{a \, \a}(p_1)j_{loc}^{b \, \b}(p_2)\rangle_g 
\label{chza:adds}
\eeq
which are not part of the transverse-traceless sector, but appear in the longitudinal sector of the decomposition. These terms are not set to zero by the Slavnov-Taylor identities, as in the quark sector, or in the general conformal solution.  
At the same time, the trace sector is modified by the presence of extra terms which are proportional to the equations of motion of the gluons, here indicated as $\mathcal{B}^{\alpha \beta a b}_g, $ which are absent in the on-shell decomposition. Defining  
\begin{allowdisplaybreaks}
	\begin{equation}\label{chza:aig}
		A_{i }^{(g) a b}= -\,   \frac{g_s^2}{16\pi^2}\, \bar{A}_i^{{(g)}a b}, \qquad i=1,2\ldots 4
	\end{equation}
their explicit expressions are given by 
	\begin{eqnarray}
		\label{chza:Abar_g}
		\bar{A}_1^{(g) ab}  &=& -\frac{C_A\delta^{ab}}{48 \left(p_1^2 p_2^2-(p_1 \cdot p_2)^2 \right)^4} \Big[ A_{10} + A_{11}  B_0 (p_1^2) + A_{12} B_0(p_2^2) + A_{13} B_0(q^2) + A_{14} C_0 (p_1^2 , p_2^2 , q^2 ) \Big] \notag \\
		\bar{A}_2^{(g) ab}  &=& -\frac{\delta ^{ab}}{144 \left(p_1^2 p_2^2-(p_1 \cdot p_2)^2 \right)^3} \Big[ A^{(g)}_{20} + A^{(g)}_{21}  B_0 (p_1^2) + A^{(g)}_{22} B_0(p_2^2)  + A^{(g)}_{23} B_0(q^2) + A^{(g)}_{24} C_0 (p_1^2 , p_2^2 , q^2 ) \Big] \notag \\
		\bar{A}_3^{(g) ab}  &=& \frac{\delta ^{ab}}{72 \left(p_1^2 p_2^2-(p_1 \cdot p_2)^2 \right)^3} \Big[A^{(g)}_{31}  B_0 (p_1^2) + A^{(g)}_{32} B_0(p_2^2)  + A^{(g)}_{33} B_0(q^2) + A^{(g)}_{34} C_0 (p_1^2 , p_2^2 , q^2 ) \Big] \notag \\
		\bar{A}_4^{(g) ab}  &=& -\frac{ \delta^{ab}}{72 \left(p_1^2 p_2^2-(p_1 \cdot p_2)^2 \right)^2} \Big[A^{(g)}_{40} + A^{(g)}_{41}  B_0 (p_1^2) + A^{(g)}_{42} B_0(p_2^2) + A^{(g)}_{43} B_0(q^2) + A^{(g)}_{44} C_0 (p_1^2 , p_2^2 , q^2 ) \Big], \notag \\ 
	\end{eqnarray}
\end{allowdisplaybreaks}
where the functions $A_{ij}^{(g)}$ are extracted from  \secref{chza:ff} by selecting the part of the $A_{ij}$ proportional to the Casimir $C_A$.
The new longitudinal terms \eqref{chza:adds} take the form 
	\begin{align}
		\langle t^{\mu \nu }(q)j^{a \, \a}(p_1)j_{loc}^{b \, \b}(p_2)\rangle_g & =
		{\Pi}^{\mu \nu}_{\m_1 \n_1}(q) \, \pi^\a _{\a_1} (p_1)\, {{p_2}_{\beta_1} p_2^\beta \over p_2^2}
		\left( 	B^{ab}_1 \, p_1^{\m_1} \, p_1^{\nu_1} \, p_2^{\a_1} \, p_2^{\b_1} + B_2^{ab} \, p_1^{\m_1} \, p_2^{\b_1} \, \delta^{\a_1 \n_1}  \right) 
	\end{align}
which is orthogonal to the trace sector. Notice that these local contributions vanish when the gluons are on-shell. The $B_i, \,\, i=1,2$ are given by  

\begin{eqnarray}
	B^{ab}_1 & = &  \frac{ \,  \,  \, C_A \,  g_s^2 \,  \delta^{ab} \, p_1^2}{64\pi^2 \, p_2^2 \, \big(p_1^2 p_2^2 - (p_1 \cdot p_2)^2 \big)^2} \, \Big( -\Big[2 \, (p_1 \cdot p_2)^2 + p_1^2 p_2^2 + 3 \, p_1^2 p_1 \cdot p_2 \Big]\, B_0(p_1^2)  \notag \\
	&& -\Big[2\, (p_1 \cdot p_2)^2 + p_1^2 p_2^2 + 3 \, p_2^2 p_1 \cdot p_2 \Big] \, B_0(p_2^2) +  \Big[ 4 \, (p_1 \cdot p_2)^2 + 2 \, p_1^2 p_2^2 + 3\, (p_1^2 + p_2^2) p_1 \cdot p_2 \Big]\, B_0(q^2)  \notag \\
	&&  +\Big[  q^2 \, (p_1 \cdot p_2)^2 + 2 \, q^2 \, (p_1 \cdot p_2)^2  \Big] \,C_0(p_1^2, p_2^2, q^2) - 2 \, (p_1 \cdot p_2)^2 - 2 \, p_1^2 p_2^2 \Big) \\
	B^{ab}_2 & = & \frac{ \,  \,  \, C_A \,  g_s^2 \,  \delta^{ab} \, p_1^2}{32\pi^2 \, p_2^2 \, \big(p_1^2 p_2^2 - (p_1 \cdot p_2)^2 \big)^2} \Big( - p_1 \cdot p_2 \, B_0(p_1^2)  - p_2^2 \cdot p_2  \, B_0(p_2^2) + (p_2^2+p_1 \cdot p_2) \, B_0(q^2) \notag \\ &&
	+ \Big[p_1^2 p_2^2 + p_2^2 \, (p_1 \cdot p_2) \Big] \, C_0(p_1^2, p_2^2 , q^2) \Big).
\end{eqnarray}

The trace sector is also affected by terms that vanish for 
on-shell gluons, indicated as $\mathcal{B}_g^{\alpha \beta a b}$, as well as the genuine anomaly term 
$\mathcal{A}_g^{\alpha \beta a b}$. Defining 
	\beq
	\mathcal{B}_g^{\alpha \beta a b}=\left[ C_1^{ab} \, p_1^\alpha \, p_1 ^\beta + C_2^{ab} \, p_1^\alpha \, p_2^\beta + C_3^{ab}  \, p_1^\beta \, p_2^\alpha + C_4^{ab} \, p_2^\alpha \, p_2^\beta + C_5^{ab} \, \delta^{\alpha \beta} \right]  
	\label{chza:bb}
	\eeq
the trace sector is characterised by the anomaly contributions from the gluon sector $\mathcal{A}_g$, plus the 
$\mathcal{B}_g$ terms proportional to the equations of motion of the gluons 
\begin{equation}
	\mathcal{A}_g^{\alpha \beta a b}+\mathcal{B}_g^{\alpha \beta a b}= g_{\mu \nu} \langle T^{\mu \nu} (q) \, J^{a \a} (p_1) \, J^{b \b} (p_2)\, \rangle_g 
\end{equation}
with the gluon contribution to the anomaly given by
\begin{eqnarray}
	\mathcal{A}_g^{\alpha\beta ab} &=& {11 \over 3} \,  \,  \,  \, \frac{g_s^2}{16\pi^2} \, C_A  \delta^{ab}u^{\alpha\beta}(p_1,p_2)
\end{eqnarray}
and $u^{\alpha\beta}$ defined in \eqref{chza:defu}.
The form factors proportional to the equations of motion in \eqref{chza:bb} take the form
{\allowdisplaybreaks
\begin{eqnarray}
	C_1^{ab} &=& -\frac{ \,  \,  \, C_A \, g_s^2 \, \delta^{ab} }{ 32\pi^2 \big( (p_1 \cdot p_2)^2 - p_1^2 \, p_2^2 \big)} \, \Big(
	2 \, (p_1 \cdot p_2)^2 - 2 \, p_1^2 \, p_2^2 
	- p_1^2 \, (p_1 \cdot p_2 + p_2^2 ) B_0 (p_1^2)  \notag \\ &&
	+ \big[ p_2^2 \, (p_1^2 - p_1 \cdot p_2) - 2 \, (p_1 \cdot p_2)^2 \big]  B_0(p_2^2)
	+ (p_1 \cdot p_2) \, (2 \, p_1 \cdot p_2 + p_1^2 + p_2^2) \, B_0 (q^2)  \notag \\ &&
	+ \big[ 2\, (p_1 \cdot p_2)^2 \, (p_1 \cdot p_2)^2 - p_2^2 \, \big(4 (p_1 \cdot p_2)^2   + p_1^4\big) + 5 \, p_1^2 \, p_4^4  \big] C_0(p_1^2, p_2^2 , q^2)  \Big) \\
	C_2^{ab} &=& -2  \,  \,  \, C_A \, \frac{g_s^2}{16\pi^2} \, \delta^{ab} \, (p_1 \cdot p_2) \, C_0 (p_1^2 , p_2^2 , q^2) \\
	C_3^{ab} &=& -\frac{ \,  \,  \, g_s^2 \, C_A \, (p_1^2 + p_2^2) \, \delta^{ab}}{32\pi^2 \big(p_1^2 \, p_2^2 - (p_1 \cdot p_2)^2 \big)} \Big( 
	(-p_1\cdot p_2-p_1^2)  B_0(p_1^2)+(-(p_1\cdot p_2)-p_2^2) \, B_0(p_2^2) \notag \\ &&
	+q^2 \, B_0(q^2)+\big(p_1^2 \, (p_1\cdot p_2-2 p_2^2) +(p_1\cdot p_2) \, \big(4 \, (p_1\cdot p_2)+p_2^2) \big) \, C_0(p_1^2,p_2^2, q^2)  \Big)  \\
	C_4^{ab} &=&  -\frac{ \, \,  \, g_s^2 \, C_A \, g_s^2 \,  \delta^{ab}}{32\pi^2\big((p_1 \cdot p_2)^2 - p_1^2 \, p_2^2 \big)} \Big( 
	(p_1^2 \, (p_2^2-p_1\cdot p_2)\notag \\ &&  -2 (p_1\cdot p_2)^2) \, B_0(p_1^2) -p_2^2 \, (p_1\cdot p_2+p_1^2) \, B_0(p_2^2)+(p_1\cdot p_2) \, q^2 \, B_0(q^2) \notag \\ &&  
	+ \big[-p_1^2 \, (4 \, (p_1\cdot p_2)^2+p_2^4)+2 \, (p_1\cdot p_2+p_2^2) \, (p_1\cdot p_2)^2+5 \, p_1^4 \, p_2^2 \, \big] \, C_0(p_1^2,p_2^2,q^2) \, \Big) \\
	C_5^{ab} &=& \, \,  \, \frac{g_s^2}{32\pi^2} \, C_A \,  \delta^{ab} \Big( (p_1^2 - p_2^2) \, \big[ B_0 (p_1^2) - B_0(p_2^2) \big] 
	+ \big[ p_1^4 + p_2^4 - 2 \, (p_1^2 + p_2^2) \, p_1 \cdot p_2 - 6 \, p_1^2 \, p_2^2 \big] \, C_0(p_1^2, p_2^2, q^2) \Big). \nonumber \\\label{chza:cig}
\end{eqnarray}
}

\section{The contribution of the $TJJ$ and of the trace anomaly in the hard kernel}\label{chza:contjjtra}

In this section we discuss the insertion of the non-Abelian \(TJJ\) correlator into the hard-scattering factorization formula for the pion GFFs, with particular emphasis on the relation between the off-shell trace sector of the correlator and the anomaly-induced contribution to the scalar gravitational structure of the pion. We also clarify the scaling of the various terms entering the hard kernel, the role of the two-point functions generated by the loop expansion, and the behavior of the convolution integrals with the pion DA \(\tilde{\phi}_\pi(\alpha)\).

\subsection{Factorization structure and notation}

Working in momentum space and in the hard regime
\begin{equation}\label{chza:eq:factorisation_regime}
q^2\equiv -Q^2,\qquad Q^2\gg \Lambda_{\rm QCD}^2,\; m_\pi^2,
\end{equation}
the correction to the tree-level calculation in Eq. \eqref{kerfact} of the matrix element of the energy-momentum tensor between pion states in the collinear factorization framework due to the $TJJ$ insertion is given by
\begin{equation}\label{chza:eq:factorisation}
	\frac{i\,f_\pi^2 C_F g_s^2}{N_C}\int_0^1 d\alpha_1 \int_0^1 d\alpha_2 \;
	\tilde{\phi}_\pi(\alpha_1)\, K_{TJJ}^{\mu\nu}(\alpha_1,\alpha_2,q)\, \tilde{\phi}_\pi(\alpha_2).
\end{equation}
Here \(K_{TJJ}^{\mu\nu}\) is the perturbative hard kernel obtained by inserting the correlator \(TJJ\) at leading nontrivial order in the strong coupling. The light-cone momentum fractions \(\alpha_{1,2}\in[0,1]\) label the momentum carried by the quark in each pion DA.

We decompose the kernel in the tensor basis constructed from the pion momenta \(P_1\), \(P_2\) and the metric,
\begin{equation}\label{chza:eq:Kdecomp}
	K_\pi^{\mu\nu}(\alpha_1,\alpha_2,q)
	= \sum_{i=1}^5 F_i^{TJJ}(\alpha_1,\alpha_2,q)\, T_i^{\mu\nu},
\end{equation}
following the same decomposition previously used for the tree-level kernel in Eq. \eqref{decompkpi}
\begin{equation}\label{chza:eq:tensorbasis}
T_1^{\mu\nu}=P_1^\mu P_2^\nu,\qquad
T_2^{\mu\nu}=P_2^\mu P_1^\nu,\qquad
T_3^{\mu\nu}=P_1^\mu P_1^\nu,\qquad
T_4^{\mu\nu}=P_2^\mu P_2^\nu,\qquad
T_5^{\mu\nu}=g^{\mu\nu}.
\end{equation}
The form-factor coefficients \(F_i^{TJJ}\) encode the full dependence on the partonic fractions and the hard scale \(q^2\). In the expressions below we keep the superscript \(TJJ\) to emphasize that these are the pieces originating from the non-Abelian \(TJJ\) insertion, while the tree-level kernel pieces will be denoted by \(F_i^{\rm tree}\).

\subsection{Explicit form of the kernel coefficients and their interpretation}

From the perturbative calculation one obtains the following compact form for the anomaly-induced pieces. For the momentum-space form factors one finds\allowdisplaybreaks
\begin{align}\label{chza:eq:F12}
	F_1^{TJJ}=F_2^{TJJ}=\frac{1}{3(\alpha_1-1)\alpha_1(\alpha_2-1)\alpha_2 q^2}\  \Biggl[&{q}^2 \bigg(4  \left(6 \text{$\alpha_1$} \text{$\alpha_2$}+(2 \text{$\alpha_1 $}-5) \text{$\alpha_1 $}+2 \text{$\alpha_2 $}^2-5 \text{$\alpha_2$}+2\right){A_2}\nn\\&+3  (3 \text{$\alpha_1$}+\text{$\alpha_2$}-2){A_3}+2  (-2 \text{$\alpha_1$} \text{$\alpha_2$}+\text{$\alpha_1$}+\text{$\alpha_2$}-2)\mathcal{A}\bigg)\nn\\&+4 \bigg((8 \text{$\alpha_1$}+8 \text{$\alpha_2$}-4) \mathcal{JJ}\left[(\text{$\alpha_2$}-1) {{P_2}}-(\text{$\alpha_1$}-1){{P_1}}\right]\nn\\&
	-4 (2 \text{$\alpha_1$}+2 \text{$\alpha_2$}-3) \mathcal{JJ}\left[\text{$\alpha_1$} {{P_1}}-\text{$\alpha_2$} P_2\right]+4 {A_4}-{C_5}\bigg)\Biggl]
\end{align}
\begin{align}\label{chza:eq:F3}
	F_3^{TJJ}
	=-\frac{2}{3(\alpha_1-1)\alpha_1(\alpha_2-1)\alpha_2 q^2}\  \Biggl[& {q}^2 \left( \left(13 \text{$\alpha_1$}^2+2 \text{$\alpha_1$} (3 \text{$\alpha_2$}-8)+(\text{$\alpha_2$}-2)^2\right){A_2}+(-2 \text{$\alpha_1$} \text{$\alpha_2$}+\text{$\alpha_1$}+\text{$\alpha_2$}-2)\mathcal{A} \right)\nn\\&+2 \bigg(-2 (7 \text{$\alpha_1$}+\text{$\alpha_2$}-6) \mathcal{JJ} \left[\text{$\alpha_1$} {P_1}-\text{$\alpha_2 $} {P_2}\right]\nn\\&+2 (7 \text{$\alpha_1$}+\text{$\alpha_2$}-2) \left(\mathcal{JJ} \left[(\text{$\alpha_2$}-1) {P_2}-(\text{$\alpha_1$}-1) {P_1}\right]\right)+{A_4}-{C_5}\bigg) \Biggl],
\end{align}
\begin{equation}\label{chza:eq:F4}
	F_4^{TJJ}
	=F_3^{TJJ}(\a_1\leftrightarrow\a_2,P_1\leftrightarrow P_2),
\end{equation}
and the scalar, trace-sector coefficient
\begin{align}\label{chza:eq:F5}
	F_5^{TJJ}=&-\frac{1}{3(\alpha_1-1)\alpha_1(\alpha_2-1)\alpha_2 q^2}\Biggl[
	q^2\Bigl(\bigl((\alpha_2-2)^2+\alpha_1^2+(6\alpha_2-4)\alpha_1\bigr)A_2 \nonumber\\
	&\qquad\qquad\qquad\qquad\qquad\qquad
	+2\bigl((2\alpha_2-1)\alpha_1-\alpha_2+2\bigr)\mathcal{A}\Bigr)\nonumber\\
	&\qquad\qquad\qquad\qquad\qquad\qquad
	+2\Bigl(2(\alpha_1+\alpha_2-2)\,\mathcal{JJ}[(\alpha_2-1)P_2-(\alpha_1-1)P_1]\nonumber\\
	&\qquad\qquad\qquad\qquad\qquad\qquad
	-2(\alpha_1+\alpha_2)\,\mathcal{JJ}[\alpha_1 P_1-\alpha_2 P_2]+A_4+2C_5\Bigr)\Biggr].
\end{align}
The coefficient \(\mathcal{A}\) is proportional to the one-loop QCD trace-anomaly coefficient, and it can be written as
\begin{equation}\label{chza:eq:Adef}
\mathcal{A}=\frac{g_s^2}{16\pi^2}\,\beta_0
\end{equation}
in terms of the leading coefficient of the QCD beta function in Eq. \eqref{chza:eq:beta0def}.
Its appearance in \(F_i^{TJJ}\) signals that the anomaly insertion reweights the trace structure and generates a contribution proportional to the running of the coupling.

The remaining quantities \(A_2\), \(A_3\), \(A_4\), and \(C_5\) are scalar coefficients originating from the loop integrals and from the tensor reduction of the triangle and bubble diagrams contributing to the \(TJJ\) correlator. They are finite functions of the renormalization scale \(\mu\) and of the gluon momenta \(p_1\) and \(p_2\). Their relation to the pion external momenta is fixed by the collinear kinematics,
\begin{equation}\label{chza:eq:p1p2def}
	p_1=\alpha_2 P_2-\alpha_1 P_1,\qquad
	p_2=\bar{\alpha}_2 P_2-\bar{\alpha}_1 P_1,
\end{equation}
with \(\bar{\alpha}_i=1-\alpha_i\). The coefficients \(A_2\), \(A_3\), \(A_4\), and \(C_5\) are understood here as the scalar functions obtained from the off-shell decomposition of the \(TJJ\) correlator after combining the quark and gluon sectors.

The symbol \(\mathcal{JJ}[p]\) denotes the scalar two-point function generated by the vacuum-polarisation substructure of the correlator. In the present normalization it reads
\begin{equation}\label{chza:eq:JJdef}
	\mathcal{JJ}[p]=\frac{g_s^2}{16\pi^2}\left(\frac{2}{3}n_f-\frac{5}{3}C_A\right)p^2 B_0(p^2),
\end{equation}
At large Euclidean momentum one has the expected logarithmic behavior
\begin{equation}\label{chza:eq:B0asympt}
B_0(p^2)\sim \ln\!\left(\frac{-p^2}{\mu^2}\right)+\text{const.},
\end{equation}
so that \(\mathcal{JJ}[p]\sim g_s^2 p^2 \ln(-p^2/\mu^2)\).

For the large-\(Q^2\) analysis it is important to observe that \(A_2\) and \(A_3\) do not introduce additional parametric powers of \(Q^2\) beyond those displayed explicitly in Eqs.~\eqref{chza:eq:F12}--\eqref{chza:eq:F5}, whereas \(A_4\), \(C_5\), and \(\mathcal{JJ}[p]\) scale as \(Q^2\), but this growth is compensated by the overall factor \(1/q^2\) in the hard coefficients. As a consequence, the anomaly-induced kernel contributes at order \(1/Q^2\), up to logarithmic corrections.

\subsection{From the kernel to the GFFs \(A_\pi\) and \(D_\pi\)}

Taking the trace of Eq.~\eqref{chza:eq:GFF_definition_improved} gives
\begin{equation}\label{chza:eq:trace_spinzero}
g_{\mu\nu}\langle \pi(P_2)|T^{\mu\nu}|\pi(P_1)\rangle
=2P^2 A_\pi(q^2)-\frac{3}{2}q^2D_\pi(q^2)
=2\left(m_\pi^2-\frac{q^2}{4}\right)A_\pi(q^2)-\frac{3}{2}q^2D_\pi(q^2),
\end{equation}
where in the last equality we used \(P_1^2=P_2^2=m_\pi^2\). In the hard, chiral limit used in the factorized calculation, \(q^2=-Q^2\) and \(m_\pi^2\ll Q^2\), so
\begin{equation}\label{chza:eq:trace_hard_chiral}
g_{\mu\nu}\langle \pi(P_2)|T^{\mu\nu}|\pi(P_1)\rangle
\simeq -\frac{q^2}{2}\left[A_\pi(q^2)+3D_\pi(q^2)\right]
=\frac{Q^2}{2}\left[A_\pi(-Q^2)+3D_\pi(-Q^2)\right].
\end{equation}
This relation makes explicit why the anomaly insertion is naturally tied to the scalar projection of the pion matrix element. Since the full QCD energy-momentum tensor is conserved, no independent form factor multiplying \(g^{\mu\nu}\) is needed in Eq.~\eqref{chza:eq:GFF_definition_improved}. Terms of this type may appear in separated quark and gluon sectors, or in the off-shell decomposition of the \(TJJ\) correlator, but after imposing the Ward or Slavnov--Taylor identities and matching onto the on-shell pion matrix element they are absorbed into the two conserved form factors \(A_\pi\) and \(D_\pi\). In the present leading-power matching the anomaly coefficient \(\mathcal{A}\) enters through the trace-sector coefficient \(F_5^{TJJ}\), and therefore gives its most direct correction to \(D_\pi\).

For the numerical interpretation it is important to keep the full \(TJJ\) insertion distinct from its beta-function part.  In the \(A_\pi\) projection the term proportional to the anomaly coefficient cancels, so the isolated anomaly does not contribute to the momentum form factor.  The full \(TJJ\) correction can nevertheless shift the leading Sudakov curve downward at low \(Q^2\), because it also contains non-anomalous tensor structures.  In the \(D_\pi\) projection the anomaly survives and gives an important scalar contribution.  The plotted \(TJJ\) term, however, is not the complete order-\(\alpha_s^2\) hard kernel; additional non-anomalous radiative terms may still modify the comparison with the lattice data.

Using this decomposition, one projects onto the two pion GFFs as
\begin{align}
	A_\pi(q^2)
	&= \frac{f_\pi^2 C_F g_s^2}{N_C} \int_0^1 d\alpha_1 \int_0^1 d\alpha_2\, \tilde{\phi}_\pi(\alpha_1)\,
	\big(F_1^{\rm tree}+F_1^{TJJ}+F_3^{\rm tree}+F_3^{TJJ}\big)\,
	\tilde{\phi}_\pi(\alpha_2),\label{chza:eq:Api}\\[6pt]
	D_\pi(q^2)
	&= - \frac{f_\pi^2 C_F g_s^2}{N_C}\int_0^1 d\alpha_1 \int_0^1 d\alpha_2\, \tilde{\phi}_\pi(\alpha_1)\,
	\frac{2\big(F_5^{\rm tree}+F_5^{TJJ}\big)}{q^2}\,
	\tilde{\phi}_\pi(\alpha_2).\label{chza:eq:Dpi}
\end{align}
Here the \(F_i^{\rm tree}\) denote the tree-level contributions of the hard kernel, while the \(F_i^{TJJ}\) encode the one-loop \(TJJ\) corrections, including the anomaly-controlled trace sector of the non-Abelian correlator.

These expressions make the role of the trace anomaly particularly transparent. First, the \(D\)-term is directly sensitive to the coefficient \(\mathcal{A}\), and therefore to the QCD beta-function coefficient \(\beta_0\). This is natural, since the scalar part of the matrix element probes the trace sector of the stress tensor, whose anomalous contribution is proportional to \(\beta(g_s)\). The result is a correction to \(D_\pi\) proportional to \(\beta_0 g_s^2\), which distinguishes it from the non-anomalous contributions already present at tree level.

Second, the dependence on the two-point function \(B_0(p^2)\), appearing through \(\mathcal{JJ}\) and implicitly in the remaining scalar coefficients, generates the expected logarithmic dependence on the hard scale. After convolution with the pion DA, these logarithms translate into scale-dependent terms of the form \(\ln(Q^2/\mu^2)\), superimposed on the overall \(1/Q^2\) suppression. The renormalization scale \(\mu\) therefore enters both through the hard kernel and through the evolution of the pion DA \(\tilde{\phi}_\pi(\alpha;\mu)\).

Finally, one must examine the endpoint behavior of the convolution integrals. The denominators \((\alpha_i-1)\alpha_i\) appearing in the coefficients suggest possible singularities as \(\alpha_i\to 0,1\). In practice, these are regulated by the pion DA, which vanishes linearly at the endpoints. 
Therefore the convolution integrals in Eqs.~\eqref{chza:eq:Api} and \eqref{chza:eq:Dpi} are finite at leading twist. 
Only when one goes beyond the strict collinear approximation, for instance by including transverse-momentum effects or power-suppressed endpoint-enhanced terms, does one expect a stronger sensitivity to the endpoint region and the need for an extended factorization framework.

To verify the gauge independence of the physical pion GFFs, let us examine the structure of the three-point Green's function embedded within the hard-scattering kernel. The off-shell Slavnov-Taylor identities dictate that all gauge-fixing dependent modifications are strictly confined to two isolated structures: the non-local longitudinal sector $\mathcal{B}_g$ and the local operator sector denoted by $\langle t^{\mu\nu}(q)j^{a\,\alpha}(p_1)j_{\text{loc}}^{b\,\beta}(p_2)\rangle$. At the diagrammatic level, any unphysical variation contributing to this Green's function yields structures that scale explicitly with the inverse gluon propagators, and are thus proportional to the virtualities $p_1^2$ and $p_2^2$. In collinear kinematics, as stated previously, these off-shell lines correspond to $p_1^2 = -\alpha_1 \alpha_2 Q^2, \quad \text{and} \quad p_2^2 = -(1-\alpha_1)(1-\alpha_2)Q^2.$ Crucially, every form factor residing within the $\langle t^{\mu\nu}(q)j^{a\,\alpha}(p_1)j_{\text{loc}}^{b\,\beta}(p_2)\rangle$ sector is automatically cancelled out upon algebraic contraction with the valence quark projector of the pion state. The only remaining contribution from this gauge-dependent sector is the coefficient $C_5$ which only appears in $D_\pi$. It tracks the remnant off-shell equations of motion and vanishes identically once the full phase-space integration over the hard kernel is performed against the pion DAs. In other words, while individual off-shell lines inside the unintegrated loop parameterize gauge variations, the collinear projection mapping acts as a natural gauge filter. Since these unphysical terms vanish automatically upon convolution, complete gauge independence is fully restored to the hard-scattering kernel. This ensures that our final extraction of the $D_\pi$ form factor safely isolates the physical contribution.

In summary, the insertion of the one-loop non-Abelian \(TJJ\) correlator into the hard kernel generates a calculable correction to the pion GFFs. Its trace sector is proportional to the QCD beta function and therefore represents the perturbative realization of the conformal anomaly inside the hard-scattering description. After convolution with the pion DA, this produces a contribution of order \(1/Q^2\), modulated by logarithmic dependence on the hard scale.

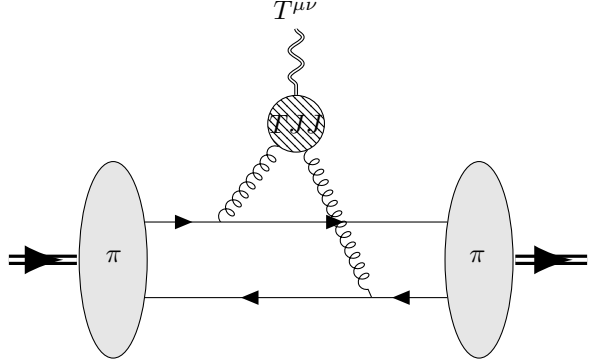
\begin{figure}
	\centering
	\begin{tikzpicture}
		\begin{feynman}
			\vertex (i1);
			\vertex[right=1cm of i1] (a1);
			\vertex[right=3cm of a1] (o1) ;
			\vertex[below=1cm of i1] (i2) ;
			\vertex[right=3cm of i2] (a2);
			\vertex[right=1cm of a2] (o2) ;
			
			\vertex[above=1cm of a1] (t2);
			\vertex[above=2cm of a2] (t3);
			\vertex[blob] (t1) at (2,1.3) {{$TJJ$}};
			\vertex[above=1.5cm of t1] (i3) {$T^{\mu\nu}$};
			
			\diagram* { 
				(i1)  -- [fermion] (a1)   --[fermion] (o1),
				
				(i2)  -- [anti fermion] (a2)  --[anti fermion] (o2),

				(i3)  -- [graviton] (t1),
				(t1) -- [gluon] (a1) ,
				(t1) -- [gluon] (a2),
			};
			
			\node[] at (-0.35,-0.45) {$\pi\ $};
			\node[] at (4.46,-0.45) {$\pi\ $};

			\filldraw[fill=black, fill opacity=0.1] (-0.42,-0.5) ellipse (.45cm and 1.3cm);
			\filldraw[fill=black, fill opacity=0.1] (4.42 ,-0.5) ellipse (.45cm and 1.3cm);
			
			\draw[double, double distance=1.5,line width=1.2, postaction={decorate, decoration={
					markings,
					mark=at position 0.8 with  {\arrow{ latex} } } } ] (-1.8,-0.5) -- node [above] {} (-0.9,-0.5);
			
			\draw[double, double distance=1.5,line width=1.2, postaction={decorate, decoration={
					markings,
					mark=at position 0.8 with  {\arrow{ latex} } } } ] (4.9,-0.5) -- node [above] {} (5.85,-0.5);
		\end{feynman}
	\end{tikzpicture}
	\caption{TJJ contribution to the GFF of the pion.}
	\label{chza:fig:1}
\end{figure}

\section{Sudakov corrections and modification of the pion wave function}\label{chza:phenomesec1}

In the previous sections the factorized description of the pion GFFs has been formulated in the collinear approximation, where the transverse momenta of the partons are neglected and the nonperturbative dynamics is entirely encoded in the DA \(\tilde{\phi}_\pi(\alpha)\). While this framework correctly captures the leading-power structure at large momentum transfer, it is well known that radiative corrections generate large logarithms associated with soft and collinear gluon emission. These logarithms must be resummed in order to obtain a reliable description of the hard amplitude, and their resummation leads to the emergence of Sudakov suppression factors.

The inclusion of Sudakov effects requires a refinement of the factorization scheme, in which transverse degrees of freedom are retained and the pion DA is promoted from a purely collinear DA to a transverse-momentum-dependent object. In this section we describe how Sudakov resummation modifies the structure of the pion DA and how it affects the convolution formula for the GFFs.

The perturbative corrections to the hard kernel involve loop integrals in which virtual gluons become soft or collinear with respect to the external partons. In these regions the amplitude develops double logarithmic enhancements of the form \cite{Sterman:1986aj} \cite{Botts:1989kf}\cite{Li:1992nu,Li:1994cka}
\begin{equation}\label{chza:eq:sudakov_logs}
\alpha_s \ln^2\!\left(\frac{Q^2}{k_\perp^2}\right),
\end{equation}
where \(k_\perp\) is the transverse momentum of the parton inside the pion. Such contributions arise from the overlap of soft and collinear divergences and appear to all orders in perturbation theory. In the collinear approximation these logarithms are absent.

The resummation of these logarithms leads to an exponential suppression factor, which damps the contributions from regions of phase space where soft-gluon emission would otherwise enhance the amplitude. This suppression plays a crucial role in regulating endpoint configurations and ensuring the self-consistency of the perturbative expansion.

To incorporate Sudakov effects, one introduces a transverse-momentum-dependent (TMD) pion wave function
\begin{equation}\label{chza:eq:TMDwf}
\Psi_\pi(\alpha,k_\perp,\mu),
\end{equation}
which generalizes the collinear DA according to
\begin{equation}\label{chza:eq:LCDA_reduction}
\tilde{\phi}_\pi(\alpha,\mu)=\int d^2 k_\perp\, \Psi_\pi(\alpha,k_\perp,\mu).
\end{equation}
It is convenient to work in impact-parameter space, defined by the Fourier transform
\begin{equation}\label{chza:eq:bspace}
\tilde{\Psi}_\pi(\alpha,b,\mu)=\int \frac{d^2 k_\perp}{(2\pi)^2}\,
e^{-i \vec{k}_\perp\cdot \vec{b}}\,
\Psi_\pi(\alpha,k_\perp,\mu),
\end{equation}
where \(b\) is conjugate to \(k_\perp\). In this representation, the Sudakov factor exponentiates in a simple form.

The factorized expression for the matrix element \eqref{chza:eq:factorisation} is then modified into
\begin{align}\label{chza:eq:factorisation_sudakov}
\langle\pi(P_2)|T^{\mu\nu}|\pi(P_1)\rangle
= \frac{if_\pi^2 C_Fg_s^2}{N_C}
\int_0^1 d\alpha_1 \int_0^1 d\alpha_2
\int d^2 b\;
\tilde{\Psi}_\pi(\alpha_1,b,\mu)\,
K_\pi^{\mu\nu}(\alpha_1,\alpha_2,b,Q,\mu)\,
\tilde{\Psi}_\pi(\alpha_2,b,\mu),
\end{align}
where the hard kernel now depends explicitly on the transverse separation \(b\), and the convolution includes the Sudakov suppression factor.

The resummation of soft and collinear logarithms leads to a universal exponential factor modifying the wave function,
\begin{equation}\label{chza:eq:sudakov_factor}
\tilde{\Psi}_\pi(\alpha,b,Q,\mu)
=
\exp\!\left[-S(\alpha,b,Q,\mu)\right]\,
\tilde{\Psi}_\pi^{(0)}(\alpha,b,\mu),
\end{equation}
where \(\tilde{\Psi}_\pi^{(0)}\) is a nonperturbative input function and \(S(\alpha,b,Q,\mu)\) is the Sudakov exponent.

At next-to-leading logarithmic accuracy, the Sudakov exponent can be written as
\begin{equation}\label{chza:eq:sudakov_explicit}
S(\alpha,b,Q,\mu)
=
s(\alpha, Q,b)+s(1-\alpha,Q,b)
+2\int_{1/b}^{\mu}\frac{d\mu'}{\mu'}\gamma_q(\alpha_s(\mu')),
\end{equation}
where \(\gamma_q\) is the quark anomalous dimension and the function \(s(\alpha,Q,b)\) resums the double logarithms,
\begin{equation}\label{chza:eq:s_function}
s(\alpha,Q,b)
=
\int_{1/b}^{\alpha\,Q}\frac{d\mu'}{\mu'}\left[
\ln\!\left(\frac{\alpha\,Q}{\mu'}\right) A(\alpha_s(\mu'))
+ B(\alpha_s(\mu'))
\right].
\end{equation}
The coefficient \(A(\alpha_s)\) is proportional to the cusp anomalous dimension and controls the double logarithmic behavior, while \(B(\alpha_s)\) encodes single logarithmic corrections.
The Sudakov suppression factor can be analytically computed and expressed as 
\begin{equation}\label{chza:eq:sudakov_explicit_results}
	S(\alpha,b,Q,\mu)
	= s(\alpha, Q,b)+s(1-\alpha,Q,b)-\frac{1}{\b_1}\log \frac{ \log (t/\Lambda_{ \rm QCD})}{\log (1/b\Lambda_{\rm QCD})}
\end{equation}
with
\begin{align}
	s(\alpha, Q,b) = &\frac{A^{(1)}}{2\beta_1} \hat{q} \ln \left( \frac{\hat{q}}{\hat{b}} \right) + \frac{A^{(2)}}{4\beta_1^2} \left( \frac{\hat{q}}{\hat{b}} - 1 \right) - \frac{A^{(1)}}{2\beta_1} (\hat{q} - \hat{b})-\frac{A^{(1)}\beta_2}{16\beta_1^3}\hat{q}\left[\frac{\ln(2\hat{b})+1}{\hat{b}} - \frac{\ln(2\hat{q})+1}{\hat{q}}\right]\nn\\
	&- \left[\frac{A^{(2)}}{4\beta_1^2} - \frac{A^{(1)}}{4\beta_1}\ln\left(\frac{e^{2\gamma-1}}{2}\right)\right]\ln\left(\frac{\hat{q}}{\hat{b}}\right)- \frac{A^{(1)}\beta_2}{32\beta_1^3}\left[\ln^2(2\hat{q}) - \ln^2(2\hat{b})\right]
\end{align}
and
\begin{equation}
	\hat{q} \equiv \ln \left(\alpha Q / (\sqrt{2} \Lambda_{ \rm QCD}) \right)
\end{equation}
\begin{equation}
	\hat{b} \equiv \ln (1 / b \Lambda_{\rm QCD})
\end{equation}
\begin{equation}
	t=\text{max}\left(Q\sqrt{\a_1\a_2},{1/b}\right)
\end{equation}
The coefficients $A^{(i)}$ and $\beta_i$ are
\begin{equation}
	\beta_1 = \frac{33 - 2n_f}{12}\,, \quad \beta_2 = \frac{153 - 19n_f}{24}\,,
\end{equation}
\begin{equation}
	A^{(1)} = \frac{4}{3}\,, \quad A^{(2)} = \frac{67}{9} - \frac{\pi^2}{3} - \frac{10}{27}n_f + \frac{8}{3}\beta_1 \ln\left(\frac{e^\gamma}{2}\right)\,,
\end{equation}
where $n_f = 3$ in this case is the number of quark flavors and $\gamma$ is the Euler constant.

The exponential factor \(\exp[-S]\) suppresses configurations with large transverse separation \(b\), corresponding to soft gluon exchange and long-distance dynamics. As a result, the dominant contribution to the convolution integral comes from the perturbative region \(b\lesssim 1/\Lambda_{\rm QCD}\), where the use of the hard kernel remains justified.

\subsection{Impact on the anomaly-induced amplitude}

The inclusion of Sudakov effects modifies the convolution structure of the anomaly-induced contributions. In particular, the form factors \(F_i^{TJJ}\) are now convoluted with the Sudakov-suppressed wave functions rather than with the purely collinear DA. Symbolically, one has
\begin{equation}\label{chza:eq:F_sudakov}
F_i^{TJJ} \;\longrightarrow\;
\int d^2 b \;
\tilde{\Psi}_\pi(\alpha_1,b,Q,\mu)\,
F_i^{TJJ}(\alpha_1,\alpha_2,q,b)\,
\tilde{\Psi}_\pi(\alpha_2,b,Q,\mu).
\end{equation}
The most important consequence of this modification is the suppression of endpoint regions \(\alpha\to 0,1\), where the denominators in Eqs.~\eqref{chza:eq:F12}--\eqref{chza:eq:F5} would otherwise enhance the integrand. The Sudakov factor therefore stabilizes the perturbative expansion and reduces the sensitivity to nonperturbative configurations.

Furthermore, the logarithmic dependence introduced by the two-point function \(B_0(p^2)\), which generates terms of the form \(\ln(Q^2/\mu^2)\), is now consistently combined with the Sudakov logarithms arising from soft-gluon resummation. The resulting structure reflects the interplay between ultraviolet renormalization, governed by the QCD beta function and the anomaly coefficient \(\mathcal{A}\), and infrared dynamics, encoded in the Sudakov exponent.

The Sudakov suppression provides a dynamical mechanism that enforces the dominance of short-distance configurations in exclusive processes at large momentum transfer. In the present context, it ensures that the anomaly-induced contribution to the GFFs remains under perturbative control, even in the presence of potentially dangerous endpoint configurations.

At the same time, the modification of the pion wave function reflects the fact that the hadronic structure probed in the hard process is not purely collinear, but involves a nontrivial interplay between longitudinal momentum fractions and transverse spatial distributions. The Sudakov factor effectively reshapes the wave function, suppressing large transverse separations and enhancing the contribution from compact configurations.

In summary, the inclusion of Sudakov corrections leads to a consistent extension of the factorization framework in which both perturbative and nonperturbative effects are properly organized. This refinement is essential for a quantitatively reliable description of the anomaly-induced contributions to the pion GFFs at large but finite momentum transfer.

\section{A phenomenological model for the Sudakov-improved pion wave function and its impact on the anomaly contribution}\label{chza:phenomesec2}

In order to make the Sudakov-resummation-improved factorization formula operational, one needs a concrete model for the pion wave function in transverse coordinate space. The purpose of this section is to introduce a simple but widely used phenomenological parametrization, to combine it with the Sudakov exponential, and to show how the anomaly-induced contribution to the pion GFFs is modified once the transverse degrees of freedom are taken into account explicitly. The discussion also makes clear how the endpoint suppression induced by Sudakov resummation stabilizes the convolution integrals that define the hard amplitude.

The starting point is the transverse-momentum-dependent pion wave function \(\Psi_\pi(\alpha,k_\perp,\mu)\), whose Fourier transform to impact-parameter space has been introduced in Eq.~\eqref{chza:eq:bspace}. A natural phenomenological ansatz consists in factorizing the longitudinal and transverse dependence of the nonperturbative input at a low scale \(\mu_0\),
\begin{equation}\label{chza:eq:psi_model_k}
\Psi_\pi^{(0)}(\alpha,k_\perp,\mu_0)
=
\phi_\pi(\alpha,\mu_0)\,\Sigma_\pi(\alpha,k_\perp),
\end{equation}
where \(\phi_\pi(\alpha,\mu_0)\) is the usual leading-twist DA and \(\Sigma_\pi\) describes the intrinsic transverse momentum profile \cite{Lepage:1980fj,Efremov:1979qk,Chernyak:1983ej}\cite{Braun:1989iv,Ball:2006wn,Jakob:1993iw}\cite{Huang:1994dy}

A commonly adopted Gaussian form is
\begin{equation}\label{chza:eq:gaussian_k}
\Sigma_\pi(\alpha,k_\perp)
=
\frac{\beta_\pi^2}{\pi\,\alpha(1-\alpha)}
\exp\!\left[-\beta_\pi^2\frac{k_\perp^2+m_q^2}{\alpha(1-\alpha)}\right],
\end{equation}
where \(\beta_\pi\) is a hadronic size parameter and \(m_q\) is an effective constituent-like quark mass introduced to regulate the infrared behavior of the model. The normalization is chosen so that
\begin{equation}\label{chza:eq:sigma_norm}
\int d^2k_\perp\,\Sigma_\pi(\alpha,k_\perp)
=
\exp\!\left[-\beta_\pi^2\frac{m_q^2}{\alpha(1-\alpha)}\right].
\end{equation}
When \(m_q\to 0\), the transverse profile integrates to unity, and the collinear DA is recovered directly from the transverse-momentum-dependent wave function. For finite \(m_q\), the exponential factor produces an additional suppression near the endpoints, which is phenomenologically useful in the modeling of exclusive amplitudes.

Passing to impact-parameter space, the Fourier transform of Eq.~\eqref{chza:eq:gaussian_k} gives
\begin{equation}\label{chza:eq:gaussian_b}
\tilde{\Sigma}_\pi(\alpha,b)
=
\exp\!\left[-\beta_\pi^2\frac{m_q^2}{\alpha(1-\alpha)}\right]
\exp\!\left[-\frac{\alpha(1-\alpha)b^2}{4\beta_\pi^2}\right],
\end{equation}
up to the conventional overall normalization associated with the definition of the Fourier transform. The corresponding pion wave function in \(b\)-space may then be written as
\begin{equation}\label{chza:eq:psi0_b}
\tilde{\Psi}_\pi^{(0)}(\alpha,b,\mu_0)
=
\phi_\pi(\alpha,\mu_0)\,
\exp\!\left[-\beta_\pi^2\frac{m_q^2}{\alpha(1-\alpha)}\right]
\exp\!\left[-\frac{\alpha(1-\alpha)b^2}{4\beta_\pi^2}\right].
\end{equation}
The Sudakov-improved wave function is obtained by multiplying this nonperturbative input by the resummed exponential factor introduced in Eq.~\eqref{chza:eq:sudakov_factor},
\begin{equation}\label{chza:eq:psi_b_sud}
\tilde{\Psi}_\pi(\alpha,b,Q,\mu)
=
\exp\!\left[-S(\alpha,b,Q,\mu)\right]\,
\tilde{\Psi}_\pi^{(0)}(\alpha,b,\mu_0).
\end{equation}
Substituting Eq.~\eqref{chza:eq:psi0_b} into Eq.~\eqref{chza:eq:psi_b_sud}, one finds the explicit form
\begin{equation}\label{chza:eq:psi_b_full}
\tilde{\Psi}_\pi(\alpha,b,Q,\mu)
=
\phi_\pi(\alpha,\mu_0)\,
\exp\!\left[-S(\alpha,b,Q,\mu)\right]
\exp\!\left[-\beta_\pi^2\frac{m_q^2}{\alpha(1-\alpha)}\right]
\exp\!\left[-\frac{\alpha(1-\alpha)b^2}{4\beta_\pi^2}\right].
\end{equation}
This expression makes the structure of the improved wave function transparent. The first exponential in Eq.~\eqref{chza:eq:psi_b_full} is perturbative and universal, encoding the resummation of soft and collinear logarithms. The second and third exponentials represent the intrinsic nonperturbative transverse structure of the pion. Their combined effect is to suppress both large transverse separations and endpoint regions, thus improving the convergence of the hard convolution.

The DA entering Eq.~\eqref{chza:eq:psi_b_full} may itself be modelled at the input scale \(\mu\) by a truncated Gegenbauer expansion keeping only the first nontrivial term,
\begin{equation}\label{chza:eq:phi_model_simple}
\phi_\pi(\alpha,\mu)
=
6\alpha(1-\alpha)
\left[
1+a_2(\mu)\,C_2^{3/2}(2\alpha-1)
\right],
\end{equation}
which already captures the main departure from the asymptotic shape and is sufficient for a first numerical analysis.
The Gegenbauer moment $a_2(\mu)$ is given by
\begin{equation}
	a_2(\mu)
	=
	a_2(\mu_0)
	\left(
	\frac{\alpha_s(\mu)}{\alpha_s(\mu_0)}
	\right)^{\frac{25\,C_F}{6}},
\end{equation}
and the running coupling is evaluated using Eq.~\eqref{chza:evolalph} at the largest hard scale in the convolution. The parameter $a_2(\mu_0)=0.2$ is chosen at the scale $\mu_0=1$ GeV following \cite{BAKULEV2001279}.

The hard amplitude for the pion GFFs is then rewritten as a threefold convolution in the longitudinal fractions and in the transverse separation,
\begin{align}\label{chza:eq:full_conv_sud}
\langle\pi(P_2)|T^{\mu\nu}|\pi(P_1)\rangle
=
\frac{if_\pi^2 C_Fg_s^2}{N_C}
\int_0^1 d\alpha_1
\int_0^1 d\alpha_2
\int_0^\infty 2\pi b\,db\;
\tilde{\Psi}_\pi(\alpha_1,b,Q,\mu)\,
K_\pi^{\mu\nu}(\alpha_1,\alpha_2,b,Q,\mu)\,
\tilde{\Psi}_\pi(\alpha_2,b,Q,\mu).
\end{align}
Here the hard kernel includes both the tree-level and the anomaly-induced components,
\begin{equation}\label{chza:eq:kernel_split_sud}
K_\pi^{\mu\nu}(\alpha_1,\alpha_2,b,Q,\mu)
=
K_{{\rm tree}}^{\mu\nu}(\alpha_1,\alpha_2,b,Q,\mu)
+
K_{TJJ}^{\mu\nu}(\alpha_1,\alpha_2,b,Q,\mu).
\end{equation}
The kernels appearing in Eq.~\eqref{chza:eq:kernel_split_sud} are obtained from the corresponding momentum-space amplitudes through a Fourier transformation with respect to the transverse momentum exchanged by the hard gluon,
\begin{equation}
	K_\pi^{\mu\nu}(\alpha_1,\alpha_2,b,Q,\mu)
	=
	\int \frac{d^2{k}_T}{(2\pi)^2}\,
	e^{-i{k}_T\cdot{b}}\,
	K_\pi^{\mu\nu}(\alpha_1,\alpha_2,{k}_T,Q,\mu),
\end{equation}
where ${b}$ denotes the transverse separation between the valence quarks in impact-parameter space. The same definition applies separately to the tree-level and to the $TJJ$ contributions.

The transverse momentum dependence is retained exclusively in the propagator of the exchanged gluon, while the intrinsic transverse momenta carried by the external constituent quarks are neglected in the quark propagators and in the Dirac numerators.
This approximation isolates the dominant non-collinear contribution associated with the exchanged hard gluon and allows the Fourier transform to be performed analytically in terms of modified Bessel functions.

The simplification becomes particularly important for the anomaly-induced $TJJ$ contribution. Indeed, the insertion generates non-local momentum structures and logarithmic terms in the gluon virtuality, making the exact two-dimensional Fourier transform analytically intractable if the full transverse dependence of all propagators is retained. Restricting the transverse dynamics to the exchanged gluon propagator preserves the factorized structure of the kernel and leads to a computable representation in impact-parameter space. In this way, both the tree-level and the anomaly-corrected amplitudes can be expressed in terms of one-dimensional Bessel transforms, which are suitable for stable numerical evaluation.

The explicit expressions of the tree-level kernel, the modified propagator structure induced by the $TJJ$ insertion, and the corresponding impact-parameter-space kernels are collected in Appendix~\ref{chza:kernelexpb}, together with a detailed discussion of the approximations employed in the treatment of the transverse momentum dependence.

Projecting again onto the two independent form factors, one obtains the Sudakov-improved expressions
\begin{align}
A_\pi(q^2)
&= \frac{f_\pi^2 C_F g_s^2}{N_C}
\int_0^1 d\alpha_1
\int_0^1 d\alpha_2
\int_0^\infty 2\pi b\,db\;
\tilde{\Psi}_\pi(\alpha_1,b,Q,\mu)
\Bigl[
F_1^{\rm tree}+F_1^{TJJ}+F_3^{\rm tree}+F_3^{TJJ}
\Bigr]
\tilde{\Psi}_\pi(\alpha_2,b,Q,\mu),
\label{chza:eq:Api_sud}
\\[6pt]
D_\pi(q^2)
&=
- \frac{f_\pi^2 C_F g_s^2}{N_C}
\int_0^1 d\alpha_1
\int_0^1 d\alpha_2
\int_0^\infty 2\pi b\,db\;
\tilde{\Psi}_\pi(\alpha_1,b,Q,\mu)
\frac{2\bigl(F_5^{\rm tree}+F_5^{TJJ}\bigr)}{q^2}
\tilde{\Psi}_\pi(\alpha_2,b,Q,\mu),
\label{chza:eq:Dpi_sud}
\end{align}
where the \(F_i\) now depend implicitly on \(b\) through the transverse-momentum dependence retained in the hard scattering kernel, and are explicitly given in Appendix~\ref{chza:kernelexpb}. The anomaly-induced correction to the \(D\)-term is therefore given by
\begin{equation}\label{chza:eq:Dpi_anom_sud}
D_\pi^{\rm anom}(q^2)
=
-\frac{f_\pi^2 C_F g_s^2}{N_C}
\int_0^1 d\alpha_1
\int_0^1 d\alpha_2
\int_0^\infty 2\pi b\,db\;
\tilde{\Psi}_\pi(\alpha_1,b,Q,\mu)
\frac{2F_5^{TJJ}(\alpha_1,\alpha_2,b,q,\mu)}{q^2}
\tilde{\Psi}_\pi(\alpha_2,b,Q,\mu).
\end{equation}

At this stage it is useful to examine the effect of the Sudakov improvement on the anomaly contribution. In the purely collinear treatment, the form factor \(F_5^{TJJ}\) contains denominators proportional to
\begin{equation}\label{chza:eq:endpoint_den}
\alpha_1(1-\alpha_1)\alpha_2(1-\alpha_2),
\end{equation}
which enhance the integrand near the endpoints \(\alpha_i\to 0,1\). Although the asymptotic DA \(\phi_\pi^{\rm as}(\alpha)=6\alpha(1-\alpha)\) already renders the convolution finite, the endpoint region may still contribute significantly once logarithmic corrections and non-asymptotic models are included. The Sudakov factor cures this problem dynamically. Indeed, for small \(\alpha\) or \(1-\alpha\), the exponent \(S(\alpha,b,Q,\mu)\) grows, so that
\begin{equation}\label{chza:eq:endpoint_supp}
\exp[-S(\alpha,b,Q,\mu)]\to 0,
\qquad
\alpha\to 0,1.
\end{equation}
This implies a strong suppression of the endpoint configurations that would otherwise endanger the perturbative stability of the result.

A second important effect concerns the \(b\)-integration. Without Sudakov suppression, the convolution in Eq.~\eqref{chza:eq:full_conv_sud} would receive substantial contributions from large \(b\), corresponding to long-distance transverse separations where the perturbative hard kernel cannot be trusted. The Sudakov exponent suppresses precisely this region,
\begin{equation}\label{chza:eq:large_b_supp}
\exp[-S(\alpha,b,Q,\mu)]\to 0,
\qquad
b\to \infty,
\end{equation}
and therefore ensures that the dominant contribution arises from a finite window of transverse separations. In this way the improved factorization formula becomes internally consistent: the nonperturbative information remains encoded in the model wave function, while the hard kernel is sampled mainly in the region where perturbation theory is applicable.

For practical numerical applications, one needs to specify the scales entering the Sudakov exponent. A standard choice is to set the factorization and renormalization scale to the largest perturbative scale in the problem,
\begin{equation}\label{chza:eq:hard_scale_choice}
t=\max\!\left(\sqrt{\alpha_1\alpha_2}\,Q,\frac{1}{b}\right),
\end{equation}
and to evaluate the running coupling \(\alpha_s(t)\) at this scale. The Sudakov exponent is then written in terms of \(t\), while the DA is evolved from the input scale \(\mu_0\) to the same hard scale by the usual ERBL evolution. In a first phenomenological analysis, however, one may keep the evolution of \(\phi_\pi\) fixed at a representative scale and focus on the dominant effect of the Sudakov exponential itself.

Within such a setup, the principal effect of the Sudakov improvement on the anomaly-induced amplitude can be summarized schematically as
\begin{equation}\label{chza:eq:supp_scheme}
D_\pi^{\rm anom}(Q^2)
=
D_{\pi,\,{\rm coll}}^{\rm anom}(Q^2)\,
\mathcal{S}_D(Q^2),
\end{equation}
where \(D_{\pi,\,{\rm coll}}^{\rm anom}\) is the result obtained in the collinear approximation and \(\mathcal{S}_D(Q^2)\) is an effective suppression factor generated by the \(b\)-space convolution with the Sudakov-improved wave functions. Although \(\mathcal{S}_D(Q^2)\) is not universal and depends on the chosen model parameters, it is typically smaller than unity and decreases the relative importance of the endpoint regions. In the \(A_\pi(Q^2)\) projection the anomalous part cancels, while the full \(TJJ\) insertion still affects the curve through its non-anomalous tensor components.  The effect is more directly controlled in the \(D\)-term and in the trace form factor because the trace-sector coefficient \(F_5^{TJJ}\) is sensitive to the scalar and logarithmic pieces of the kernel.

The phenomenological parameters entering the model, such as \(\beta_\pi\), \(m_q\), and the Gegenbauer moments \(a_n(\mu_0)\), may be fixed by comparison with other exclusive observables, for instance the electromagnetic pion form factor or the \(\gamma\gamma^*\to\pi\) transition form factor. Once these parameters are chosen, the same Sudakov-improved wave function can be used consistently in the computation of the GFFs. This makes the framework predictive and allows one to assess whether the anomaly-induced correction survives as a numerically appreciable effect once the realistic suppression of long-distance configurations is taken into account.

In summary, the introduction of a phenomenological \(b\)-space model for the pion wave function provides the missing ingredient needed to turn the Sudakov-improved factorization formula into a calculable scheme. The wave function is modified in two essential ways: first, by the intrinsic transverse Gaussian profile, which encodes the finite transverse size of the pion, and second, by the Sudakov exponent, which suppresses large transverse separations and endpoint regions. When these effects are incorporated into the convolution for \(D_\pi(q^2)\), the anomaly-induced contribution remains present but is reshaped and typically reduced relative to the purely collinear estimate. This mechanism is essential for a realistic perturbative analysis of the pion GFFs at finite momentum transfer.

\section{Numerical Results}\label{chza:numericalsec}

\begin{figure}
	\centering
	\includegraphics[scale=0.5]{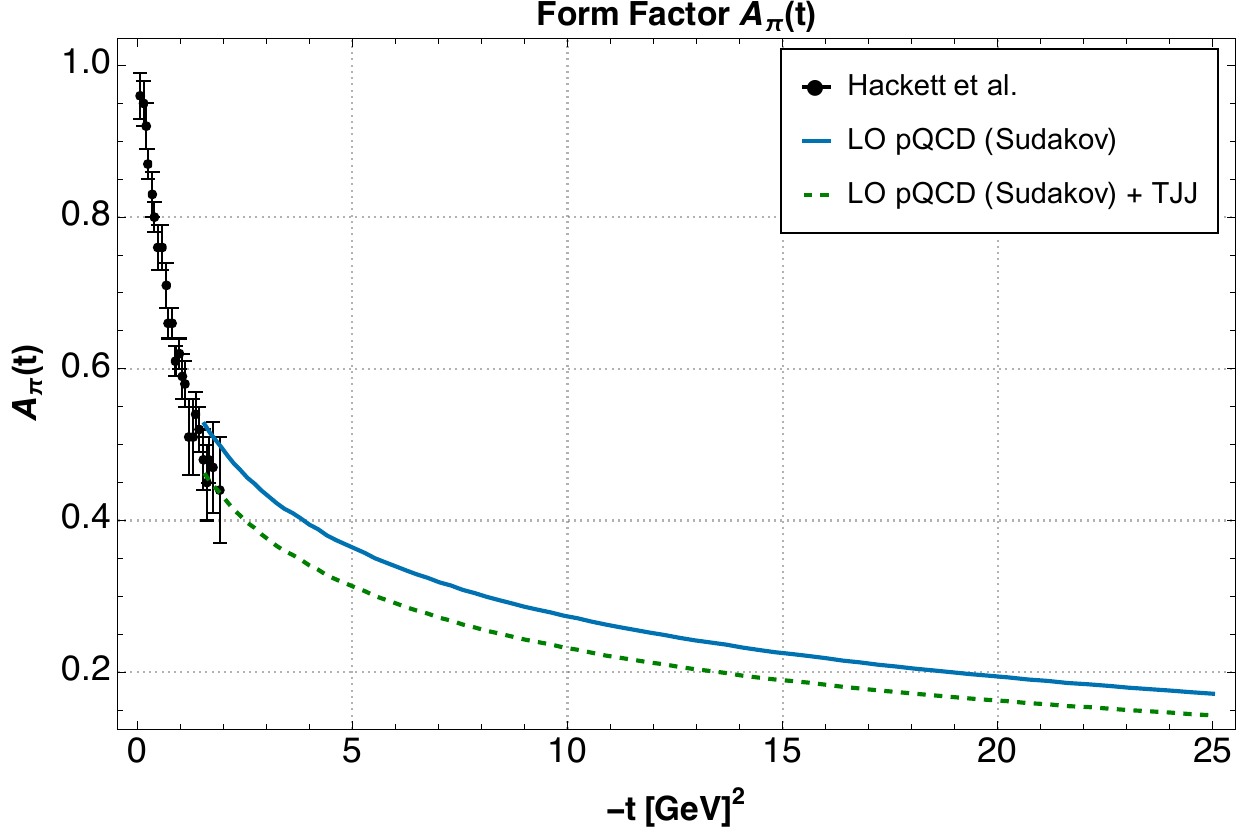}
	\caption{$Q^2=-t$. In the momentum form factor \(A_\pi\), the isolated anomaly cancels and does not generate a separate contribution.  The full \(TJJ\) insertion nevertheless lowers the leading Sudakov curve at small \(Q^2\), improving the local overlap with the lattice data in the first kinematic window. The curves are obtained for $a_2(\mu_0)=0.2$, $\mu_0=1$ GeV, $m_q=0.33\,\, \text{GeV}$ and $\beta_\pi=0.24 \,\,\text{GeV}^{-1}$.}
	\label{chza:ap}
\end{figure}
\begin{figure}
	\centering
	\includegraphics[scale=0.5]{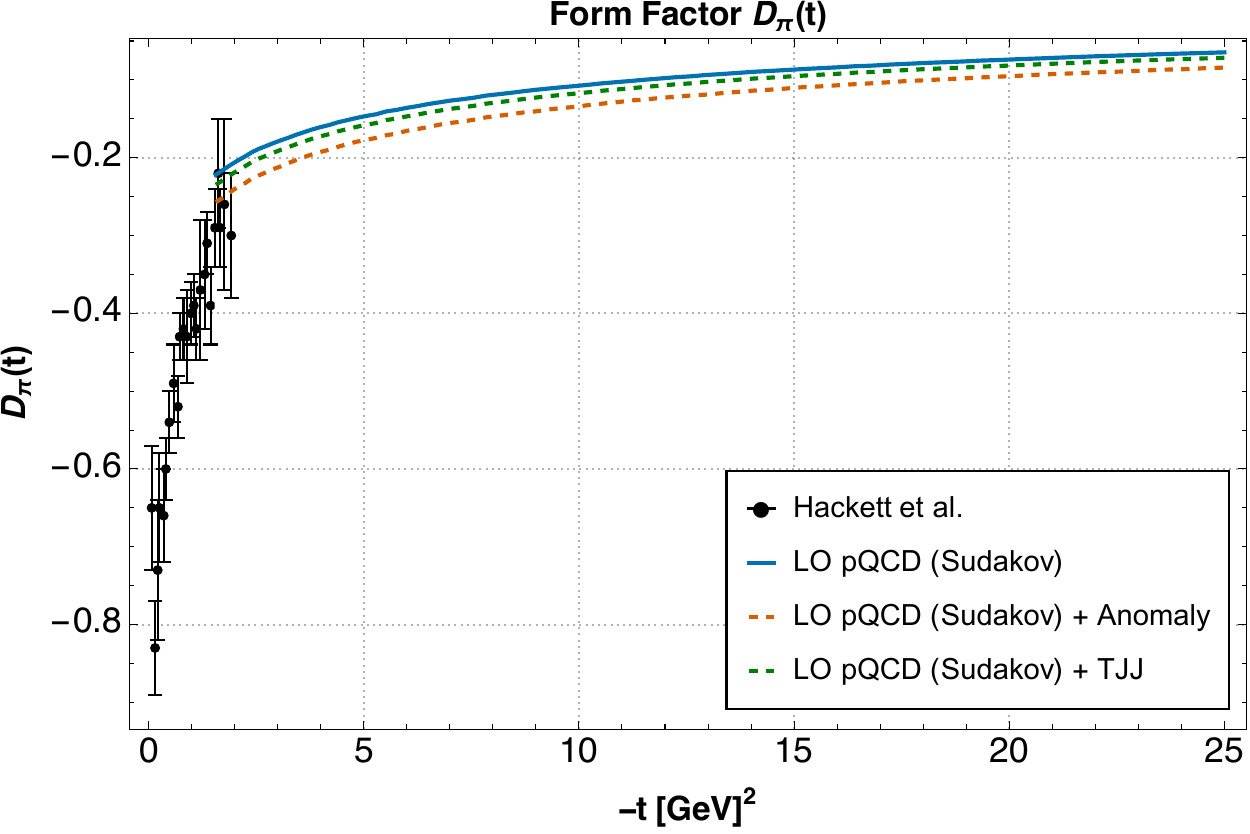}
	\caption{In the \(D_\pi\) form factor the anomaly gives an important contribution.  The displayed full \(TJJ\) correction is more modest and should not be read as the complete order-\(\alpha_s^2\) hard kernel; additional non-anomalous terms may improve the comparison with the data. The curves are obtained for $a_2(\mu_0)=0.2$, $\mu_0=1$ GeV, $m_q=0.33\,\, \text{GeV}$ and $\beta_\pi=0.24 \,\,\text{GeV}^{-1}$.}
	\label{chza:dp}
\end{figure}
\begin{figure}
	\centering
	\includegraphics[scale=0.65]{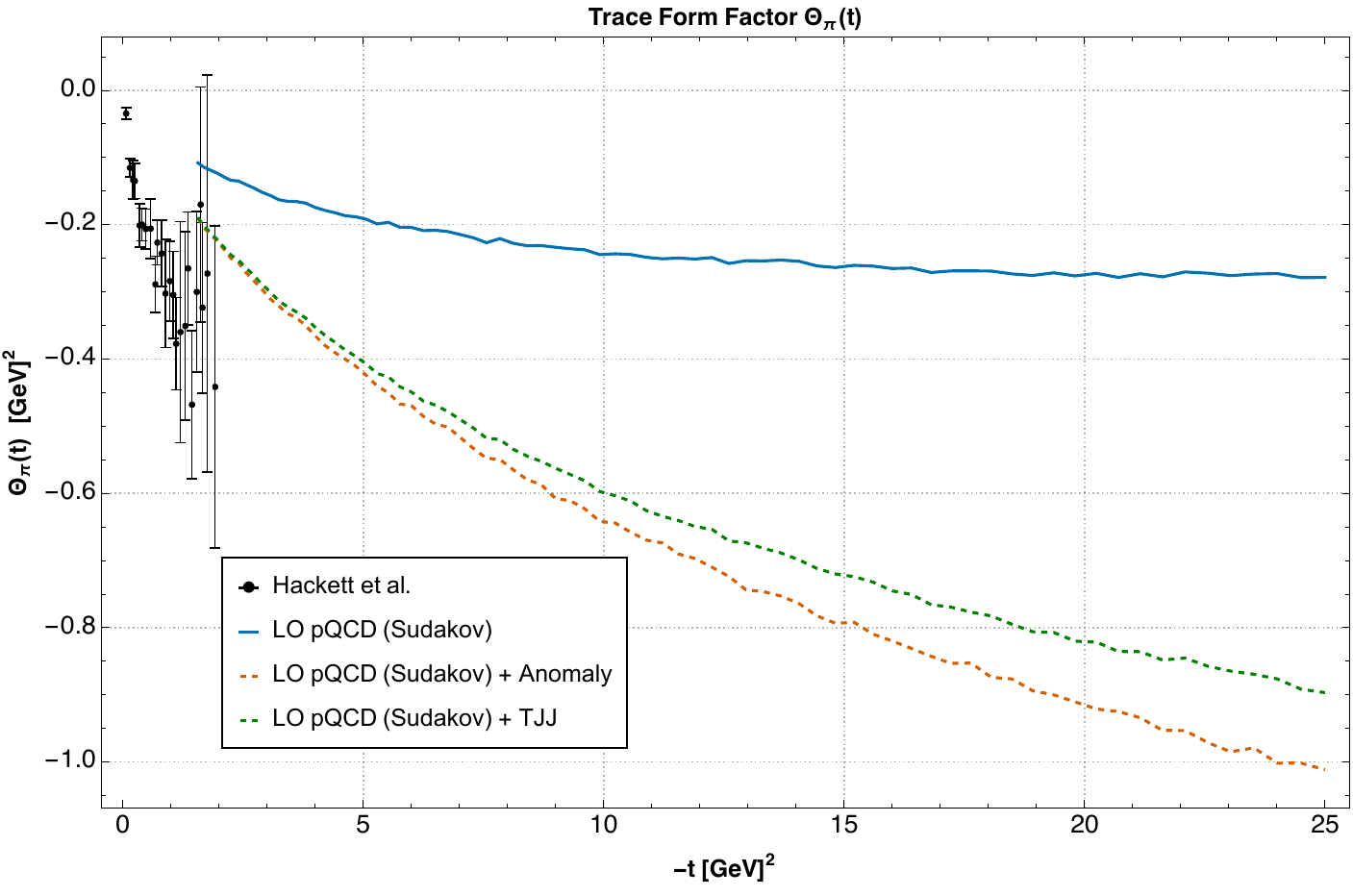}
	\caption{The trace form factor \(\Theta_\pi\) is the channel in which the anomaly is most significant: it dominates the \(TJJ\) contribution at both low and high momentum transfer.  Treating the anomaly as a constrained contribution in the hard part improves the trace prediction and helps the interpretation of the \(D_\pi\) form factor. The curves are obtained for $a_2(\mu_0)=0.2$, $\mu_0=1$ GeV, $m_q=0.33\,\, \text{GeV}$ and $\beta_\pi=0.24 \,\,\text{GeV}^{-1}$.}
	\label{chza:trp}
\end{figure}

In this section we discuss the numerical behavior of the pion GFFs obtained from the Sudakov-improved factorization framework described above, including the contribution generated by the non-Abelian $TJJ$ insertion.

A crucial aspect of the numerical calculation concerns the treatment of the impact-parameter integration.  The transverse separation $b$ is restricted to the perturbative region,
\begin{equation}
	0 \le b \lesssim \frac{1}{\Lambda_{\rm QCD}},
\end{equation}
since larger transverse distances correspond to configurations dominated by genuinely nonperturbative soft-gluon dynamics. In practice, the Sudakov exponent suppresses the integrand strongly as $b$ approaches $1/\Lambda_{\rm QCD}$, leading to a natural dynamical cutoff of the convolution integral. As emphasized in Ref.~\cite{Li:1992nu}, the perturbative contribution becomes increasingly dominated by small transverse separations as the hard scale grows.

In the present analysis we implement this prescription explicitly by truncating the numerical integration over $b$ at
\begin{equation}
	b_{\rm max}\sim \frac{1}{\Lambda_{\rm QCD}},
\end{equation}
which corresponds approximately to the onset of the nonperturbative region. This cutoff is particularly important for the stability of the anomaly-induced $TJJ$ contribution. Indeed, the $TJJ$ insertion generates additional logarithmic dependence on the gluon virtuality, enhancing the sensitivity of the Fourier transform to large transverse separations. Without Sudakov suppression and the restriction to the perturbative domain, the convolution would receive sizable contributions from regions where the perturbative treatment of the hard kernel is no longer reliable.

The numerical results for the two GFFs are shown in Fig.~\ref{chza:ap} and in Fig.~\ref{chza:dp}, while the trace form factor is shown in Fig.~\ref{chza:trp}, where the leading-order Sudakov-improved perturbative prediction is compared with the corresponding curves including the isolated anomaly contribution and the full $TJJ$ insertion. The calculations are also compared with the lattice-QCD determination of Hackett \textit{et al.}  in the low- and intermediate-$Q^2$ region \cite{Hackett:2023pyn}.

The resulting form factors exhibit a characteristic projection hierarchy. The momentum form factor $A_\pi(Q^2)$ receives only a moderate correction from the complete $TJJ$ insertion, while the isolated anomaly contribution cancels in the leading-twist projection, consistently with the tensor decomposition of the stress-tensor correlator. The cancellation is straightforward from Eqs \eqref{chza:eq:F12}  and \eqref{chza:eq:F3}. Nevertheless, the non-anomalous tensor structures contained in the full $TJJ$ kernel lower the leading-order prediction at small momentum transfer, improving the qualitative agreement with the lattice-QCD data.

A different behavior is observed for the D-term form factor $D_\pi(Q^2)$. In this channel the scalar structures associated with the trace anomaly project much more efficiently, producing a sizable correction already at moderate values of $Q^2$. This enhancement reflects the sensitivity of the D-term to the internal stress distribution and to the scalar sector of the energy-momentum tensor.

The clearest signal of the anomaly emerges in the trace form factor,
\begin{equation}
	\Theta_\pi(Q^2)
	=
	g_{\mu\nu}
	\langle \pi(P_2)|T^{\mu\nu}|\pi(P_1)\rangle ,
\end{equation}
where the scalar projection directly selects the trace component of the $TJJ$ correlator. In this case the anomaly-induced contribution dominates the full $TJJ$ correction over the entire momentum-transfer region considered in the analysis. This behavior confirms that the trace form factor is the most direct observable sensitive to the beta-function-controlled scalar channel generated by the conformal anomaly.

The hierarchy observed among the three form factors therefore provides a nontrivial consistency check of the tensor decomposition of the $TJJ$ vertex. A universal shift of all form factors would simply correspond to a global normalization effect, whereas the selective enhancement of the scalar observables demonstrates that the anomaly contribution is correctly projected onto the trace sector of the pion energy-momentum tensor.

Finally, it should be emphasized that the present calculation is not intended as a complete low-energy description of the pion GFFs. At low momentum transfer, additional contributions associated with soft overlap, higher-twist effects, chiral dynamics, and quark-mass corrections to the trace become relevant. The purpose of the present analysis is instead to isolate the perturbative trace-anomaly contribution within a controlled Sudakov-improved factorization framework and to study its projection onto the various GFFs.

Although the momentum-transfer range explored by the lattice data is not strictly within the asymptotic perturbative regime, the comparison is still useful in order to investigate the effect of the trace-anomaly contribution on the various GFFs. In particular, both the isolated anomaly contribution and the complete $TJJ$ correction tend to lower the pure leading-order perturbative prediction in the low/intermediate-$Q^2$ region, producing a visibly improved overlap with the available lattice points.

This effect is especially evident in the trace form factor $\Theta_\pi(Q^2)$ and, to a lesser extent, in the D-term form factor $D_\pi(Q^2)$, where the scalar component associated with the conformal anomaly contributes more efficiently through the tensor decomposition of the $TJJ$ correlator. In the case of $A_\pi(Q^2)$, the isolated anomaly contribution cancels at leading twist, but the full $TJJ$ insertion still generates a moderate downward shift of the perturbative curve at low momentum transfer.

At the same time, the comparison should be interpreted with some caution. In the low-$Q^2$ region the pion dynamics is expected to receive important contributions from soft overlap, chiral effects, higher-twist corrections, and genuinely nonperturbative physics \cite{Liu:2024zahed}, which are not fully captured within a perturbative factorization framework. Consequently, the present calculation should not be viewed as a precision description of the low-energy pion GFFs. Rather, its purpose is to isolate the perturbative trace-anomaly contribution and to study how the non-Abelian $TJJ$ insertion modifies the hierarchy of the various tensor projections.

From this perspective, the qualitative agreement obtained after including the anomaly-related corrections is encouraging. It indicates that the scalar channel associated with the QCD beta function contributes in the expected direction and improves the phenomenological behavior of the perturbative prediction even in a kinematic region where strict collinear factorization is only partially applicable.

\section{Dilaton Exchange as a Correlated Pair Channel}
\label{sec:dilaton-correlated-pair}

The perturbative dilaton discussed in the analysis of GFFs should not be interpreted, in the first instance, as an elementary scalar added to QCD.  It is a scalar projection of the renormalized \(TJJ\) vertex.  More precisely, it is the part of the stress-tensor insertion which is selected by the trace sector and whose normalization is fixed by the beta function.  This interpretation follows the perturbative anomaly-pole analysis of the \(TJJ\) correlator and its gravitational effective action representation developed in Refs.~\cite{Armillis:2009pq,Armillis:2010qk,Coriano:2018zdo}. The corresponding scalar form factor in the \(TJJ\) correlator is therefore the perturbative realization of a dilaton-like exchange.  Its role in hard exclusive processes is to provide a definite trace-channel component of the hard kernel, rather than a universal rescaling of all GFFs.  The same mechanism underlies the recent formulation of hadron GFFs from momentum-space CFT and perturbative QCD~\cite{Coriano:2024qbr}.

The point can be made precise by using the momentum-space decomposition of the \(TJJ\) vertex developed by Bzowski, McFadden and Skenderis and later applied to anomaly form factors in QCD~\cite{Bzowski:2013sza,Bzowski:2018fql,Coriano:2018bbe,Coriano:2020ees,Coriano:2024qbr,Coriano:2025sumrule}.  In the conformal limit the trace sector contains a form factor of the form
\begin{equation}
	\Phi_{\rm anom}(q^2)\sim \frac{\mathcal A}{q^2},
	\qquad
	\mathcal A =
	\frac{g_s^2}{16\pi^2}\,\beta_0 ,
\end{equation}
with \(q\) the momentum carried by the stress tensor.  This pole is the momentum-space image of the nonlocal anomaly action. In a local phenomenological language one may introduce a dilaton field coupled to \(F^2\).  In the perturbative construction, however, the same object is already present in the one-particle-irreducible \(TJJ\) correlator as the scalar anomaly pole.  The connection between this pole and the scalar component of the gravitational exchange has also been emphasized in the anomaly-mediated description of conformal sectors~\cite{Coriano:2026vya}.

\begin{figure}[t]
	\centering
	\includegraphics[width=0.22\textwidth]{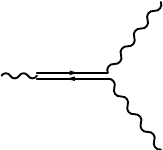}
	\caption{Effective pole representation of the anomalous \(TJJ\) interaction. The diagram gives a graphical representation of the nonlocal \(R^{(1)}\Box^{-1}F^2\) exchange generated by the anomaly form factor, as in the Abelian and non-Abelian analyses of Refs.~\cite{Armillis:2009pq,Armillis:2010qk,Coriano:2025sumrule}.}
	\label{fig:dilaton-pole-exchange}
\end{figure}

This interpretation is sharpened by Duff's definition of the anomaly, namely the failure of the trace operation to commute with the quantum average,
\begin{equation}
	\mathcal A
	=
	g_{\mu\nu}\langle T^{\mu\nu}\rangle
	-
	\langle T^\mu_{\ \mu}\rangle .
\end{equation}
In dimensional regularization the \(TJJ\) tensor basis shows that the anomaly pole is not produced by an arbitrary pole term inserted by hand.  The singularity is traced to a unique renormalized form factor, originally introduced in \cite{Giannotti:2008cv}.  The trace Ward identities imply that, when \(d\to4\), the \(1/(d-4)\) singularity of this form factor feeds the traced tensor structure and generates the \(1/q^2\) pole \cite{Coriano:2018zdo}.  All other form factors are finite in this limit.  Thus the dilaton pole is the residual effect of renormalization in the trace sector, consistently with the CFT reconstruction of anomalous graviton vertices and with the anomaly-action analysis of higher stress-tensor correlators~\cite{Coriano:2017mux,Coriano:2018bbe,Coriano:2020ees}.

The same conclusion has a dispersive formulation.  For general virtualities and nonzero fermion mass the scalar form factor can be represented as
\begin{equation}
	\Phi_{\rm anom}(q^2;p_1^2,p_2^2,m^2)
	=
	\frac{1}{\pi}
	\int_0^\infty ds\,
	\frac{\rho_{\rm anom}(s;p_1^2,p_2^2,m^2)}
	{s-q^2-i0},
	\label{eq:anom-dispersion}
\end{equation}
with a spectral density obeying the anomaly sum rule~\cite{Coriano:2014gja,Coriano:2025sumrule}
\begin{equation}
	\frac{1}{\pi}
	\int_0^\infty ds\,
	\rho_{\rm anom}(s;p_1^2,p_2^2,m^2)
	=
	\mathcal A .
	\label{eq:anom-area-law}
\end{equation}
The right-hand side is independent of the mass and of the external virtualities.  The spectral distribution may change, but the total area is fixed.  This is the sense in which the dilaton exchange can be viewed as a correlated exchange of intermediate pairs.  The spectral density is generated by the intermediate two-particle states of the \(TJJ\) subgraph, such as quark-antiquark pairs in the fermion loop and gluon pairs in the non-Abelian sector.  These states are not exchanged independently; they are projected onto the color-singlet scalar channel selected by the trace of the stress tensor.  The dilaton is therefore the coherent scalar channel carried by these correlated partonic pairs.  This spectral-flow interpretation extends the earlier anomaly sum-rule analyses of dilaton and superconformal channels~\cite{Coriano:2014gja,Coriano:2025sumrule}.

After the \(TJJ\) correlator is embedded in the pion matrix element, the same sum rule should be interpreted as a matching constraint on the scalar part of the hard kernel.  It does not become a separate universal sum rule for each individual pion GFF.  Rather, the fixed anomaly strength in Eq.~\eqref{eq:anom-area-law} is projected onto the hadronic amplitude by the pion DA and, at finite momentum transfer, by the Sudakov-improved wave function.  The convolution redistributes this scalar strength among the observable projections: it may cancel, or be numerically suppressed, in the anomalous part of \(A_\pi\), while it remains visible in \(D_\pi\) and is selected most directly by the trace form factor.

At lower momentum transfer the same statement admits a hadronic spectral interpretation.  The partonic two-particle cut of the \(TJJ\) subgraph is then replaced by scalar hadronic intermediate states, including pion continua, scalar resonances, and gluonic components.  The detailed saturation of the sum rule is nonperturbative, but the total anomaly strength remains fixed by the trace identity.  In this sense the perturbative dilaton pole is the short-distance representative of a scalar spectral channel which, at hadron level, is redistributed over physical intermediate states.

In the massive theory the same strength is distributed over the cut, beginning at the physical threshold \(s=4m^2\) in the fermion sector.  As the conformal limit is approached, the sequence of spectral densities is squeezed toward \(s=0\).  In distributional form one obtains a spectral flow of the type
\begin{equation}
	\rho_{\rm anom}(s;m^2)
	\longrightarrow
	\pi\,\mathcal A\,\delta(s),
	\qquad
	m\to0,
\end{equation}
so that the continuum representation collapses into the anomaly pole.  The pole is therefore not a separate assumption; it is the limiting form of a correlated two-particle spectral distribution whose area is fixed by the anomaly.  In the gluonic part of the non-Abelian vertex the localization at \(s=0\) is already present in the massless sector, while fermion mass effects smear the pole into a continuum before the conformal limit is taken.

There is, however, an important qualification.  The analysis of the particle-pole residue in light-cone variables shows that the anomaly pole and a genuine particle pole are not identical notions in the \(TJJ\) case~\cite{Giannotti:2008cv,Armillis:2009pq,Armillis:2010qk,Coriano:2025sumrule}.  If the external vector legs are off shell, or if the fermions in the loop are massive, the residue of \(q^2\langle TJJ\rangle\) vanishes as \(q^2\to0\).  A nonzero residue appears only in the massless, on-shell, light-cone configuration corresponding to a virtual graviton decaying into two transverse on-shell gluons or photons.  In the on-shell basis of the Abelian \(TJJ\) correlator, later extended to the non-Abelian case, the surviving residue contains two independent tensor structures.  With \(k=p_1+p_2\), \(s=k^2\), \(p_1^2=p_2^2=0\) and Eq. \eqref{chza:defu},
one may write the light-cone residue by separating the scalar anomaly pole from the transverse-traceless pole contribution as
\begin{equation}
	\lim_{s\to0}
	s\,\langle T^{\mu\nu}(k)J^{a \alpha}(p_1)J^{b \beta}(p_2)\rangle
	=
	\frac{g_s^2}{48\pi^2}\delta^{a b}\left(\frac{11}{3}C_A-\frac{2}{3}n_f\right)
	\tilde\phi_1^{\mu\nu\alpha\beta}
	-\frac{g_s^2}{288\pi^2}\delta^{a b}(n_f - C_A)\,
	\tilde\phi_2^{\mu\nu\alpha\beta}.
		\label{eq:tjj-lightcone-two-poles}
\end{equation}
The second term denotes a traceless contribution, already present in the Abelian case.\\
The first tensor is the scalar anomaly structure,
\begin{equation}
	\tilde\phi_1^{\mu\nu\alpha\beta}
	=
	\left(sg^{\mu\nu}-k^\mu k^\nu\right)
	u^{\alpha\beta}(p_1,p_2),
\end{equation}
and is the residue associated with the \(R^{(1)}\Box^{-1}F^2\) exchange represented in Fig.~\ref{fig:dilaton-pole-exchange}.\\  The second tensor is
\begin{equation}
	\tilde\phi_2^{\mu\nu\alpha\beta}
	=
	-2\,u^{\alpha\beta}(p_1,p_2)
	\left[
	sg^{\mu\nu}
	+2\left(p_1^\mu p_1^\nu+p_2^\mu p_2^\nu\right)
	-4\left(p_1^\mu p_2^\nu+p_2^\mu p_1^\nu\right)
	\right],
	\label{eq:traceless-pole-tensor}
\end{equation}
which obeys \(g_{\mu\nu}\tilde\phi_2^{\mu\nu\alpha\beta}=0\) in four dimensions and is therefore not part of the trace anomaly. In the Abelian case (QED), discussed in \cite{Giannotti:2008cv}, the additional massless pole associated to this structure was interpreted as a plasmon-like mode of the anomaly vertex, proportional to the number of fermion flavors $n_f$.  As discussed in \cite{Coriano:2025sumrule}, it is a light-cone pole which in the context of a gravitational exchange is mediated by the spin--2 part of the graviton propagator rather than by the scalar trace projection \cite{Coriano:2026vya}. It plays a role in the scattering of two gluons $(g g  \to gg)$ once two $TJJ$ vertices are coupled by the exchange of an intermediate graviton. The factor $(n_f - C_A)$ in front of $\tilde\phi_2$ in \eqref{eq:tjj-lightcone-two-poles} accounts for both the quark and the gluon/ghost contributions.\\
This second traceless structure contributes to the anomaly effective action describing the interaction around the light-cone, of the form
\begin{align}\label{expoleac}
	S_{\text{extra pole}} = \frac{g_s^2}{72 \pi^2} \left(n_f - C_A \right) \int &d^4x \sqrt{-g} \int d^4x' \sqrt{-g'}\, h_{\mu\nu}(x) \Box^{-1}_{x,x'} 
	\nonumber\\&
	\times \left[ 
	3 (\partial^\mu \partial^\nu F^a_{\alpha\beta}) F^{\alpha\beta a}
	+ \frac{1}{4} \left( g^{\mu\nu} \Box - 4 \partial^\mu \partial^\nu \right) F^a_{\alpha\beta} F^{\alpha\beta a}
	\right]_{x'},
\end{align}
while the first tensor structures is associated with the well-known effective action of the form \cite{Armillis:2009pq}
\bea
S_{\text{anom}} &=&\frac{1}{3} \, \frac{g^3}{16 \pi^2} \left (  - \frac{11}{3} \, C_A + \frac{2}{3} \, n_f \right)  \, \int d^4 x \, d^4 y \,R^{(1)}(x)\, \square^{-1}(x,y) \, F^a_{\alpha \beta}F^{\alpha \beta a},
\eea
that generalizes the expression derived in \cite{Giannotti:2008cv} for QED. $R^{(1)}(x)$ is the linearized scalar curvature.

This sector has the same massless propagator as the scalar anomaly exchange, but its residue is traceless.  It should therefore be kept conceptually separate from the dilaton channel: the dilaton is the scalar projection fixed by the trace Ward identity and by the anomaly sum rule, while the plasmon-like pole is a light-cone excitation of the transverse-traceless sector.  This separation is precisely the reason why the CFT momentum-space decomposition and the explicit perturbative \(TJJ\) calculation have to be used together when the hard kernel is projected onto pion GFFs.

For the pion GFFs this distinction is essential.  The hard kernel contains the one-loop \(TJJ\) insertion, and the pion DAs attach to the external partonic lines.  The trace-anomaly contribution is then the scalar projection of the correlated intermediate pair exchanged inside this hard subgraph.  It contributes most directly to the trace form factor and to the scalar part of the \(D_\pi\)-term, while it cancels in the anomalous part of the \(A_\pi\) projection.  This is precisely what one expects if the dilaton is not an elementary external state, but the coherent trace-channel component of the two-parton intermediate spectrum.

\section{Conclusions}

We have analyzed the pion GFFs in the hard-scattering regime by combining the standard collinear factorization picture with the one-loop non-Abelian \(TJJ\) correlator. The calculation isolates the trace-anomaly part of the correlator and shows how it enters the short-distance kernel through the coefficient \(\mathcal{A}=g_s^2\beta_0/(16\pi^2)\). This coefficient is fixed by the QCD beta function and therefore provides a perturbative realization of the scale anomaly inside the hard kernel.

For a spin-zero external state the trace projection relates the scalar matrix element to the combination \(A_\pi+3D_\pi\) in the chiral large-\(Q^2\) limit. In the explicit matching performed here, the anomaly contribution appears most directly in the coefficient \(F_5^{TJJ}\), and hence in the projection onto the \(D\)-term. The isolated anomaly cancels in \(A_\pi\), while the full \(TJJ\) insertion still lowers the leading curve at small \(Q^2\) through its non-anomalous tensor components.  In \(D_\pi\) the anomaly gives an important contribution, although a complete order-\(\alpha_s^2\) hard kernel would also require the remaining non-anomalous radiative terms.

The Sudakov-improved formulation extends the collinear result by including transverse separation and the resummation of soft-collinear logarithms. The resulting suppression of large \(b\) and endpoint configurations stabilizes the anomaly-induced convolution and makes the perturbative description more reliable at finite momentum transfer. The numerical analysis therefore supports a picture in which the trace anomaly gives a distinct, beta-function-controlled correction to the mechanical structure of the pion, with its clearest impact in the trace form factor and an important contribution to \(D_\pi\).

The comparison with the instanton-vacuum analysis of Ref. \cite{Liu:2024zahed} clarifies the domain of validity of this result. Their calculation supplies a nonperturbative soft and semi-hard realization of chiral and scale breaking, including twist-3 pion DAs and pressure/shear distributions. The present calculation supplies the short-distance $TJJ$
anomaly kernel. A complete phenomenological description of pion GFFs should combine these ingredients, matching the
nonperturbative instanton or chiral component to the perturbative TJJ-improved kernel while avoiding double counting
in the scalar trace sector.

\section*{Acknowledgements}

We dedicate this work to Prof. George Sterman. This work is partially supported by INFN under Iniziativa Specifica QG-sky, and by National Science and Technology Council of the Republic of China under Grant No. NSTC-113-2112-M-001-024-MY3. C.C. thanks the Yang Institute and the Simons Center at Stony Brook for hospitality while completing this work.

\appendix

\section{Feynman rules}
\label{chza:rules}
The Feynman rules used throughout the paper are collected here
\begin{itemize}
	\item {Graviton - fermion - fermion vertex}
	\\
	\\
	\bmi{95pt}
	\includegraphics[scale=1.0]{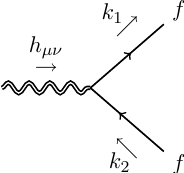}
	\emi
	\bmi{70pt}
	\begin{eqnarray*}
	&=& - i \, \frac{\kappa}{2} \, V^{\prime}_{\mu\nu}(k_1,k_2) \nn \\
	&=& - i \, \frac{\kappa}{2} \, \left\{ \frac{1}{4} \left[\gamma_\mu (k_1 + k_2)_\nu
	+\gamma_\nu (k_1 + k_2)_\mu \right] - \frac{1}{2} g_{\mu \nu}
	[\gamma^{\lambda}(k_1 + k_2)_{\lambda} - 2 m]  \right\} \nn \\
	\end{eqnarray*}
	\emi
	\begin{eqnarray}
	\label{chza:VGff}
	\end{eqnarray}
	\item{Graviton - gluon - gluon vertex}
	\\ \\
	\bmi{110pt}
	\includegraphics[scale=1.0]{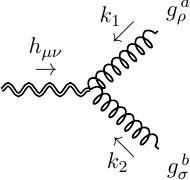}
	\emi
	\bmi{100pt}
	\begin{eqnarray*}
	&=& - i \, \frac{\kappa}{2} \, \delta_{a b} \, V^{Ggg}_{\mu\nu\rho\sigma}(k_1,k_2) \nn \\
	&= &- i \, \frac{\kappa}{2} \, \delta_{a b} \left\{ k_1\cdot k_2 \, C_{\mu\nu\rho\si} + D_{\mu\nu\rho\si}(k_1,k_2) + \frac{1}{\xi} \, E_{\mu\nu\rho\si}(k_1,k_2)  \right\}
	\end{eqnarray*}
	\emi
	\begin{eqnarray}
	\end{eqnarray}
	\item{Graviton - ghost - ghost vertex}
	\\ \\
	\bmi{110pt}
	\includegraphics[scale=1.0]{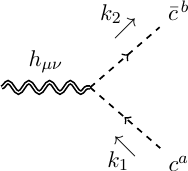}
	\emi
	\bmi{70pt}
	\begin{eqnarray*}
	& = & - i \,  \frac{\kappa}{2}\,  \delta^{a b} \, C_{\mu\nu\rho\sigma} \, k_{1\,\rho} \, k_{2\,\sigma}
	\end{eqnarray*}
	\emi
	\begin{eqnarray}
	\end{eqnarray}
	%
	%
	\item{Graviton - fermion - fermion - gauge boson vertex}
	\\ \\
	\bmi{110pt}
	\includegraphics[scale=1.0]{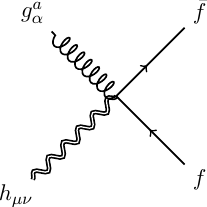}
	\emi
	\bmi{70pt}
	\begin{eqnarray*}
	&=& i g \, \frac{\kappa}{2} \,T^a\, W^{\prime}_{\mu\nu\alpha}
	= i g \, \frac{\kappa}{2} \, T^a \left\{ -\frac{1}{2} (\gamma_\mu \, g_{\nu\alpha}
	+\gamma_\nu \, g_{\mu\alpha}) +  g_{\mu \nu} \, \gamma_{\alpha} \right \}
	\end{eqnarray*}
	\emi
	\begin{eqnarray}
	\label{chza:WGffg}
	\end{eqnarray}
	%
	%
	%
	\item{Graviton - gluon - gluon - gluon  vertex}
	\\ \\
	\bmi{110pt}
	\includegraphics[scale=1.0]{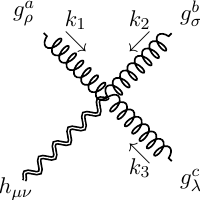}
	\emi
	\bmi{110pt}
	\begin{eqnarray*}
		&=&- g \frac{\kappa}{2} f^{a b c} V^{Gggg}_{\mu\nu\rho\sigma\lambda}(k_1,k_2,k_3) \nn \\
		&=&  - g \frac{\kappa}{2} f^{a b c} \left\{ C_{\mu\nu\rho\sigma}(k_1-k_2)_{\lambda} + C_{\mu\nu\rho\lambda}(k_3-k_1)_{\sigma}   \right. \nn \\
		&& \hspace{2.5cm}  + \left.  C_{\mu\nu\sigma\lambda}(k_2-k_3)_{\rho} + F_{\mu\nu\rho\sigma\lambda}(k_1,k_2,k_3)  \right\}
		\hspace{1.7cm}
	\end{eqnarray*}
	\emi
	
	%
	%
	\item{Graviton - ghost - ghost - gauge boson vertex}
	\\ \\
	\bmi{110pt}
	\includegraphics[scale=1.0]{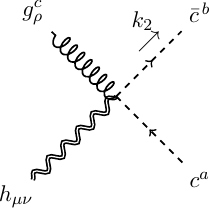}
	\emi
	\bmi{70pt}
	\begin{eqnarray*}
	&=&- \frac{\kappa}{2} \, g \, f^{a b c} \, C_{\mu\nu\rho\sigma} \, k_{2}^{\sigma}
	\end{eqnarray*}
	\emi
	\begin{eqnarray}
	\end{eqnarray}
	%
	%
	\begin{eqnarray}
	&& C_{\mu\nu\rho\sigma} = g_{\mu\rho}\, g_{\nu\sigma}
	+g_{\mu\sigma} \, g_{\nu\rho}
	-g_{\mu\nu} \, g_{\rho\sigma}\,
	\\
	&& D_{\mu\nu\rho\sigma} (k_1, k_2) =
	g_{\mu\nu} \, k_{1 \, \sigma}\, k_{2 \, \rho}
	- \biggl[g^{\mu\sigma} k_1^{\nu} k_2^{\rho}
	+ g_{\mu\rho} \, k_{1 \, \sigma} \, k_{2 \, \nu}
	- g_{\rho\sigma} \, k_{1 \, \mu} \, k_{2 \, \nu}
	+ (\mu\leftrightarrow\nu)\biggr]\, \\
	&& E_{\mu\nu\rho\sigma} (k_1, k_2) = g_{\mu\nu} \, (k_{1 \, \rho} \, k_{1 \, \sigma}
	+k_{2 \, \rho} \, k_{2 \, \sigma} +k_{1 \, \rho} \, k_{2 \, \sigma})
	-\biggl[g_{\nu\sigma} \, k_{1 \, \mu} \, k_{1 \, \rho}
	+g_{\nu\rho} \, k_{2 \, \mu} \, k_{2 \, \sigma}
	+(\mu\leftrightarrow\nu)\biggr]\ , \nn  \\ \\
	&& F_{\mu\nu\rho\sigma\lambda} (k_1,k_2,k_3) =
	g_{\mu\rho} \,  g_{\sigma\lambda} \, (k_2-k_3)_{\nu}
	+g_{\mu\sigma} \, g_{\rho\lambda} \, (k_3-k_1)_{\nu}
	+g_{\mu\lambda} \, g_{\rho\sigma}(k_1-k_2)_{\nu}
	+ (\mu\leftrightarrow\nu) \nn \\
	\end{eqnarray}
	\end{itemize}
	
	\section{Form factors of the transverse-traceless sector: $A_1,A_2,A_3,A_4$ for massless quarks} 
	\label{chza:ff}
	We present here the explicit expressions of the coefficient functions $A_{ij}$ in \eqref{chza:Abar}  characterizing the transverse-traceless sector. The quark contributions are identified by the factor $n_f$ and the gluon contributions by the factor $C_A$:
	
	{\allowdisplaybreaks
		\begin{eqnarray*}
			A_{10} &=& 16\, (p_1\cdot p_2)^7+36 \, p_1^2 \, (p_1\cdot p_2)^6+36 \, p_2^2 \, (p_1\cdot p_2)^6+12 \, p_1^4 \, (p_1\cdot p_2)^5+12 \, p_2^4 \, (p_1\cdot p_2)^5+240 \, p_1^2 \, p_2^2 \, (p_1\cdot p_2)^5 \notag \\ &&
			+188 \, p_1^2 \, p_2^4 \, (p_1\cdot p_2)^4+188 \, p_1^4 \, p_2^2 \, (p_1\cdot p_2)^4+46 \, p_1^2 \, p_2^6 \, (p_1\cdot p_2)^3-108 \, p_1^4 \, p_2^4 \, (p_1\cdot p_2)^3+46 \, p_1^6 \, p_2^2 \, (p_1\cdot p_2)^3 \notag \\ &&
			-204 \, p_1^4 \, p_2^6 \, (p_1\cdot p_2)^2-204 \, p_1^6 \, p_2^4 \, (p_1\cdot p_2)^2-20 \, p_1^6 \, p_2^8 \,-20 \, p_1^8 \, p_2^6 \,-58 \, p_1^4 \, p_2^8 \, (p_1\cdot p_2) \notag \\ &&
			-148 \, p_1^6 \, p_2^6 \, (p_1\cdot p_2)-58 \, p_1^8 \, p_2^4 \, (p_1\cdot p_2) \\
			A_{11} &=& -24 \, p_1^2 \,  \, (p_1\cdot p_2)^6-24 \, p_1^4 \,  \,\, (p_1\cdot p_2)^5-224 \, p_1^2 \,  \, p_2^2 \, (p_1\cdot p_2)^5-6\, p_1^6 \,  \,\, (p_1\cdot p_2)^4-228 \, p_1^2 \,  \, p_2^4 \, (p_1\cdot p_2)^4 \notag \\ &&
			-522 \, p_1^4 \,  \, p_2^2 \, (p_1\cdot p_2)^4- 60 \, p_1^2 \,  \, p_2^6 \, (p_1\cdot p_2)^3- 652 \, p_1^4 \,  \, p_2^4 \, (p_1\cdot p_2)^3-372 \, p_1^6 \,  \, p_2^2 \, (p_1\cdot p_2)^3  \notag \\ &&
			-279 \, p_1^4 \,  \, p_2^6 \, (p_1\cdot p_2)^2-474 \, p_1^6 \,  \, p_2^4 \, (p_1\cdot p_2)^2-83 \, p_1^8 \,  \, p_2^2 \, (p_1\cdot p_2)^2-18 \, p_1^6 \,  \, p_2^8 \,  \notag \\ &&
			-30 \, p_1^8 \,  \, p_2^6 \,-16 p_1^{10} \, p_2^4 \,-45 \, p_1^4 \,  \, p_2^8 \, (p_1\cdot p_2)-174 \, p_1^6 \,  \, p_2^6 \, (p_1\cdot p_2)-129 \, p_1^8 \,  \, p_2^4 \, (p_1\cdot p_2) \\
			A_{12} &=& -24 \, p_2^2 \, (p_1\cdot p_2)^6-24 \, p_2^4 \, (p_1\cdot p_2)^5-224 \, p_1^2 \,  \, p_2^2 \, (p_1\cdot p_2)^5-6 \, p_2^6 \, (p_1\cdot p_2)^4-522 \, p_1^2 \,  \, p_2^4 \, (p_1\cdot p_2)^4   \notag \\ &&
			-228 \, p_1^4 \,  \, p_2^2 \, (p_1\cdot p_2)^4-372 \, p_1^2 \,  \, p_2^6 \, (p_1\cdot p_2)^3-652 \, p_1^4 \,  \, p_2^4 \, (p_1\cdot p_2)^3-60 \, p_1^6 \,  \, p_2^2 \, (p_1\cdot p_2)^3-83 \, p_1^2 \,  \, p_2^8 \, (p_1\cdot p_2)^2   \notag \\ &&
			-474 \, p_1^4 \,  \, p_2^6 \, (p_1\cdot p_2)^2-279 \, p_1^6 \,  \, p_2^4 \, (p_1\cdot p_2)^2-16 \, p_1^4 \,  \, p_2^{10} \,-30 \, p_1^6 \,  \, p_2^8 \,-18 \, p_1^8 \,  \, p_2^6 \,-129 \, p_1^4 \,  \, p_2^8 \, (p_1\cdot p_2)   \notag \\ &&
			-174 \, p_1^6 \,  \, p_2^6 \, (p_1\cdot p_2)-45 \, p_1^8 \,  \, p_2^4 \, (p_1\cdot p_2) \\
			A_{13} &=& 24 \, p_1^2 \,  \, (p_1\cdot p_2)^6+24 \, p_2^2 \, (p_1\cdot p_2)^6+24 \, p_1^4 \,  \, (p_1\cdot p_2)^5+24 \, p_2^4 \, (p_1\cdot p_2)^5 + 448 \, p_1^2 \,  \, p_2^2 \, (p_1\cdot p_2)^5  \notag \\ &&
			+6 \, p_1^6 \,  \, (p_1\cdot p_2)^4+6 \, p_2^6 \, (p_1\cdot p_2)^4+750 \, p_1^2 \,  \, p_2^4 \, (p_1\cdot p_2)^4  +750 \, p_1^4 \,  \, p_2^2 \, (p_1\cdot p_2)^4+432 \, p_1^2 \,  \, p_2^6 \, (p_1\cdot p_2)^3  \notag \\ &&
			+ 1304 \, p_1^4 \,  \, p_2^4 \, (p_1\cdot p_2)^3+432 \, p_1^6 \,  \, p_2^2 \, (p_1\cdot p_2)^3  + 83 \, p_1^2 \,  \, p_2^8 \, (p_1\cdot p_2)^2+753 \, p_1^4 \,  \, p_2^6 \, (p_1\cdot p_2)^2+753 \, p_1^6 \,  \, p_2^4 \, (p_1\cdot p_2)^2  \notag \\ &&
			+83 \, p_1^8 \,  \, p_2^2 \, (p_1\cdot p_2)^2 + 16 \, p_1^4 \,  \, p_2^{10}+48 \, p_1^6 \,  \, p_2^8 \,+48 \, p_1^8 \,  \, p_2^6 \,+16 p_1^{10} \, p_2^4 \,  \notag \\ &&
			+174 \, p_1^4 \,  \, p_2^8 \, (p_1\cdot p_2)+348 \, p_1^6 \,  \, p_2^6 \, (p_1\cdot p_2)+174 \, p_1^8 \,  \, p_2^4 \, (p_1\cdot p_2)	\\
			A_{14} &=& 192 \, p_1^2 \,  \, p_2^2 \, (p_1\cdot p_2 )^6+432 \, p_1^2 \,  \, p_2^4 \, (p_1\cdot p_2 )^5+432 \, p_1^4 \,  \, p_2^2 \, (p_1\cdot p_2 )^5+288 \, p_1^2 \,  \, p_2^6 \, (p_1\cdot p_2 )^4+1152 \, p_1^4 \,  \, p_2^4 \, (p_1\cdot p_2 )^4  \notag \\ &&
			+288 \, p_1^6 \,  \, p_2^2 \, (p_1\cdot p_2 )^4+60 \, p_1^2 \,  \, p_2^8 \, (p_1\cdot p_2 )^3+936 \, p_1^4 \,  \, p_2^6 \, (p_1\cdot p_2 )^3+936 \, p_1^6 \,  \, p_2^4 \, (p_1\cdot p_2 )^3+60 \, p_1^8 \,  \, p_2^2 \, (p_1\cdot p_2 )^3  \notag \\ &&
			+324 \, p_1^4 \,  \, p_2^8 \, (p_1\cdot p_2 )^2+720 \, p_1^6 \,  \, p_2^6 \, (p_1\cdot p_2 )^2+324 \, p_1^8 \,  \, p_2^4 \, (p_1\cdot p_2 )^2+18 \, p_1^6 \,  \, p_2^{10} \,+36 \, p_1^8 \,  \, p_2^8 \notag \\ &&
			+18 p_1^{10} \, p_2^6 \,+45 \, p_1^4 \,  \, p_2^{10} (p_1\cdot p_2) +207 \, p_1^6 \,  \, p_2^8 \, (p_1\cdot p_2)+207 \, p_1^8 \,  \, p_2^6 \, (p_1\cdot p_2)+45 p_1^{10} \, p_2^4 \, (p_1\cdot p_2)
		\end{eqnarray*}
		
		\begin{eqnarray*}
			A_{20} &=& C_A \Big[-520 \, (p_1\cdot p_2)^6-372 \, p_1^2 \, (p_1\cdot p_2)^5-372 \, p_2^2 \, (p_1\cdot p_2)^5-60 \, p_1^4 \, (p_1\cdot p_2)^4-60 \, p_2^4 \, (p_1\cdot p_2)^4 \notag \\ &&
			+1056 \, p_1^2 \, p_2^2 \, (p_1\cdot p_2)^4+864 \, p_1^2 \, p_2^4 \, (p_1\cdot p_2)^3+864 \, p_1^4 \, p_2^2 \, (p_1\cdot p_2)^3+150 \, p_1^2 \, p_2^6 \, (p_1\cdot p_2)^2 \notag \\ &&
			-372 \, p_1^4 \, p_2^4 \, (p_1\cdot p_2)^2+150 \, p_1^6 \, p_2^2 \, (p_1\cdot p_2)^2-90 \, p_1^4 \, p_2^8 \,-164 \, p_1^6 \, p_2^6 \,-90 \, p_1^8 \, p_2^4 \notag \\ &&
			-492 \, p_1^4 \, p_2^6 \, (p_1\cdot p_2)-492 \, p_1^6 \, p_2^4 \, (p_1\cdot p_2) \Big]+ \notag \\ &&
			n_f\Big[ 40 (p_1\cdot p_2)^6+12 \, p_1^2 \, (p_1\cdot p_2)^5+12 \, p_2^2 \, (p_1\cdot p_2)^5-12 \, p_1^4 \, (p_1\cdot p_2)^4-12 \, p_2^4 \, (p_1\cdot p_2)^4 \notag \\ &&
			-192 \, p_1^2 \, p_2^2 \, (p_1\cdot p_2)^4-144 \, p_1^2 \, p_2^4 \, (p_1\cdot p_2)^3-144 \, p_1^4 \, p_2^2 \, (p_1\cdot p_2)^3-6 \, p_1^2 \, p_2^6 \, (p_1\cdot p_2)^2 \notag \\ &&
			+84 \, p_1^4 \, p_2^4 \, (p_1\cdot p_2)^2-6 \, p_1^6 \, p_2^2 \, (p_1\cdot p_2)^2+18 \, p_1^4 \, p_2^8 \,+68 \, p_1^6 \, p_2^6 \,+18 \, p_1^8 \, p_2^4 \notag \\ &&
			+132 \, p_1^4 \, p_2^6 \, (p_1\cdot p_2)+132 \, p_1^6 \, p_2^4 \, (p_1\cdot p_2) \Big] \\
			A_{21} &=& 3 \, p_1^2 \,  \, n_f \Big[-24 (p_1\cdot p_2)^5-20 \, p_2^2 \, (p_1\cdot p_2)^4+2 \, p_1^4 \,  \, (p_1\cdot p_2)^3+24 \, p_2^4 \, (p_1\cdot p_2)^3+90 \, p_1^2 \,  \, p_2^2 \, (p_1\cdot p_2)^3  \notag \\ &&
			+12 \, p_2^6 \, (p_1\cdot p_2)^2+162 \, p_1^2 \,  \, p_2^4 \, (p_1\cdot p_2)^2+54 \, p_1^4 \,  \, p_2^2 \, (p_1\cdot p_2)^2+3 \, p_1^2 \,  \, p_2^8 \,+8 \, p_1^4 \,  \, p_2^6 \,+21 \, p_1^6 \,  \, p_2^4 \,+51 \, p_1^2 \,  \, p_2^6 \, (p_1\cdot p_2)   \notag \\ &&
			+84 \, p_1^4 \,  \, p_2^4 \, (p_1\cdot p_2)+13 \, p_1^6 \,  \, p_2^2 \, (p_1\cdot p_2) \Big] \notag \\ &&
			-3 C_A \Big[ -48 (p_1\cdot p_2)^6-240 \, p_1^2 \,  \, (p_1\cdot p_2)^5-72 \, p_2^2 \, (p_1\cdot p_2)^5  \notag \\ &&
			-156 \, p_1^4 \,  \, (p_1\cdot p_2)^4 -24 \, p_2^4 \, (p_1\cdot p_2)^4-344 \, p_1^2 \,  \, p_2^2 \, (p_1\cdot p_2)^4-34 \, p_1^6 \,  \, (p_1\cdot p_2)^3-48 \, p_1^2 \,  \, p_2^4 \, (p_1\cdot p_2)^3   \notag \\ &&
			+54 \, p_1^4 \,  \, p_2^2 \, (p_1\cdot p_2)^3 +24 \, p_1^2 \,  \, p_2^6 \, (p_1\cdot p_2)^2+486 \, p_1^4 \,  \, p_2^4 \, (p_1\cdot p_2)^2+150 \, p_1^6 \,  \, p_2^2 \, (p_1\cdot p_2)^2+15 \, p_1^4 \,  \, p_2^8 \, \notag \\ &&
			+56 \, p_1^6 \,  \, p_2^6 \, +81 \, p_1^8 \,  \, p_2^4  +195 \, p_1^4 \,  \, p_2^6 \, (p_1\cdot p_2)+336 \, p_1^6 \,  \, p_2^4 \, (p_1\cdot p_2)+49 \, p_1^8 \,  \, p_2^2 \, (p_1\cdot p_2) \Big] \\
			A_{22} &=&  3 \, p_2^2 \, n_f \Big[-24 (p_1\cdot p_2)^5-20 \, p_1^2 \,  \, (p_1\cdot p_2)^4+24 \, p_1^4 \,  \, (p_1\cdot p_2)^3+2 \, p_2^4 \, (p_1\cdot p_2)^3+90 \, p_1^2 \,  \, p_2^2 \, (p_1\cdot p_2)^3  \notag \\ &&
			+12 \, p_1^6 \,  \, (p_1\cdot p_2)^2+54 \, p_1^2 \,  \, p_2^4 \, (p_1\cdot p_2)^2+162 \, p_1^4 \,  \, p_2^2 \, (p_1\cdot p_2)^2+21 \, p_1^4 \,  \, p_2^6 \,+8 \, p_1^6 \,  \, p_2^4 \,+3 \, p_1^8 \,  \, p_2^2 \,+13 \, p_1^2 \,  \, p_2^6 \, (p_1\cdot p_2)  \notag \\ &&
			+84 \, p_1^4 \,  \, p_2^4 \, (p_1\cdot p_2)+51 \, p_1^6 \,  \, p_2^2 \, (p_1\cdot p_2)\Big]-3 C_A \Big[-48 (p_1\cdot p_2)^6-72 \, p_1^2 \,  \, (p_1\cdot p_2)^5-240 \, p_2^2 \, (p_1\cdot p_2)^5  \notag \\ &&
			-24 \, p_1^4 \,  \, (p_1\cdot p_2)^4-156 \, p_2^4 \, (p_1\cdot p_2)^4-344 \, p_1^2 \,  \, p_2^2 \, (p_1\cdot p_2)^4-34 \, p_2^6 \, (p_1\cdot p_2)^3+54 \, p_1^2 \,  \, p_2^4 \, (p_1\cdot p_2)^3  \notag \\ &&
			-48 \, p_1^4 \,  \, p_2^2 \, (p_1\cdot p_2)^3+150 \, p_1^2 \,  \, p_2^6 \, (p_1\cdot p_2)^2+486 \, p_1^4 \,  \, p_2^4 \, (p_1\cdot p_2)^2+24 \, p_1^6 \,  \, p_2^2 \, (p_1\cdot p_2)^2+81 \, p_1^4 \,  \, p_2^8 \,  \notag \\ &&
			+56 \, p_1^6 \,  \, p_2^6 \,+15 \, p_1^8 \,  \, p_2^4 \,+49 \, p_1^2 \,  \, p_2^8 \, (p_1\cdot p_2)+336 \, p_1^4 \,  \, p_2^6 \, (p_1\cdot p_2)+195 \, p_1^6 \,  \, p_2^4 \, (p_1\cdot p_2)\Big] \\
			A_{23} &=& 3 C_A \, \Big(2 (p_1\cdot p_2)+\, p_1^2 \,  \,+\, p_2^2 \Big)^2 \, \Big[-44 (p_1\cdot p_2)^4-34 \, p_1^2 \,  \, (p_1\cdot p_2)^3-34 \, p_2^2 \, (p_1\cdot p_2)^3-22 \, p_1^2 \,  \, p_2^2 \, (p_1\cdot p_2)^2  \notag \\ &&
			+96 \, p_1^4 \,  \, p_2^4 \,+49 \, p_1^2 \,  \, p_2^4 \, (p_1\cdot p_2)+49 \, p_1^4 \,  \, p_2^2 \, (p_1\cdot p_2) \Big]  \notag \\ &&
			-3 n_f (2 (p_1\cdot p_2)+\, p_1^2 \,  \,+\, p_2^2 \,)^2  \Big[ -8 (p_1\cdot p_2)^4+2 \, p_1^2 \,  \, (p_1\cdot p_2)^3+2 \, p_2^2 \, (p_1\cdot p_2)^3+14 \, p_1^2 \,  \, p_2^2 \, (p_1\cdot p_2)^2  \notag \\ &&
			+24 \, p_1^4 \,  \, p_2^4 \,+13 \, p_1^2 \,  \, p_2^4 \, (p_1\cdot p_2)+13 \, p_1^4 \,  \, p_2^2 \, (p_1\cdot p_2) \Big]  \\
			A_{24} &=& 9\, C_A \Big(2 (p_1\cdot p_2)+\, p_1^2 \,  \,+\, p_2^2 \Big) \, \Big[ -32 (p_1\cdot p_2)^6-16 \, p_1^2 \,  \, (p_1\cdot p_2)^5-16 \, p_2^2 \, (p_1\cdot p_2)^5-8 \, p_1^4 \,  \, (p_1\cdot p_2)^4  \notag \\ &&
			-8 \, p_2^4 \, (p_1\cdot p_2)^4-24 \, p_1^2 \,  \, p_2^4 \, (p_1\cdot p_2)^3-24 \, p_1^4 \,  \, p_2^2 \, (p_1\cdot p_2)^3+8 \, p_1^2 \,  \, p_2^6 \, (p_1\cdot p_2)^2+20 \, p_1^4 \,  \, p_2^4 \, (p_1\cdot p_2)^2  \notag \\ &&
			+8 \, p_1^6 \,  \, p_2^2 \, (p_1\cdot p_2)^2+5 \, p_1^4 \,  \, p_2^8 \,+42 \, p_1^6 \,  \, p_2^6 \,+5 \, p_1^8 \,  \, p_2^4 \,+60 \, p_1^4 \,  \, p_2^6 \, (p_1\cdot p_2)+60 \, p_1^6 \,  \, p_2^4 \, (p_1\cdot p_2) \Big]  \notag \\ &&
			-9 \, p_1^2 \,  \, p_2^2 \, n_f  \Big( (2 (p_1\cdot p_2)+\, p_1^2 \,  \,+\, p_2^2 \, \Big)^2 \Big[-4 (p_1\cdot p_2)^3+4 \, p_1^2 \,  \, (p_1\cdot p_2)^2+4 \, p_2^2 \, (p_1\cdot p_2)^2   \notag \\ &&
			+\, p_1^2 \,  \, p_2^4 \,+\, p_1^4 \,  \, p_2^2 \,+14 \, p_1^2 \,  \, p_2^2 \, (p_1\cdot p_2) \Big]
		\end{eqnarray*}

		\begin{eqnarray*}
			A_{31} &=& 	3 \, p_1^2 \,  \, n_f \Big[-16 (p_1\cdot p_2)^5-4 \, p_1^2 \,  \, (p_1\cdot p_2)^4-70 \, p_2^2 \, (p_1\cdot p_2)^4-54 \, p_2^4 \, (p_1\cdot p_2)^3-44 \, p_1^2 \,  \, p_2^2 \, (p_1\cdot p_2)^3  \notag \\ &&
			-12 \, p_2^6 \, (p_1\cdot p_2)^2-15 \, p_1^2 \,  \, p_2^4 \, (p_1\cdot p_2)^2-15 \, p_1^4 \,  \, p_2^2 \, (p_1\cdot p_2)^2-3 \, p_1^2 \,  \, p_2^8 \,-5 \, p_1^4 \,  \, p_2^6 \,+4 \, p_1^6 \,  \, p_2^4 \,-6 \, p_1^2 \,  \, p_2^6 \, (p_1\cdot p_2)\Big]  \notag \\ &&
			-3 C_A\Big[-24 (p_1\cdot p_2)^6-100 \, p_1^2 \,  \, (p_1\cdot p_2)^5-12 \, p_2^2 \, (p_1\cdot p_2)^5-34 \, p_1^4 \,  \, (p_1\cdot p_2)^4-172 \, p_1^2 \,  \, p_2^2 \, (p_1\cdot p_2)^4  \notag \\ && 
			-84 \, p_1^2 \,  \, p_2^4 \, (p_1\cdot p_2)^3 -38 \, p_1^4 \,  \, p_2^2 \, (p_1\cdot p_2)^3-12 \, p_1^2 \,  \, p_2^6 \, (p_1\cdot p_2)^2+99 \, p_1^4 \,  \, p_2^4 \, (p_1\cdot p_2)^2-9 \, p_1^6 \,  \, p_2^2 \, (p_1\cdot p_2)^2-3 \, p_1^4 \,  \, p_2^8 \,  \notag \\ &&
			+7 \, p_1^6 \,  \, p_2^6 \,+28 \, p_1^8 \,  \, p_2^4 \,+36 \, p_1^4 \,  \, p_2^6 \, (p_1\cdot p_2)+78 \, p_1^6 \,  \, p_2^4 \, (p_1\cdot p_2)\Big] \\
			A_{32} &=& 3 \, p_2^2 \, n_f \Big[-24 (p_1\cdot p_2)^5-54 \, p_1^2 \,  \, (p_1\cdot p_2)^4-12 \, p_2^2 \, (p_1\cdot p_2)^4-18 \, p_1^4 \,  \, (p_1\cdot p_2)^3-2 \, p_2^4 \, (p_1\cdot p_2)^3 \notag \\ &&
			-62 \, p_1^2 \,  \, p_2^2 \, (p_1\cdot p_2)^3-55 \, p_1^2 \,  \, p_2^4 \, (p_1\cdot p_2)^2-3 \, p_1^4 \,  \, p_2^2 \, (p_1\cdot p_2)^2+7 \, p_1^4 \,  \, p_2^6 \,-3 \, p_1^6 \,  \, p_2^4 \,-13 \, p_1^2 \,  \, p_2^6 \, (p_1\cdot p_2) \notag \\ &&
			-4 \, p_1^4 \,  \, p_2^4 \, (p_1\cdot p_2)+3 \, p_1^6 \,  \, p_2^2 \, (p_1\cdot p_2)\Big] \notag \\ &&
			-3  C_A \Big[-24 (p_1\cdot p_2)^6-12 \, p_1^2 \,  \, (p_1\cdot p_2)^5-108 \, p_2^2 \, (p_1\cdot p_2)^5-42 \, p_2^4 \, (p_1\cdot p_2)^4 \notag \\ &&
			-156 \, p_1^2 \,  \, p_2^2 \, (p_1\cdot p_2)^4-2 \, p_2^6 \, (p_1\cdot p_2)^3-56 \, p_1^2 \,  \, p_2^4 \, (p_1\cdot p_2)^3-48 \, p_1^4 \,  \, p_2^2 \, (p_1\cdot p_2)^3-49 \, p_1^2 \,  \, p_2^6 \, (p_1\cdot p_2)^2 \notag \\ &&
			+111 \, p_1^4 \,  \, p_2^4 \, (p_1\cdot p_2)^2+31 \, p_1^4 \,  \, p_2^8 \,+9 \, p_1^6 \,  \, p_2^6 \,-13 \, p_1^2 \,  \, p_2^8 \, (p_1\cdot p_2)+74 \, p_1^4 \,  \, p_2^6 \, (p_1\cdot p_2)+45 \, p_1^6 \,  \, p_2^4 \, (p_1\cdot p_2)\Big] \notag \\ 
			A_{33} &=& 3 \, C_A \, \Big(2 (p_1\cdot p_2)+\, p_1^2 \,  \,+\, p_2^2 \Big) \, \Big[ -44 (p_1\cdot p_2)^5-34 \, p_1^2 \,  \, (p_1\cdot p_2)^4-38 \, p_2^2 \, (p_1\cdot p_2)^4-2 \, p_2^4 \, (p_1\cdot p_2)^3 \notag \\ &&
			-68 \, p_1^2 \,  \, p_2^2 \, (p_1\cdot p_2)^3-35 \, p_1^2 \,  \, p_2^4 \, (p_1\cdot p_2)^2-9 \, p_1^4 \,  \, p_2^2 \, (p_1\cdot p_2)^2+28 \, p_1^4 \,  \, p_2^6 \,+28 \, p_1^6 \,  \, p_2^4  \notag \\ &&
			-13 \, p_1^2 \,  \, p_2^6 \, (p_1\cdot p_2) +67 \, p_1^4 \,  \, p_2^4 \, (p_1\cdot p_2)\Big] \notag \\ &&
			-3 \, n_f \, \Big(2 (p_1\cdot p_2)+\, p_1^2 \,  \,+\, p_2^2 \Big)^2 \, \Big[-4 (p_1\cdot p_2)^4-2 \, p_2^2 \, (p_1\cdot p_2)^3-15 \, p_1^2 \,  \, p_2^2 \, (p_1\cdot p_2)^2+4 \, p_1^4 \,  \, p_2^4 \,-13 \, p_1^2 \,  \, p_2^4 \, (p_1\cdot p_2) \Big] \notag \\ 
			A_{34} &=& 9 \, C_A \, \Big(2 (p_1\cdot p_2)+\, p_1^2 \,  \,+\, p_2^2 \Big) \, \Big[-16 (p_1\cdot p_2)^6-4 \, p_1^2 \,  \, (p_1\cdot p_2)^5-4 \, p_2^2 \, (p_1\cdot p_2)^5-24 \, p_1^2 \,  \, p_2^4 \, (p_1\cdot p_2)^3 \notag \\ &&
			-16 \, p_1^4 \,  \, p_2^2 \, (p_1\cdot p_2)^3-4 \, p_1^2 \,  \, p_2^6 \, (p_1\cdot p_2)^2-14 \, p_1^4 \,  \, p_2^4 \, (p_1\cdot p_2)^2-\, p_1^4 \,  \, p_2^8 \,\notag \\ &&
			+15 \, p_1^6 \,  \, p_2^6 \,+13 \, p_1^4 \,  \, p_2^6 \, (p_1\cdot p_2)+15 \, p_1^6 \,  \, p_2^4 \, (p_1\cdot p_2)\Big] \notag \\ &&
			-9 \, p_1^2 \,  \, p_2^2 \, n_f \, \Big(2 (p_1\cdot p_2)+\, p_1^2 \,  \,+\, p_2^2 \Big)^2 \, \Big[-6 (p_1\cdot p_2)^3-4 \, p_2^2 \, (p_1\cdot p_2)^2-\, p_1^2 \,  \, p_2^4 \,+\, p_1^2 \,  \, p_2^2 \, (p_1\cdot p_2) \Big] \notag \\
		\end{eqnarray*}
		
		\begin{eqnarray*}
			A_{40} &=& C_A \Big[ -284 \, (p_1\cdot p_2)^5-132 \, p_1^2 \, (p_1\cdot p_2)^4-132 \, p_2^2 \, (p_1\cdot p_2)^4+556 \, p_1^2 \, p_2^2 \, (p_1\cdot p_2)^3 \notag \\ &&
			+258 \, p_1^2 \, p_2^4 \, (p_1\cdot p_2)^2+258 \, p_1^4 \, p_2^2 \, (p_1\cdot p_2)^2-126 \, p_1^4 \, p_2^6 \,-126 \, p_1^6 \, p_2^4 \,-272 \, p_1^4 \, p_2^4 \, (p_1\cdot p_2) \Big]  \notag \\ &&
			+ n_f \Big[ 44 (p_1\cdot p_2)^5+24 \, p_1^2 \, (p_1\cdot p_2)^4+24 \, p_2^2 \, (p_1\cdot p_2)^4-76 \, p_1^2 \, p_2^2 \, (p_1\cdot p_2)^3-42 \, p_1^2 \, p_2^4 \, (p_1\cdot p_2)^2 \notag \\ &&
			-42 \, p_1^4 \, p_2^2 \, (p_1\cdot p_2)^2+18 \, p_1^4 \, p_2^6 \,+18 \, p_1^6 \, p_2^4 \,+32 \, p_1^4 \, p_2^4 \, (p_1\cdot p_2) \Big] \\
			A_{41} &=& 3 \, p_1^2 \,  \, n_f \Big[-16 \, (p_1\cdot p_2)^4-4 \, p_1^2 \,  \, (p_1\cdot p_2)^3-18 \, p_2^2 \, (p_1\cdot p_2)^3-6 \, p_2^4 \, (p_1\cdot p_2)^2+10 \, p_1^2 \,  \, p_2^2 \, (p_1\cdot p_2)^2 \notag \\ &&
			+3 \, p_1^2 \,  \, p_2^6 \,-3 \, p_1^4 \,  \, p_2^4 \,+9 \, p_1^2 \,  \, p_2^4 \, (p_1\cdot p_2)+\, p_1^4 \,  \, p_2^2 \, (p_1\cdot p_2)\Big] \notag \\ &&
			-3 C_A\Big[-24 \, (p_1\cdot p_2)^5-76 \, p_1^2 \,  \, (p_1\cdot p_2)^4-12 \, p_2^2 \, (p_1\cdot p_2)^4-22 \, p_1^4 \,  \, (p_1\cdot p_2)^3-24 \, p_1^2 \,  \, p_2^2 \, (p_1\cdot p_2)^3 \notag \\ &&
			+76 \, p_1^4 \,  \, p_2^2 \, (p_1\cdot p_2)^2+9 \, p_1^4 \,  \, p_2^6 \,-9 \, p_1^6 \,  \, p_2^4 \,+39 \, p_1^4 \,  \, p_2^4 \, (p_1\cdot p_2)+19 \, p_1^6 \,  \, p_2^2 \, (p_1\cdot p_2)\Big] \\
			A_{42} &=&  3 \, p_2^2 \, n_f \Big[-16 (p_1\cdot p_2)^4-18 \, p_1^2 \,  \, (p_1\cdot p_2)^3-4 \, p_2^2 \, (p_1\cdot p_2)^3-6 \, p_1^4 \,  \, (p_1\cdot p_2)^2+10 \, p_1^2 \,  \, p_2^2 \, (p_1\cdot p_2)^2 \notag \\ &&
			-3 \, p_1^4 \,  \, p_2^4 \,+3 \, p_1^6 \,  \, p_2^2 \,+\, p_1^2 \,  \, p_2^4 \, (p_1\cdot p_2)+9 \, p_1^4 \,  \, p_2^2 \, (p_1\cdot p_2)\Big] \notag \\ &&
			-3 C_A\Big[-24 \, (p_1\cdot p_2)^5-12 \, p_1^2 \,  \, (p_1\cdot p_2)^4-76 \, p_2^2 \, (p_1\cdot p_2)^4-22 \, p_2^4 \, (p_1\cdot p_2)^3-24 \, p_1^2 \,  \, p_2^2 \, (p_1\cdot p_2)^3 \notag \\ &&
			+76 \, p_1^2 \,  \, p_2^4 \, (p_1\cdot p_2)^2-9 \, p_1^4 \,  \, p_2^6 \,+9 \, p_1^6 \,  \, p_2^4 \,+19 \, p_1^2 \,  \, p_2^6 \, (p_1\cdot p_2)+39 \, p_1^4 \,  \, p_2^4 \, (p_1\cdot p_2)\Big] \\
			A_{43} &=& 3 C_A \, (p_1\cdot p_2) \, \Big(2 (p_1\cdot p_2)+\, p_1^2 \,  \,+\, p_2^2 \Big)^2 \Big[ 19 \, p_1^2 \,  \, p_2^2 \,-22 (p_1\cdot p_2)^2 \Big] \notag \\ &&
			-3 n_f \, (p_1\cdot p_2) \, \Big(2 (p_1\cdot p_2)+\, p_1^2 \,  \,+\, p_2^2 \Big)^2 \Big[ p_1^2 \,  \, p_2^2 \,-4 (p_1\cdot p_2)^2 \Big] \\
			A_{44} &=& 9 C_A \Big(2 (p_1\cdot p_2)+\, p_1^2 \,  \,+\, p_2^2 \Big) \Big[-16 (p_1\cdot p_2)^5-4 \, p_1^2 \,  \, (p_1\cdot p_2)^4-4 \, p_2^2 \, (p_1\cdot p_2)^4+16 \, p_1^2 \,  \, p_2^2 \, (p_1\cdot p_2)^3 \notag \\ &&
			+3 \, p_1^4 \,  \, p_2^6 \,+3 \, p_1^6 \,  \, p_2^4 \,-2 \, p_1^4 \,  \, p_2^4 \, (p_1\cdot p_2) -9 \, p_1^2 \,  \, p_2^2 \, n_f (2 (p_1\cdot p_2)+\, p_1^2 \,  \,+\, p_2^2 \,)^2 \, p_1^2 \,  \, p_2^2 \,-2 (p_1\cdot p_2)^2 \Big]
		\end{eqnarray*}
	}

\section{Explicit kernel expressions in the impact parameter space}\label{chza:kernelexpb}
In this appendix we provide the explicit expression of the hard-scattering kernel at tree level together with the corrections induced by the $TJJ$ insertion. The transverse momentum dependence is retained only in the gluon propagator connecting the constituent quarks, both for the leading-order contribution and for the $TJJ$ correction. This approximation is motivated by the fact that the dominant transverse-distance effects originate from the exchanged hard gluon, while the intrinsic transverse momentum dependence of the external valence quarks gives rise only to subleading contributions in the kinematic regime considered here.

More precisely, the quark propagators carry an off-shell virtuality of order $\alpha Q^2$, which 
is larger than the virtuality $\alpha_1\alpha_2 Q^2$ 
of the exchanged gluon, particularly in the endpoint region. As a consequence, the transverse momentum 
dependence is negligible in the quark propagators compared to their 
large collinear off-shellness, while it plays a crucial role in 
regulating the endpoint singularities arising from the gluon propagator. 
The transverse momentum appearing in the Dirac numerators is neglected 
separately, as it constitutes a higher-power effect suppressed relative 
to the leading-twist contribution. The full $k_T$ dependence is 
therefore retained exclusively in the gluon propagator, where it 
provides the essential infrared regulator and allows for a natural 
implementation of Sudakov suppression in impact-parameter space.

The approximation considerably simplifies the structure of the hard kernel while preserving the essential non-collinear dynamics. In particular, after Fourier transformation to impact-parameter space, the dependence on the transverse momentum is encoded in modified Bessel functions, leading to compact analytical expressions suitable for numerical evaluation. In impact-parameter space the tree-level kernel reads

\begin{equation}
	K_{\text{tree}}^{\mu\nu}(\alpha_1,\alpha_2) = F_1^{\text{tree}}\, P_1^{\mu } P_2^{\nu } 
	+ F_2^{\text{tree}}\, P_2^{\mu } P_1^{\nu } 
	+ F_3^{\text{tree}}\, P_1^{\mu } P_1^{\nu } 
	+ F_4^{\text{tree}}\, P_2^{\mu } P_2^{\nu } 
	+ F_5^{\text{tree}}\, g^{\mu\nu},
\end{equation}
with
\begin{equation}
	F_1^{\text{tree}}=F_2^{\text{tree}}= 4\left(K_0\left({b{Q} \sqrt{(\alpha_1-1) (\alpha_2-1)}}\right)+K_0\left({b{Q} \sqrt{\alpha_1 \alpha_2}}\right)\right),
\end{equation}

\begin{equation}
	F_3^{\text{tree}}=F_4^{\text{tree}}(\alpha_1\leftrightarrow\alpha_2)=2\frac{\alpha_1 (2 \alpha_2-1)-\alpha_2+1}{\alpha_1+\alpha_2-1}\left(K_0\left({b{Q} \sqrt{(\alpha_1-1) (\alpha_2-1)}}\right)-K_0\left({b{Q} \sqrt{\alpha_1 \alpha_2}}\right)\right),
\end{equation}

\begin{equation}
	F_5^{\text{tree}}=-{Q}^2 \left(K_0\left({b{Q} \sqrt{(\alpha_1-1) (\alpha_2-1)}}\right)+K_0\left({b{Q} \sqrt{\alpha_1 \alpha_2}}\right)\right).
\end{equation}
where $K_0$ is the modified Bessel function of order zero. 

The same treatment of transverse momentum is adopted for the $TJJ$ correction. In this case the insertion modifies the gluon propagator structure while preserving the factorized dependence on the impact parameter. The resulting correction kernel can therefore be written in a form analogous to the tree-level contribution, with the transverse dynamics entirely governed by the Fourier transform of the corrected gluon propagator. In evaluating the hard-scattering kernel corrected by the $TJJ$, we employ the $\overline{\text{MS}}$ scheme. Accordingly, the scale $\mu$ featuring in the loop integrals and logarithms represents the renormalization scale, to be distinguished from the factorization scale governing the DAs.
In this framework the corrections to the kernel due to the $TJJ$ insertion are given by
\begin{equation}
	K_{{TJJ}}^{\mu\nu}(\alpha_1,\alpha_2) = F_1^{{TJJ}}\, P_1^{\mu } P_2^{\nu } 
	+ F_2^{{TJJ}}\, P_2^{\mu } P_1^{\nu } 
	+ F_3^{{TJJ}}\, P_1^{\mu } P_1^{\nu } 
	+ F_4^{{TJJ}}\, P_2^{\mu } P_2^{\nu } 
	+ F_5^{{TJJ}}\, g^{\mu\nu},
\end{equation}
\begin{align}
	F_1^{TJJ}=F_2^{TJJ}
	=&
	\frac{\KBessel}{\Den}\nonumber\\&
	\biggl[
	\frac{\bar C_{11}\, g_s^2}
	{144\pi^2(\alpha_1-\alpha_2)^2}
	-
	\frac{g_s^2}
	{\pi^2(\alpha_1-\alpha_2)^4}
	\biggl(
	\frac{\bar C_{12}}{8}\,\Cz\, Q^2
	\nonumber\\&
	+\frac{\bar C_{13}}{24}\,\Lq
	+\frac{\bar C_{14}}{24}\,\La
	+\frac{\bar C_{15}}{24}\,\Lb
	\biggl)
	\biggl]
\end{align}

where
\begin{align}
	\bar C_{11}
	&=
	-(65C_A-23n_f)(1-2\alpha_1)^2\alpha_2^4
	\nonumber\\
	&\quad
	+
	(2\alpha_1-1)
	\Bigl[
	(4(3-26\alpha_1)\alpha_1-47)n_f
	+
	(4(68\alpha_1-3)\alpha_1+161)C_A
	\Bigr]\alpha_2^3
	\nonumber\\
	&\quad
	+
	\Bigl[
	n_f\alpha_1\bigl(2(46\alpha_1^2+64\alpha_1+29)\alpha_1+1\bigr)
	\nonumber\\
	&\qquad
	-
	C_A\bigl((2\alpha_1(2(65\alpha_1+74)\alpha_1+215)-113)\alpha_1+54\bigr)
	\Bigr]\alpha_2^2
	\nonumber\\
	&\quad
	+
	\Bigl[
	(1-2\alpha_1(46\alpha_1+53))\alpha_1 n_f
	-36n_f
	\nonumber\\
	&\qquad
	+
	\bigl((260\alpha_1^2+334\alpha_1+113)\alpha_1+144\bigr)C_A
	\Bigr]\alpha_1\alpha_2
	\nonumber\\
	&\quad
	+
	\Bigl[
	n_f\alpha_1(23\alpha_1+47)
	-
	C_A(5\alpha_1+2)(13\alpha_1+27)
	\Bigr]\alpha_1^2 ,
\end{align}

\begin{align}
	\bar C_{12}
	&=
	2(\alpha_1-1)(\alpha_2-1)n_f\alpha_1\alpha_2
	\Bigl(
	(2\alpha_2((2\alpha_2-3)\alpha_2+4)-3)\alpha_1^3
	\nonumber\\
	&\qquad
	+\left((-6\alpha_2^2+2\alpha_2-5)\alpha_2+1\right)\alpha_1^2
	\nonumber\\
	&\qquad
	+((8\alpha_2-5)\alpha_2+4)\alpha_2\alpha_1
	-3\alpha_2^3+\alpha_2^2
	\Bigr)
	\nonumber\\
	&\quad
	+C_A\Bigl(
	(4(\alpha_2-1)\alpha_2+3)\alpha_1^6
	\nonumber\\
	&\qquad
	-(4((\alpha_2((2\alpha_2-5)\alpha_2+16)-16)\alpha_2+7)\alpha_2+1)\alpha_1^5
	\nonumber\\
	&\qquad
	+(((4\alpha_2((5\alpha_2+7)\alpha_2+1)-23)\alpha_2+11)\alpha_2+2)\alpha_1^4
	\nonumber\\
	&\qquad
	+2(2(-16\alpha_2^2+\alpha_2-4)\alpha_2+9)\alpha_2^2\alpha_1^3
	\nonumber\\
	&\qquad
	+(\alpha_2((4\alpha_2(\alpha_2+16)-23)\alpha_2+18)-16)\alpha_2^2\alpha_1^2
	\nonumber\\
	&\qquad
	+(11-4\alpha_2(\alpha_2+7))\alpha_2^4\alpha_1
	+\left((3\alpha_2-1)\alpha_2+2\right)\alpha_2^4
	\Bigr),
\end{align}

\begin{align}
	\bar C_{13}
	&=
	2n_f\Bigl(
	((\alpha_2-1)(12(\alpha_2-1)\alpha_2+19)\alpha_2+2)\alpha_1^4
	\nonumber\\
	&\qquad
	-4\alpha_2(\alpha_2((6\alpha_2-7)\alpha_2+8)-5)\alpha_1^3
	\nonumber\\
	&\qquad
	+(\alpha_2((31\alpha_2-32)\alpha_2+22)-9)\alpha_2\alpha_1^2
	\nonumber\\
	&\qquad
	+((20-19\alpha_2)\alpha_2-9)\alpha_2^2\alpha_1
	+2\alpha_2^4
	\Bigr)
	\nonumber\\
	&\quad
	+C_A\Bigl(
	(-2(\alpha_2-1)(12(\alpha_2-1)\alpha_2+49)\alpha_2-7)\alpha_1^4
	\nonumber\\
	&\qquad
	+(2\alpha_2(2\alpha_2(4(3\alpha_2+4)\alpha_2+1)-29)-15)\alpha_1^3
	\nonumber\\
	&\qquad
	+(33-2\alpha_2((61\alpha_2-2)\alpha_2+1))\alpha_2\alpha_1^2
	\nonumber\\
	&\qquad
	+(98\alpha_2^2-58\alpha_2+33)\alpha_2^2\alpha_1
	-\alpha_2^3(7\alpha_2+15)
	\Bigr).
\end{align}

\begin{align}
	\bar C_{14}
	&=
	(2\alpha_1-1)
	\left(
	\alpha_1 n_f+n_f-C_A(\alpha_1+4)
	\right)
	\alpha_1\alpha_2^6
	\nonumber\\
	&\quad
	+\Bigl(
	2(((3-7\alpha_1)\alpha_1-8)\alpha_1+5)n_f
	+\left(\alpha_1(2(7\alpha_1+9)\alpha_1+37)-34\right)C_A
	\Bigr)\alpha_1\alpha_2^5
	\nonumber\\
	&\quad
	+\Bigl(
	(\alpha_1((2(41-7\alpha_1)\alpha_1-73)\alpha_1+62)-21)n_f\alpha_1
	\nonumber\\
	&\qquad
	+\Bigl(
	((\alpha_1(2(7\alpha_1-59)\alpha_1+49)-86)\alpha_1+36)\alpha_1+6
	\Bigr)C_A
	\Bigr)\alpha_2^4
	\nonumber\\
	&\quad
	+\Bigl(
	((\alpha_1((2\alpha_1(\alpha_1+3)-73)\alpha_1+60)-45)\alpha_1+6)n_f
	\nonumber\\
	&\qquad
	+\left(
	(\alpha_1(49-2(\alpha_1-9)\alpha_1)+36)\alpha_1+30
	\right)\alpha_1 C_A
	\Bigr)\alpha_1\alpha_2^3
	\nonumber\\
	&\quad
	+\Bigl(
	((\alpha_1((\alpha_1-16)\alpha_1+62)-45)\alpha_1+24)n_f
	\nonumber\\
	&\qquad
	+\left(
	\alpha_1(((37-7\alpha_1)\alpha_1-86)\alpha_1+30)-48
	\right)C_A
	\Bigr)\alpha_1^2\alpha_2^2
	\nonumber\\
	&\quad
	+\Bigl(
	2C_A\alpha_1((2\alpha_1-17)\alpha_1+18)
	\nonumber\\
	&\qquad
	-n_f((\alpha_1-7)(\alpha_1-3)\alpha_1-6)
	\Bigr)\alpha_1^3\alpha_2
	+6\alpha_1^4 C_A,
\end{align}

\begin{align}
	\bar C_{15}
	&=
	-n_f(\alpha_1-1)
	\Bigl(
	(\alpha_2-2)(\alpha_2-1)(2\alpha_2-1)\alpha_1^5
	\nonumber\\
	&\qquad
	-(\alpha_2(2(7\alpha_2-13)\alpha_2+21)-4)(\alpha_2-1)\alpha_1^4
	\nonumber\\
	&\qquad
	-(\alpha_2(2(7\alpha_2-15)\alpha_2+35)-28)(\alpha_2-1)\alpha_2\alpha_1^3
	\nonumber\\
	&\qquad
	+(\alpha_2-1)\alpha_2
	\Bigl(
	\alpha_2((2\alpha_2(\alpha_2+13)-35)\alpha_2+32)-18
	\Bigr)\alpha_1^2
	\nonumber\\
	&\qquad
	-(\alpha_2-1)
	\Bigl(
	(\alpha_2(5\alpha_2+21)-28)\alpha_2+18
	\Bigr)\alpha_2^2\alpha_1
	\nonumber\\
	&\qquad
	+2\alpha_2^4(\alpha_2^2+\alpha_2-2)
	\Bigr)
	\nonumber\\
	&\quad
	+C_A
	\Bigl(
	(\alpha_2-5)(\alpha_2-1)(2\alpha_2-1)\alpha_1^6
	\nonumber\\
	&\qquad
	-(\alpha_2-1)(\alpha_2(2(7\alpha_2-24)\alpha_2+49)-5)\alpha_1^5
	\nonumber\\
	&\qquad
	+\Bigl(
	(\alpha_2((2(11-7\alpha_2)\alpha_2-57)\alpha_2+173)-119)\alpha_2+1
	\Bigr)\alpha_1^4
	\nonumber\\
	&\qquad
	+\Bigl(
	((\alpha_2(\alpha_2(2\alpha_2(\alpha_2+31)-57)-60)-44)\alpha_2+58)\alpha_2+15
	\Bigr)\alpha_1^3
	\nonumber\\
	&\qquad
	-\alpha_2
	\Bigl(
	(\alpha_2((\alpha_2(13\alpha_2+97)-173)\alpha_2+44)-50)\alpha_2+33
	\Bigr)\alpha_1^2
	\nonumber\\
	&\qquad
	+\Bigl(
	\alpha_2((2\alpha_2(8\alpha_2+27)-119)\alpha_2+58)-33
	\Bigr)\alpha_2^2\alpha_1
	\nonumber\\
	&\qquad
	+\left(-5\alpha_2^3-5\alpha_2^2+\alpha_2+15\right)\alpha_2^3
	\Bigr).
\end{align}

\begin{align}
	F_3^{TJJ}=F_4^{TJJ}(\alpha_1\leftrightarrow\alpha_2)
	=&
	\frac{\KBessel}{\Den}\nonumber\\&
	\times\biggl[
	\frac{\bar C_{31}\, g_s^2}
	{72\pi^2(\alpha_1-\alpha_2)^2}
	-
	\frac{g_s^2}
	{\pi^2(\alpha_1-\alpha_2)^4}
	\biggl(
	\frac{\bar C_{32}}{4}\,\Cz\, Q^2
	\nonumber\\&
	+\frac{\bar C_{33}}{24}\,\Lq
	+\frac{\bar C_{34}}{24}\,\La
	+\frac{\bar C_{35}}{24}\,\Lb
	\biggl)
	\biggl]
\end{align}

with
\begin{align}
	\bar C_{31}
	&=
	(\alpha_1-1)(2\alpha_1-1)(65C_A-23n_f)\alpha_1\alpha_2^3
	\nonumber\\
	&\quad
	+
	\Bigl[
	((\alpha_1-1)\alpha_1((104\alpha_1-35)\alpha_1+20)-3)n_f
	\nonumber\\
	&\qquad
	+
	\bigl(
	12-(\alpha_1-1)\alpha_1((272\alpha_1-77)\alpha_1+116)
	\bigr)C_A
	\Bigr]\alpha_2^2
	\nonumber\\
	&\quad
	+
	\Bigl[
	(\alpha_1(((11-46\alpha_1)\alpha_1-17)\alpha_1+64)-6)n_f
	\nonumber\\
	&\qquad
	+
	(\alpha_1(\alpha_1((130\alpha_1-53)\alpha_1+167)-256)-12)C_A
	\Bigr]\alpha_1\alpha_2
	\nonumber\\
	&\quad
	+
	\Bigl[
	((\alpha_1-1)\alpha_1(23\alpha_1+32)-3)n_f
	\nonumber\\
	&\qquad
	+
	\bigl(
	12-(\alpha_1-1)\alpha_1(65\alpha_1+128)
	\bigr)C_A
	\Bigr]\alpha_1^2 ,
\end{align}
\begin{align}
	\bar C_{32}
	&=
	-\biggl((\alpha_1-1)\alpha_1
	\Bigl(
	C_A\Bigl(
	3(2\alpha_1-1)\alpha_2^5
	+\left((5-4\alpha_1((\alpha_1-2)\alpha_1+6))\alpha_1+1\right)\alpha_2^4
	\nonumber\\
	&\qquad
	+2(4(\alpha_1+2)\alpha_1^2+1)\alpha_1\alpha_2^3
	-4\alpha_1(\alpha_1+1)((5\alpha_1-3)\alpha_1+1)\alpha_2^2
	\nonumber\\
	&\qquad
	+(\alpha_1(2\alpha_1+11)-2)\alpha_1^3\alpha_2
	-\alpha_1^4(\alpha_1+1)
	\Bigr)
	\nonumber\\
	&\qquad
	-n_f(\alpha_2-1)\alpha_2
	\Bigl(
	-2\alpha_1^4+\alpha_1^3+\alpha_1^2
	+\left((4\alpha_1^2-2\alpha_1-3)\alpha_1+2\right)\alpha_2\alpha_1
	\nonumber\\
	&\qquad\qquad
	+(2\alpha_1-1)\alpha_2^3
	-\left(2(2(\alpha_1-2)\alpha_1+3)\alpha_1^2+\alpha_1-1\right)\alpha_2^2
	\Bigr)
	\Bigr)\biggl),
\end{align}

\begin{align}
	\bar C_{33}
	&=
	-\biggl(\left((-12(\alpha_1-1)\alpha_1-1)n_f
	+(24(\alpha_1-1)\alpha_1-2)C_A\right)\alpha_2^4
	\nonumber\\
	&\quad
	+\Bigl(
	2\Bigl(
	(2\alpha_1^2(3(2\alpha_1-5)\alpha_1+14)-9)n_f\alpha_1
	\nonumber\\
	&\qquad
	+\bigl((33-2\alpha_1(3(2\alpha_1-5)\alpha_1+29))\alpha_1+9\bigr)C_A\alpha_1
	+C_A
	\Bigr)+n_f
	\Bigr)\alpha_2^3
	\nonumber\\
	&\quad
	+\Bigl(
	(11-2\alpha_1(2((9\alpha_1-14)\alpha_1+4)\alpha_1+9))n_f
	\nonumber\\
	&\qquad
	+2(\alpha_1((2\alpha_1(9\alpha_1+7)-31)\alpha_1+3)-7)C_A
	\Bigr)\alpha_1\alpha_2^2
	\nonumber\\
	&\quad
	+\Bigl(
	(2(2\alpha_1(5\alpha_1-4)-7)\alpha_1+11)n_f
	\nonumber\\
	&\qquad
	+(2\alpha_1(\alpha_1(11-28\alpha_1)+25)-14)C_A
	\Bigr)\alpha_1^2\alpha_2
	\nonumber\\
	&\quad
	-(\alpha_1-1)\alpha_1^3
	\Bigl(
	4\alpha_1 n_f+n_f+(2-22\alpha_1)C_A
	\Bigr)\biggl),
\end{align}
\begin{align}
	\bar C_{34}
	&=
	\alpha_1
	\Biggl[
	n_f\alpha_2
	\Bigl(
	-2(\alpha_1^2-1)\alpha_1\alpha_2^4
	+\bigl(2\alpha_1((7(\alpha_1-1)\alpha_1-9)\alpha_1+13)-9\bigr)\alpha_2^3
	\nonumber\\
	&\qquad
	+\Bigl(
	(2\alpha_1(\alpha_1((7\alpha_1-41)\alpha_1+57)-17)-15)\alpha_1+6
	\Bigr)\alpha_2^2
	\nonumber\\
	&\qquad
	+\Bigl(
	(2\alpha_1((-\alpha_1^2+\alpha_1+5)\alpha_1+1)-27)\alpha_1+12
	\Bigr)\alpha_1\alpha_2
	\nonumber\\
	&\qquad
	+\Bigl(
	\alpha_1(3-2(\alpha_1-3)^2\alpha_1)+6
	\Bigr)\alpha_1^2
	\Bigr)
	\nonumber\\
	&\quad
	+2C_A
	\Bigl(
	(\alpha_1-1)(\alpha_1+4)\alpha_1\alpha_2^5
	+\bigl(3-\alpha_1((\alpha_1(7\alpha_1+5)-21)\alpha_1+13)\bigr)\alpha_2^4
	\nonumber\\
	&\qquad
	+\bigl(((59-7\alpha_1)\alpha_1-87)\alpha_1+38\bigr)\alpha_1^2\alpha_2^3
	\nonumber\\
	&\qquad
	+\bigl(
	\alpha_1((\alpha_1((\alpha_1-13)\alpha_1+31)-40)\alpha_1+30)-12
	\bigr)\alpha_1\alpha_2^2
	\nonumber\\
	&\qquad
	+\bigl(
	\alpha_1((4\alpha_1-21)\alpha_1+24)-6
	\bigr)\alpha_1^3\alpha_2
	+3(\alpha_1-1)\alpha_1^4
	\Bigr)
	\Biggr].
\end{align}

\begin{align}
	\bar C_{35}
	&=
	(1-\alpha_1)
	\Biggl[
	2C_A
	\Bigl(
	(\alpha_2-5)(\alpha_2-1)\alpha_1^6
	-(\alpha_2-2)(7(\alpha_2-2)\alpha_2+4)\alpha_1^5
	\nonumber\\
	&\qquad
	+\bigl(
	\alpha_2((\alpha_2(4-7\alpha_2)+11)\alpha_2+15)-11
	\bigr)\alpha_1^4
	\nonumber\\
	&\qquad
	+\bigl(
	((\alpha_2(\alpha_2(\alpha_2+28)-43)-23)\alpha_2+18)\alpha_2+1
	\bigr)\alpha_1^3
	\nonumber\\
	&\qquad
	+\bigl(
	\alpha_2((23-6\alpha_2(\alpha_2+1))\alpha_2+8)-7
	\bigr)\alpha_2\alpha_1^2
	\nonumber\\
	&\qquad
	+\bigl(
	\alpha_2((5\alpha_2-11)\alpha_2+10)-7
	\bigr)\alpha_2^2\alpha_1
	-(\alpha_2-1)\alpha_2^3
	\Bigr)
	\nonumber\\
	&\quad
	-n_f(\alpha_2-1)
	\Bigl(
	2(\alpha_2-2)\alpha_1^6
	-2(\alpha_2-1)(7\alpha_2-2)\alpha_1^5
	\nonumber\\
	&\qquad
	+\bigl(
	4-2\alpha_2((7\alpha_2-15)\alpha_2+4)
	\bigr)\alpha_1^4
	\nonumber\\
	&\qquad
	+\bigl(
	2(\alpha_2(\alpha_2(\alpha_2+17)-20)-4)\alpha_2+1
	\bigr)\alpha_1^3
	\nonumber\\
	&\qquad
	+\left(-6\alpha_2^3-8\alpha_2+11\right)\alpha_2\alpha_1^2
	\nonumber\\
	&\qquad
	+\left(4(\alpha_2-3)\alpha_2+11\right)\alpha_2^2\alpha_1
	+\alpha_2^3
	\Bigr)
	\Biggr].
\end{align}

\begin{align}
	F_5^{TJJ}=
	&\frac{\KBessel}{\Den}\nonumber\\
	&\times
	\biggl[
	\frac{	\bar C_{51}\, g_s^2 Q^2}
	{48\pi^2(\alpha_1-\alpha_2)^2}
	+
	\frac{g_s^2}
	{\pi^2(\alpha_1-\alpha_2)^4}
	\biggl(
	\frac{\bar C_{52}}{16}\,\Cz\, Q^4\nonumber\\
	&
	-\frac{\bar C_{53}}{48}\,\Lq\, Q^2
	-\frac{\bar C_{54}}{48}\,\La\, Q^2
	-\frac{\bar C_{55}}{48}\,\Lb\, Q^2
	\biggl)
	\biggl]
\end{align}

where
\begin{align}
	\bar C_{51}
	&=
	-\Bigl[
	n_f\,\bigl(
	2\alpha_1^2-(2\alpha_2+1)\alpha_1+(2\alpha_2-1)\alpha_2
	\bigr)
	\bigl(\alpha_1(2\alpha_2-1)-\alpha_2\bigr)
	\nonumber\\
	&\qquad
	+
	C_A\Bigl(
	(5-10\alpha_2)\alpha_1^3
	+\bigl((16\alpha_2-1)\alpha_2+2\bigr)\alpha_1^2
	-\alpha_2(10\alpha_2^2+\alpha_2+8)\alpha_1
	+(5\alpha_2+2)\alpha_2^2
	\Bigr)
	\Bigr],
\end{align}
\begin{align}
	\bar C_{52}
	&=
	C_A\Bigl(
	-2\alpha_1^6
	+\left(2(4\alpha_2^2-6\alpha_2+5)\alpha_2+3\right)\alpha_1^5
	\nonumber\\
	&\quad
	-\left(\left(2(4\alpha_2^2+2\alpha_2-5)\alpha_2+13\right)\alpha_2+1\right)\alpha_1^4
	\nonumber\\
	&\quad
	+2\left((2\alpha_2-1)(2\alpha_2^2+3)\alpha_2+2\right)\alpha_2\alpha_1^3
	\nonumber\\
	&\quad
	+2\left(((5-6\alpha_2)\alpha_2-3)\alpha_2+1\right)\alpha_2^2\alpha_1^2
	\nonumber\\
	&\quad
	+(5\alpha_2-4)(2\alpha_2-1)\alpha_2^3\alpha_1
	+\left((3-2\alpha_2)\alpha_2-1\right)\alpha_2^4
	\Bigr)
	\nonumber\\
	&\quad
	-2n_f(\alpha_1-1)\alpha_1(\alpha_2-1)\alpha_2
	\Bigl((\alpha_1-1)\alpha_1+(\alpha_2-1)\alpha_2\Bigr)
	\Bigl(\alpha_1(2\alpha_2-1)-\alpha_2\Bigr),
\end{align}

\begin{align}
	\bar C_{53}
	&=
	-n_f\Bigl((\alpha_1-1)\alpha_1+(\alpha_2-1)\alpha_2\Bigr)
	\Bigl(
	(12(\alpha_2-1)\alpha_2+1)\alpha_1^2
	\nonumber\\
	&\qquad
	+2\alpha_2(5-6\alpha_2)\alpha_1
	+\alpha_2^2
	\Bigr)
	\nonumber\\
	&\quad
	-2C_A\Bigl(
	(1-12(\alpha_2-1)\alpha_2)\alpha_1^4
	+\left(2\alpha_2(6\alpha_2^2+3\alpha_2-10)-1\right)\alpha_1^3
	\nonumber\\
	&\qquad
	+\left(2(-6\alpha_2^2+3\alpha_2+1)\alpha_2+7\right)\alpha_2\alpha_1^2
	\nonumber\\
	&\qquad
	+(6\alpha_2-7)(2\alpha_2-1)\alpha_2^2\alpha_1
	+(\alpha_2-1)\alpha_2^3
	\Bigr),
\end{align}

\begin{align}
	\bar C_{54}
	&=
	n_f\alpha_1\alpha_2
	\Bigl(
	4(\alpha_2-1)\alpha_1^4
	+(8(\alpha_2-2)\alpha_2+9)\alpha_1^3
	\nonumber\\
	&\qquad
	+(\alpha_2(8(\alpha_2-4)\alpha_2+27)-6)\alpha_1^2
	\nonumber\\
	&\qquad
	+(\alpha_2(4(\alpha_2-4)\alpha_2+27)-12)\alpha_2\alpha_1
	\nonumber\\
	&\qquad
	+\left((9-4\alpha_2)\alpha_2-6\right)\alpha_2^2
	\Bigr)
	\nonumber\\
	&\quad
	+C_A\Bigl(
	(3-10(\alpha_2-1)\alpha_2)\alpha_1^5
	-\left((2(\alpha_2-8)\alpha_2+21)\alpha_2+3\right)\alpha_1^4
	\nonumber\\
	&\qquad
	-2\alpha_2\left((\alpha_2-9)(\alpha_2-1)\alpha_2-6\right)\alpha_1^3
	\nonumber\\
	&\qquad
	+2\left(((8-5\alpha_2)\alpha_2-9)\alpha_2+3\right)\alpha_2^2\alpha_1^2
	\nonumber\\
	&\qquad
	+\left((10\alpha_2-21)\alpha_2+12\right)\alpha_2^3\alpha_1
	+3(\alpha_2-1)\alpha_2^4
	\Bigr),
\end{align}

\begin{align}
	\bar C_{55}
	&=
	C_A\Bigl(
	(10(\alpha_2-1)\alpha_2-3)\alpha_1^5
	+\left((2(\alpha_2-20)\alpha_2+45)\alpha_2+5\right)\alpha_1^4
	\nonumber\\
	&\qquad
	+2\left(\alpha_2(\alpha_2((\alpha_2+2)\alpha_2+15)-26)-1\right)\alpha_1^3
	\nonumber\\
	&\qquad
	+2\left((5(\alpha_2-3)(\alpha_2-1)\alpha_2-1)\alpha_2+7\right)\alpha_2\alpha_1^2
	\nonumber\\
	&\qquad
	+\left((5(9-2\alpha_2)\alpha_2-52)\alpha_2+14\right)\alpha_2^2\alpha_1
	\nonumber\\
	&\qquad
	+\left((5-3\alpha_2)\alpha_2-2\right)\alpha_2^3
	\Bigr)
	\nonumber\\
	&\quad
	-n_f(\alpha_1-1)(\alpha_2-1)
	\Bigl(
	4\alpha_2\alpha_1^4
	+(8(\alpha_2-2)\alpha_2+1)\alpha_1^3
	\nonumber\\
	&\qquad
	+(8(\alpha_2-2)\alpha_2+11)\alpha_2\alpha_1^2
	\nonumber\\
	&\qquad
	+(4(\alpha_2-4)\alpha_2+11)\alpha_2^2\alpha_1
	+\alpha_2^3
	\Bigr).
\end{align}

In every expression the massless three-point scalar function $C_0(p_1,p_2,p_3,0,0,0)$, also referred to as $C_0(p_1,p_2,p_3)$ for three arbitrary momenta $p_1$, $p_2$ and $p_3$, is given by
\begin{align}\label{chza:c0formula}
	C_0(p_1,p_2,p_3,0,0,0)=\frac{1}{\Delta}
	\Bigg[&
	Li_2\!\left(-\frac{-x_1+\Delta}{x_1+\Delta}\right)
	+
	Li_2\!\left(-\frac{x_2+\Delta}{-x_2+\Delta}\right)
	+
	Li_2\!\left(-\frac{x_3+\Delta}{-x_3+\Delta}\right)
	\nonumber\\&
	-
	Li_2\!\left(\frac{x_2-\Delta}{x_2+\Delta}\right)
	-
	Li_2\!\left(\frac{-x_1-\Delta}{-x_1+\Delta}\right)
	-
	Li_2\!\left(\frac{x_3-\Delta}{x_3+\Delta}\right)
	\Bigg],
\end{align}
where we have defined
\begin{equation}
\Delta
\equiv
\sqrt{
	p_1^2-2(p_2+p_3)p_1+(p_2-p_3)^2
}
\end{equation}
and 
\begin{equation}
x_1 \equiv p_1-p_2-p_3,
\qquad
x_2 \equiv p_1-p_2+p_3,
\qquad
x_3 \equiv p_1+p_2-p_3 .
\end{equation}

\providecommand{\href}[2]{#2}\begingroup\raggedright\endgroup


\begin{thebibliography}{10}

\bibitem{Burkert:2018bqq}
V.D.~Burkert, L.~Elouadrhiri and F.X.~Girod, \emph{{The pressure distribution
  inside the proton}},
  \href{https://doi.org/10.1038/s41586-018-0060-z}{\emph{Nature} {\bfseries
  557} (2018) 396}.

\bibitem{Duran:2022xag}
B.~Duran et~al., \emph{{Determining the gluonic gravitational form factors of
  the proton}}, \href{https://doi.org/10.1038/s41586-023-05730-4}{\emph{Nature}
  {\bfseries 615} (2023) 813}
  [\href{https://arxiv.org/abs/2207.05212}{{\ttfamily 2207.05212}}].

\bibitem{Belle:2015oin}
{\scshape Belle} collaboration, \emph{{Study of $\pi^0$ pair production in
  single-tag two-photon collisions}},
  \href{https://doi.org/10.1103/PhysRevD.93.032003}{\emph{Phys. Rev. D}
  {\bfseries 93} (2016) 032003}
  [\href{https://arxiv.org/abs/1508.06757}{{\ttfamily 1508.06757}}].

\bibitem{Savinov:2013hda}
{\scshape Belle} collaboration, \emph{{Measurement of $\gamma \gamma^{*} \to
  \pi^{0}$ transition form factor at Belle}},
  \href{https://doi.org/10.1016/j.nuclphysbps.2012.12.033}{\emph{Nucl. Phys. B
  Proc. Suppl.} {\bfseries 234} (2013) 287}.

\bibitem{Kumano:2017lhr}
S.~Kumano, Q.-T.~Song and O.V.~Teryaev, \emph{{Hadron tomography by generalized
  distribution amplitudes in pion-pair production process $\gamma^* \gamma
  \rightarrow \pi^0 \pi^0 $ and gravitational form factors for pion}},
  \href{https://doi.org/10.1103/PhysRevD.97.014020}{\emph{Phys. Rev. D}
  {\bfseries 97} (2018) 014020}
  [\href{https://arxiv.org/abs/1711.08088}{{\ttfamily 1711.08088}}].

\bibitem{JeffersonLabHallA:2022pnx}
{\scshape Jefferson Lab Hall A} collaboration, \emph{{Deeply Virtual Compton
  Scattering Cross Section at High Bjorken xB}},
  \href{https://doi.org/10.1103/PhysRevLett.128.252002}{\emph{Phys. Rev. Lett.}
  {\bfseries 128} (2022) 252002}
  [\href{https://arxiv.org/abs/2201.03714}{{\ttfamily 2201.03714}}].

\bibitem{CLAS:2022syx}
{\scshape CLAS} collaboration, \emph{{First CLAS12 Measurement of Deeply
  Virtual Compton Scattering Beam-Spin Asymmetries in the Extended Valence
  Region}}, \href{https://doi.org/10.1103/PhysRevLett.130.211902}{\emph{Phys.
  Rev. Lett.} {\bfseries 130} (2023) 211902}
  [\href{https://arxiv.org/abs/2211.11274}{{\ttfamily 2211.11274}}].

\bibitem{AbdulKhalek:2021gbh}
R.~Abdul~Khalek et~al., \emph{{Science Requirements and Detector Concepts for
  the Electron-Ion Collider}: {EIC Yellow Report}},
  \href{https://doi.org/10.1016/j.nuclphysa.2022.122447}{\emph{Nucl. Phys. A}
  {\bfseries 1026} (2022) 122447}
  [\href{https://arxiv.org/abs/2103.05419}{{\ttfamily 2103.05419}}].

\bibitem{Amore:2004ng}
P.~Amore, C.~Corian\`o and M.~Guzzi, \emph{{Deeply virtual neutrino scattering
  (DVNS)}}, \href{https://doi.org/10.1088/1126-6708/2005/02/038}{\emph{JHEP}
  {\bfseries 02} (2005) 038}
  [\href{https://arxiv.org/abs/hep-ph/0404121}{{\ttfamily hep-ph/0404121}}].

\bibitem{Tong:2021ctu}
X.-B.~Tong, J.-P.~Ma and F.~Yuan, \emph{{Gluon gravitational form factors at
  large momentum transfer}},
  \href{https://doi.org/10.1016/j.physletb.2021.136751}{\emph{Phys. Lett. B}
  {\bfseries 823} (2021) 136751}
  [\href{https://arxiv.org/abs/2101.02395}{{\ttfamily 2101.02395}}].

\bibitem{Tong:2022zax}
X.-B.~Tong, J.-P.~Ma and F.~Yuan, \emph{{Perturbative calculations of
  gravitational form factors at large momentum transfer}},
  \href{https://doi.org/10.1007/JHEP10(2022)046}{\emph{JHEP} {\bfseries 10}
  (2022) 046} [\href{https://arxiv.org/abs/2203.13493}{{\ttfamily
  2203.13493}}].

\bibitem{Coriano:1993mr}
C.~Corian\`o, A.~Radyushkin and G.F.~Sterman, \emph{{QCD sum rules and Compton
  scattering}}, \href{https://doi.org/10.1016/0550-3213(93)90556-5}{\emph{Nucl.
  Phys. B} {\bfseries 405} (1993) 481}
  [\href{https://arxiv.org/abs/hep-ph/9301274}{{\ttfamily hep-ph/9301274}}].

\bibitem{Ji:1996ek}
X.~Ji, \emph{{Gauge-Invariant Decomposition of Nucleon Spin}},
  \href{https://doi.org/10.1103/PhysRevLett.78.610}{\emph{Phys. Rev. Lett.}
  {\bfseries 78} (1997) 610}
  [\href{https://arxiv.org/abs/hep-ph/9603249}{{\ttfamily hep-ph/9603249}}].

\bibitem{Radyushkin:1996nd}
A.V.~Radyushkin, \emph{{Asymmetric gluon distributions and hard diffractive
  electroproduction}},
  \href{https://doi.org/10.1016/0370-2693(96)00528-X}{\emph{Phys. Lett. B}
  {\bfseries 385} (1996) 333}
  [\href{https://arxiv.org/abs/hep-ph/9604317}{{\ttfamily hep-ph/9604317}}].

\bibitem{Radyushkin:1996ru}
A.V.~Radyushkin, \emph{{Scaling Limit of Deeply Virtual Compton Scattering}},
  \href{https://doi.org/10.1016/0370-2693(96)00932-8}{\emph{Phys. Lett. B}
  {\bfseries 380} (1996) 417}
  [\href{https://arxiv.org/abs/hep-ph/9605431}{{\ttfamily hep-ph/9605431}}].

\bibitem{Ji:1996nm}
X.~Ji, \emph{{Deeply virtual Compton scattering}},
  \href{https://doi.org/10.1103/PhysRevD.55.7114}{\emph{Phys. Rev. D}
  {\bfseries 55} (1997) 7114}
  [\href{https://arxiv.org/abs/hep-ph/9609381}{{\ttfamily hep-ph/9609381}}].

\bibitem{Collins:1996fb}
J.C.~Collins and A.~Freund, \emph{{Proof of factorization for deeply virtual
  Compton scattering in QCD}},
  \href{https://doi.org/10.1103/PhysRevD.59.074009}{\emph{Phys. Rev. D}
  {\bfseries 59} (1999) 074009}
  [\href{https://arxiv.org/abs/hep-ph/9611369}{{\ttfamily hep-ph/9611369}}].

\bibitem{Radyushkin:1997ki}
A.V.~Radyushkin, \emph{{Nonforward parton distributions}},
  \href{https://doi.org/10.1103/PhysRevD.56.5524}{\emph{Phys. Rev. D}
  {\bfseries 56} (1997) 5524}
  [\href{https://arxiv.org/abs/hep-ph/9704207}{{\ttfamily hep-ph/9704207}}].

\bibitem{Vanderhaeghen:1998uc}
M.~Vanderhaeghen, P.A.M.~Guichon and M.~Guidal, \emph{{Hard electroproduction
  of photons and mesons on the nucleon}},
  \href{https://doi.org/10.1103/PhysRevLett.80.5064}{\emph{Phys. Rev. Lett.}
  {\bfseries 80} (1998) 5064}
  [\href{https://arxiv.org/abs/hep-ph/9806305}{{\ttfamily hep-ph/9806305}}].

\bibitem{Coriano:2024qbr}
C.~Corian\`o, S.~Lionetti, D.~Melle and R.~Tommasi, \emph{{The Gravitational
  Form Factors of Hadrons from CFT in Momentum Space and the Dilaton in
  Perturbative QCD}},
  \href{https://doi.org/10.1140/epjc/s10052-025-14104-1}{\emph{Eur. Phys. J. C}
  {\bfseries 85} (2025) 498}
  [\href{https://arxiv.org/abs/2409.05609}{{\ttfamily 2409.05609}}].

\bibitem{Bzowski:2013sza}
A.~Bzowski, P.~McFadden and K.~Skenderis, \emph{{Implications of conformal
  invariance in momentum space}},
  \href{https://doi.org/10.1007/JHEP03(2014)111}{\emph{JHEP} {\bfseries 03}
  (2014) 111} [\href{https://arxiv.org/abs/1304.7760}{{\ttfamily 1304.7760}}].

\bibitem{Bzowski:2018fql}
A.~Bzowski, P.~McFadden and K.~Skenderis, \emph{{Renormalised CFT 3-point
  functions of scalars, currents and stress tensors}},
  \href{https://doi.org/10.1007/JHEP11(2018)159}{\emph{JHEP} {\bfseries 11}
  (2018) 159} [\href{https://arxiv.org/abs/1805.12100}{{\ttfamily
  1805.12100}}].

\bibitem{Sterman:1986aj}
G.~Sterman, \emph{Summation of large corrections to short distance hadronic
  cross-sections},
  \href{https://doi.org/10.1016/0550-3213(87)90258-6}{\emph{Nucl. Phys. B}
  {\bfseries 281} (1987) 310}.

\bibitem{Botts:1989kf}
J.~Botts and G.~Sterman, \emph{Hard elastic scattering in qcd: Leading
  behavior}, \href{https://doi.org/10.1016/0550-3213(89)90372-6}{\emph{Nucl.
  Phys. B} {\bfseries 325} (1989) 62}.

\bibitem{Li:1992nu}
H.-n.~Li and G.~Sterman, \emph{The perturbative pion form factor with sudakov
  suppression}, \href{https://doi.org/10.1016/0550-3213(92)90643-P}{\emph{Nucl.
  Phys. B} {\bfseries 381} (1992) 129}.

\bibitem{Li:1994cka}
H.-n.~Li and H.-L.~Yu, \emph{Perturbative qcd analysis of pion electromagnetic
  form factor}, \href{https://doi.org/10.1103/PhysRevD.53.2480}{\emph{Phys.
  Rev. D} {\bfseries 53} (1996) 2480}
  [\href{https://arxiv.org/abs/hep-ph/9411308}{{\ttfamily hep-ph/9411308}}].

\bibitem{Lepage:1980fj}
G.P.~Lepage and S.J.~Brodsky, \emph{Exclusive processes in perturbative quantum
  chromodynamics}, \href{https://doi.org/10.1103/PhysRevD.22.2157}{\emph{Phys.
  Rev. D} {\bfseries 22} (1980) 2157}.

\bibitem{Efremov:1979qk}
A.V.~Efremov and A.V.~Radyushkin, \emph{Factorization and asymptotical behavior
  of pion form factor in qcd},
  \href{https://doi.org/10.1016/0370-2693(80)90869-2}{\emph{Phys. Lett. B}
  {\bfseries 94} (1980) 245}.

\bibitem{Chernyak:1983ej}
V.L.~Chernyak and A.R.~Zhitnitsky, \emph{Asymptotic behavior of exclusive
  processes in qcd},
  \href{https://doi.org/10.1016/0370-1573(84)90126-1}{\emph{Phys. Rept.}
  {\bfseries 112} (1984) 173}.

\bibitem{Braun:1989iv}
V.M.~Braun and I.E.~Filyanov, \emph{Qcd sum rules in exclusive kinematics and
  pion wave function}, \href{https://doi.org/10.1007/BF01548594}{\emph{Z. Phys.
  C} {\bfseries 44} (1989) 157}.

\bibitem{Ball:2006wn}
P.~Ball, V.M.~Braun and A.~Lenz, \emph{Higher-twist distribution amplitudes of
  the pion in qcd},
  \href{https://doi.org/10.1088/1126-6708/2006/05/004}{\emph{JHEP} {\bfseries
  05} (2006) 004} [\href{https://arxiv.org/abs/hep-ph/0603063}{{\ttfamily
  hep-ph/0603063}}].

\bibitem{Jakob:1993iw}
R.~Jakob and P.~Kroll, \emph{The pion form factor: Sudakov suppressions and
  intrinsic transverse momentum},
  \href{https://doi.org/10.1016/0370-2693(93)91666-S}{\emph{Phys. Lett. B}
  {\bfseries 315} (1993) 463}
  [\href{https://arxiv.org/abs/hep-ph/9306259}{{\ttfamily hep-ph/9306259}}].

\bibitem{Huang:1994dy}
T.~Huang and Q.-X.~Shen, \emph{The perturbative qcd prediction for the pion
  form factor revisited}, \href{https://doi.org/10.1007/BF01548594}{\emph{Z.
  Phys. C} {\bfseries 50} (1991) 139}.

\bibitem{BAKULEV2001279}
A.~Bakulev, S.~Mikhailov and N.~Stefanis, \emph{Qcd-based pion distribution
  amplitudes confronting experimental data},
  \href{https://doi.org/https://doi.org/10.1016/S0370-2693(01)00517-2}{\emph{Physics
  Letters B} {\bfseries 508} (2001) 279}.

\bibitem{Hackett:2023pyn}
D.C.~Hackett, P.R.~Oare, D.A.~Pefkou and P.E.~Shanahan, \emph{Gravitational
  form factors of the pion from lattice qcd},
  \href{https://doi.org/10.1103/PhysRevD.108.114504}{\emph{Phys. Rev. D}
  {\bfseries 108} (2023) 114504}
  [\href{https://arxiv.org/abs/2307.11707}{{\ttfamily 2307.11707}}].

\bibitem{Liu:2024zahed}
W.-Y.~Liu, E.~Shuryak and I.~Zahed, \emph{Pion gravitational form factors in
  the qcd instanton vacuum. ii},
  \href{https://doi.org/10.1103/PhysRevD.110.054022}{\emph{Phys. Rev. D}
  {\bfseries 110} (2024) 054022}.

\bibitem{Armillis:2009pq}
R.~Armillis, C.~Corian\`{o} and L.~Delle~Rose, \emph{{Conformal Anomalies and
  the Gravitational Effective Action: The $TJJ$ Correlator for a Dirac
  Fermion}}, \href{https://doi.org/10.1103/PhysRevD.81.085001}{\emph{Phys.
  Rev.} {\bfseries D81} (2010) 085001}
  [\href{https://arxiv.org/abs/0910.3381}{{\ttfamily 0910.3381}}].

\bibitem{Armillis:2010qk}
R.~Armillis, C.~Corian\`o and L.~Delle~Rose, \emph{{Trace Anomaly, Massless
  Scalars and the Gravitational Coupling of QCD}},
  \href{https://doi.org/10.1103/PhysRevD.82.064023}{\emph{Phys. Rev.}
  {\bfseries D82} (2010) 064023}
  [\href{https://arxiv.org/abs/1005.4173}{{\ttfamily 1005.4173}}].

\bibitem{Coriano:2018zdo}
C.~Corian\`o and M.M.~Maglio, \emph{{Renormalization, Conformal Ward Identities
  and the Origin of a Conformal Anomaly Pole}},
  \href{https://doi.org/10.1016/j.physletb.2018.04.003}{\emph{Phys. Lett.}
  {\bfseries B781} (2018) 283}
  [\href{https://arxiv.org/abs/1802.01501}{{\ttfamily 1802.01501}}].

\bibitem{Coriano:2018bbe}
C.~Corian\`o and M.M.~Maglio, \emph{{Exact Correlators from Conformal Ward
  Identities in Momentum Space and the Perturbative $TJJ$ Vertex}},
  \href{https://doi.org/10.1016/j.nuclphysb.2018.11.016}{\emph{Nucl. Phys.}
  {\bfseries B938} (2019) 440}
  [\href{https://arxiv.org/abs/1802.07675}{{\ttfamily 1802.07675}}].

\bibitem{Coriano:2020ees}
C.~Corian\`o and M.M.~Maglio, \emph{{Conformal field theory in momentum space
  and anomaly actions in gravity: The analysis of three- and four-point
  function}}, \href{https://doi.org/10.1016/j.physrep.2021.11.005}{\emph{Phys.
  Rept.} {\bfseries 952} (2022) 2198}
  [\href{https://arxiv.org/abs/2005.06873}{{\ttfamily 2005.06873}}].

\bibitem{Coriano:2025sumrule}
C.~Corian\`o, S.~Lionetti, D.~Melle and L.~Torcellini, \emph{{A dilaton sum
  rule for the conformal anomaly form factor in QCD at order $\alpha_s$}},
  \href{https://doi.org/10.1140/epjc/s10052-025-14686-w}{\emph{Eur. Phys. J. C}
  {\bfseries 85} (2025) 983}
  [\href{https://arxiv.org/abs/2504.01904}{{\ttfamily 2504.01904}}].

\bibitem{Coriano:2026vya}
C.~Corian{\`o}, S.~Lionetti, D.~Melle and L.~Torcellini,
  \emph{{Anomaly-mediated Scalar Gravitational Interactions and the Coupling of
  Conformal Sectors}},  \href{https://arxiv.org/abs/2603.28966}{{\ttfamily
  2603.28966}}.

\bibitem{Giannotti:2008cv}
M.~Giannotti and E.~Mottola, \emph{{The Trace Anomaly and Massless Scalar
  Degrees of Freedom in Gravity}},
  \href{https://doi.org/10.1103/PhysRevD.79.045014}{\emph{Phys. Rev.}
  {\bfseries D79} (2009) 045014}
  [\href{https://arxiv.org/abs/0812.0351}{{\ttfamily 0812.0351}}].

\bibitem{Coriano:2017mux}
C.~Corian\`o, M.M.~Maglio and E.~Mottola, \emph{{TTT in CFT: Trace Identities
  and the Conformal Anomaly Effective Action}},
  \href{https://doi.org/10.1016/j.nuclphysb.2019.03.019}{\emph{Nucl. Phys.}
  {\bfseries B942} (2019) 303}
  [\href{https://arxiv.org/abs/1703.08860}{{\ttfamily 1703.08860}}].

\bibitem{Coriano:2014gja}
C.~Corian\`o, A.~Costantini, L.~Delle~Rose and M.~Serino, \emph{{Superconformal
  sum rules and the spectral density flow of the composite dilaton (ADD)
  multiplet in $\mathcal{N}=1$ theories}},
  \href{https://doi.org/10.1007/JHEP06(2014)136}{\emph{JHEP} {\bfseries 06}
  (2014) 136} [\href{https://arxiv.org/abs/1402.6369}{{\ttfamily 1402.6369}}].

\end{thebibliography}
\end{document}